\documentclass{thesis}

\usepackage[english]{babel}
\usepackage{csquotes}

\usepackage[backend=biber,
            bibencoding=utf8, 
            hyperref,
            style=numeric-comp,
            sorting=none,
            urldate=short,
            giveninits=true]{biblatex}
\usepackage{lipsum}
\usepackage{siunitx}
\usepackage{amsmath,scalerel}
\usepackage{amsmath}
\usepackage{multicol}
\usepackage{multirow}
\usepackage{graphicx}
\usepackage{tikz}
\usetikzlibrary{arrows.meta,positioning,calc}
\usepackage{adjustbox}
\usepackage{float}
\usepackage{placeins}
\usepackage[arrowdel]{physics}
\usepackage{caption}
\usepackage{subcaption}
\usepackage{textcomp, gensymb}
\usepackage[official]{eurosym}
\usepackage[raiselinks=true,
            bookmarks=true,
            bookmarksopenlevel=2,
            bookmarksopen=true,
            bookmarksnumbered=true,
            hyperindex=true,
            plainpages=false,
            pdfpagelabels=true,
            pdfpagemode=UseOutlines,
            pdfborder={0 0 0.5},
            colorlinks=false,
            allcolors=blue,
            linkbordercolor={0 0.61 0.50},   
            citebordercolor={0 0.61 0.50}]{hyperref}
\DeclareMathSizes{12}{12}{12}{12}
\usepackage{environ}
\usepackage[position=top]{subcaption}

\usepackage{enumitem}
\setlist[itemize]{itemsep=2pt, topsep=4pt, parsep=0pt, partopsep=0pt}

\usepackage[toc, acronym]{glossaries}
\newcommand{\acrfullinv}[1]{\acrshort{#1} (\acrlong{#1})}
\makeglossaries

\newcommand{\upperRomannumeral}[1]{\uppercase\expandafter{\romannumeral#1}}

\author{Yan Seyffert}

\begin{document}
\pagenumbering{roman}
\begin{titlepage}
    \centering
    \includegraphics[width=0.8\textwidth]{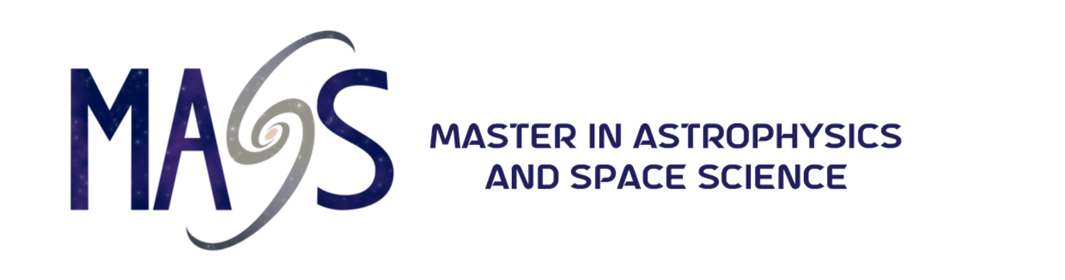}
    \vspace{1cm} % Adjust the vertical space as needed

    \large\textbf{Erasmus Mundus Master}\par
    \large\textbf{in Astrophysics and Space Science}\par
    \vspace{1cm} % Adjust the vertical space as needed

    \large\textbf{Master Thesis}\par
    
    \vspace{2cm} % Adjust the vertical space as needed
    
    \Large\textbf{Relativistic Time Modeling for \\Lunar Positioning Navigation and Timing}\par
%    \Large\textbf{Subtitle}\par %subtitle if need
    \vspace{2cm} % Adjust the vertical space as needed

    \begin{table}[!ht]
    {
    \centering
    \begin{tabular}{p{3.5in}p{2in}}
        Supervisors & Author \\
        ~ & ~ \\

        Prof. Dr. Agnès Fienga  & Yan Seyffert \\
        Geoazur, Université C\^ote d'Azur & ~ \\
        ~ & ~ \\
        
        Dr. Dennis Philipp  & ~ \\
        ZARM, Faculty of Physics, University of Bremen & ~ \\

    \end{tabular}}
    \end{table}
    
    \vspace{2cm} % Adjust the vertical space as needed

   \normalsize{Academic Year 2024/2025}
   \normalsize{\par Nice, France, September 2025}

    \vfill % Adjust the vertical space as needed
    \includegraphics[width=\textwidth]{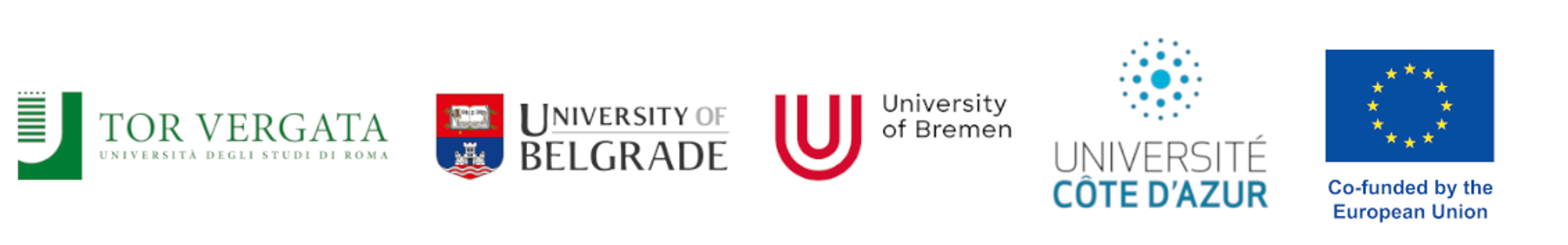}

\end{titlepage}
\newpage
\vspace*{\fill}
\begin{large}\thispagestyle{nohead}
This Master thesis is submitted in partial fulfillment of the requirements for the degree "Master Physique Fondamental et Applications, parcours MASS" as part of a multiple degree awarded in the framework of the Erasmus Mundus Joint Master in Astrophysics and Space Science -- MASS jointly delivered by a Consortium of four Universities: Tor Vergata University of Rome, University of Belgrade, University of Bremen, and Universit\'e C\^ote d’Azur, regulated by the MASS Consortium Agreement and funded by the EU under the call ERASMUS-EDU-2021-PEX-EMJM-MOB.
\end{large}
\vspace*{\fill}

\newpage
%\addcontentsline{toc}{section}{Abstract}

\begin{abstract}\thispagestyle{nohead}
Future lunar missions will depend on an internationally agreed upon timescale that remains accurate under the Moon's unique gravitational environment and its orbital dynamics. This thesis investigates the proposed \acrfull{tcl}, derived analogously to \acrfull{tcg} and thus aligned with current \acrshort{iau} proposals. We first formalise the TCL transformation and quantify its characteristics from solar system simulations. Next, we compute stationary surface-clock drifts caused by gravitational redshift and the Moon's changing orientation parameters, evaluating how accurate atomic clocks deployed on the surface of the Moon (much like for \glspl{esa} proposed NovaMoon mission) would have to be to measure these effects. Finally, we simulate relativistic proper time for \glspl{esa} Moonlight navigation satellites, identifying average drift and harmonic variations, to better understand the system that will comprise and enable a Lunar \acrfullinv{pnt} architecture. These kind of investigations are an essential step toward a sustained internationally cooperative operation at the lunar south pole and beyond.

\end{abstract}
\newpage
%\addcontentsline{toc}{section}{Contents}
{
\setlength{\parskip}{0pt}
\tableofcontents
}

\newpage
\pagenumbering{arabic}
%%============= V1 =============
%\include{chapters/Abstract}
%\include{chapters/1_Introduction}
% PART I: Foundations and State of the Art
%\include{chapters/2_History_Timekeeping}
%\include{chapters/3_Modern_Framework}
%\include{chapters/4_Emerging_Paradigms}
%\include{chapters/5_Tools_Sim_Environment}
% PART II: MY Contributions
%\include{chapters/6_Coordinate_Time}
%\include{chapters/7_Surface_Time}
%\include{chapters/8_Orbital_Time}
%PART III: Tieing it all together
%\include{chapters/9_Time_Synchronisation_Network}
%\include{chapters/Conclusion}

%============= V2 =============

%\addcontentsline{toc}{section}{Introduction}
\chapter{Introduction} \label{ch:intro}

Precise timekeeping is a cornerstone for modern space exploration, enabling accurate navigation and scientific measurement. While current systems on and around Earth, like the \acrfull{gnss}, achieve this by combining atomic clock ensembles with carefully calibrated relativistic corrections, extending these capabilities to the Moon introduces new challenges and corrections. With upcoming missions such as \acrshort{nasa}’s Artemis programme or \acrshort{esa}’s Argonaut lander and other commercial and state programs  -- aiming to establish long-term lunar infrastructure in the promising lunar south pole region -- the need for a dedicated and internationally recognized lunar time standard has become increasingly urgent.

Unlike Earth, the Moon has a weaker gravitational field, no atmosphere, and a non-trivial rotational and orbital interplay with Earth. These factors give rise to measurable relativistic effects; like a constant rate offset due to the Moon's higher position in Earth's gravitational well, periodic variations from orbital eccentricity, and perturbations from external celestial bodies. To address this, international bodies like the \acrfull{iau} are working toward a formal definition and practical implementation of a \acrfull{lcrs} with an associated \acrfull{tcl}, providing a standard for all future lunar operations.

This thesis contributes to these efforts in three ways. First (Sec.~\ref{ch:tcl}) by analyzing this proposed lunar timescale, defined analogously to \acrfull{tcg}, and comparing existing formulations in the literature.

Second (Sec.~\ref{ch:surface_clocks}) we study the behaviour of stationary clocks on the lunar surface -- exploring the impact of the Moon's gravitational field across varying selenographic heights. We assess whether the resulting redshift is measurable with current or planned atomic clock technologies, particularly in the context of proposed missions like \textit{NovaMoon}\cite{NovaMoon} on \glspl{esa} \textit{Argonaut} lander\cite{Argonaut} -- which aims to deploy a timing beacon in support of \glspl{esa} \textit{Moonlight} navigation system\cite{moonlight_video}. We also evaluate how and to what degree the Moon’s orientation, rotation and libration might influence clocks.

Third and finally (Sec.~\ref{ch:orbit_clocks}) we investigate clocks in orbit around the moon, specifically those in \acrfull{elfo} as planned for \glspl{esa} \textit{Moonlight} constellation. Using numerical simulations we implement relativistic corrections, to evaluate  how these clocks in free-fall would tick relative to other timescales.
\chapter{Revival of Lunar Exploration}\label{ch:lunar_x}
\section*{From Apollo to Artemis}
Following \acrshort{nasa}'s Apollo program (1969-1972), lunar exploration entered a decades-long hiatus. While scientific experiments like the \acrfull{llr} retroreflectors provided continuous data, crewed missions ceased entirely. China's methodical Chang'e program brought focus back to the moon, achieving milestones such as the first landing on the lunar far side (Chang'e-4, 2019) and the robotic return of samples (Chang'e-5, 2020). 

A pivotal shift occurred with the discovery of water ice in the Moon’s permanently shadowed south polar craters. \acrshort{nasa}’s \acrshort{lcross} Impactor (2009) and India’s Chandrayaan-1 (2008–2009) confirmed these deposits, transforming the Moon from a scientific curiosity into a strategic resource hub for fuel, oxygen, and life support. This revelation triggered a wave of missions targeting the lunar south pole.
\vspace{1em}
\begin{figure}[!htb]
    \centering
    \includegraphics[width=0.75\linewidth]{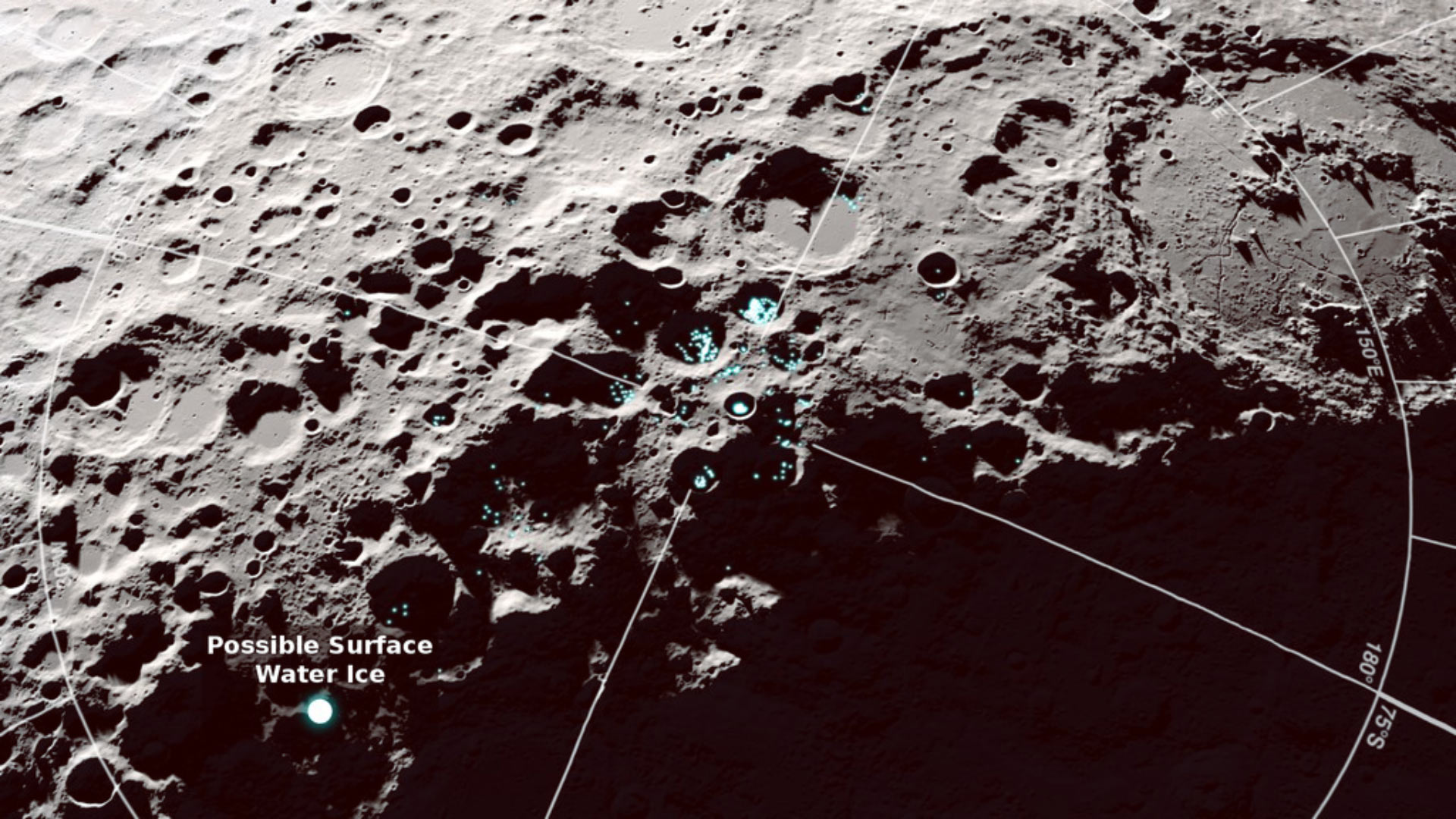}
    \caption{Render of the Moon’s south pole region showing where \acrshort{nasa}’s \acrfull{lro} had found on-going evidence in 2017, where data indicated the possible presence of surface water ice in the permanently shadowed regions of craters. Image credit: \acrshort{nasa}’s Scientific Visualization Studio.\protect\cite{nasa_waterice}}
    \label{fig:lunar_waterice}
\end{figure}

Since then, lunar missions have accelerated dramatically. Be it through national programs and missions, such as the aforementioned China (Chang'e robotic missions), India (Chandrayaan-3, 2023), Japan (SLIM, 2024), and Israel (Beresheet, 2019). Or be it private industry-driven missions like with Intuitive Machines (Nova-C lander IM1 2024 and IM2 2025) and Firefly Aerospace (Blue Ghost, 2025), who both delivered payloads under \acrshort{nasa}’s \acrfull{clps} initiative. 

\begin{figure}[htb]
    \centering
    \includegraphics[width=0.75\linewidth]{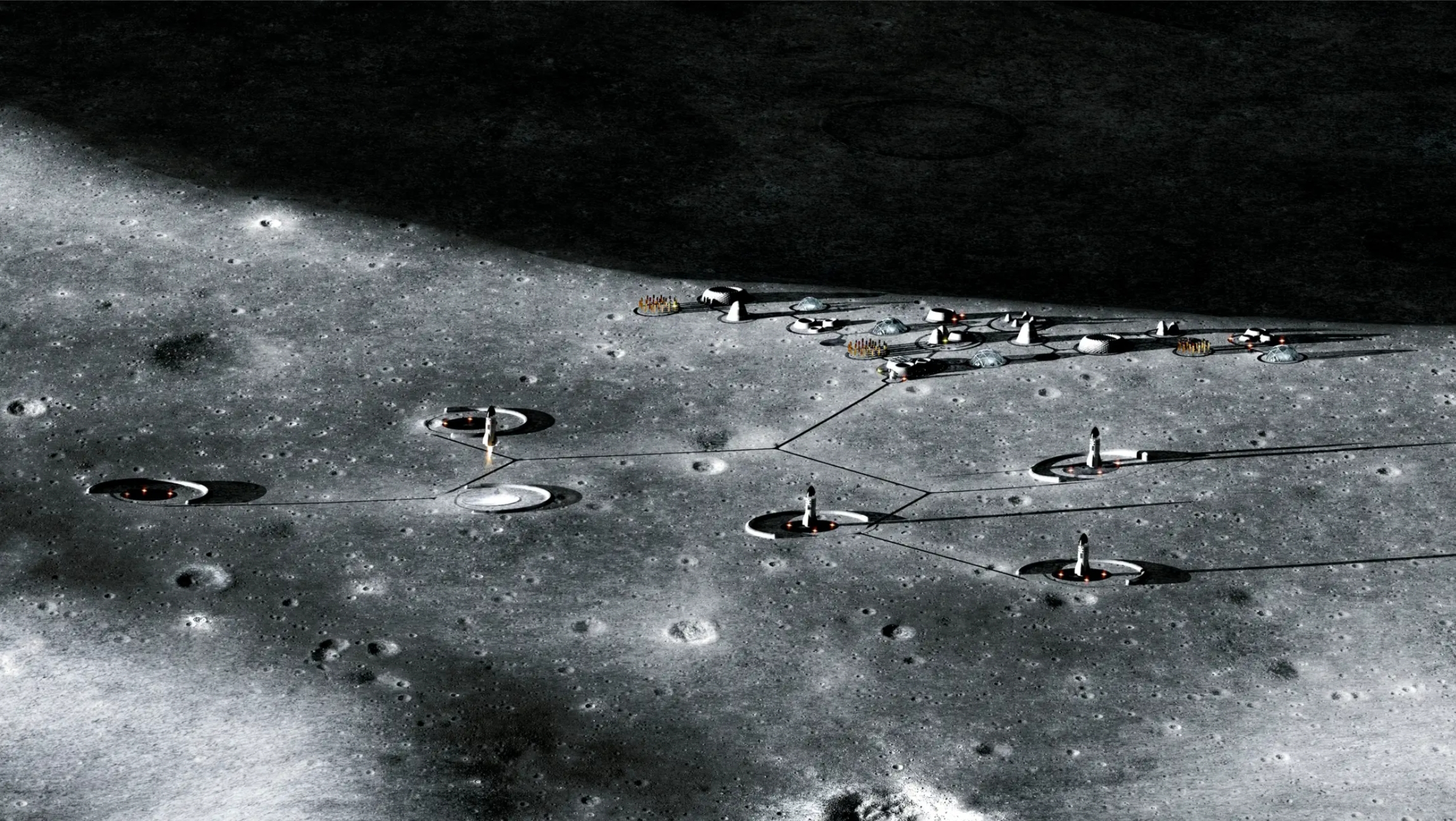}
    \caption{Project Olympus lunar habitat concept; envisaging 3D-printed structures from in-situ regolith to enable scalable and durable infrastructures for a sustained human presence on the Moon. Image Credit: BIG (Bjarke Ingels Group); client: ICON\protect\cite{ICON_Project_Olympus}; collaborators: NASA, SEArch+ (Space Exploration Architecture). Image Source: BIG project page\protect\cite{BIG_Project_Olympus}}
    \label{fig:moonbase}
\end{figure}

A new race to the Moon is said to have begun with \acrshort{nasa}, China, and private industry leading the charge. Both \acrshort{nasa} and China are committed to re-establishing a human presence on the lunar surface. \acrshort{nasa} is pursuing this goal through its \textit{Artemis} program\cite{Artemis} and the aforementioned \acrshort{clps} initiative, where private companies play a critical role in developing the landers and rockets needed to create a sustainable transportation system to the Moon. And a possible long-term lunar habitat might look like the one depicted in Fig.~\ref{fig:moonbase}.

The viability of augmenting lunar navigation with Earth-based systems was demonstrated in March 2025 when Firefly Aerospace's Blue Ghost Mission 1 successfully acquired terrestrial GPS/Galileo signals at lunar distance through \acrshort{nasa}'s \acrshort{lugre} experiment\cite{LuGRE_ION}.

Despite successes, many lander failures (Israel's Beresheet, India's Chandrayaan-2, ispace's Hakuto-R, Roscosmos's Luna 25, and Intuitive Machines partial failures), underscored the critical need for robust \acrfull{pnt} infrastructure to ensure safe operations in the Moon’s challenging environment.
\vspace{1cm}

%\FloatBarrier
\section*{LunaNet and ESA's Moonlight}
LunaNET\cite{LunaNet} -- being developed collaboratively by \acrshort{nasa}, \acrshort{esa}, and \acrshort{jaxa} -- aims to be the foundational architecture for lunar communications and navigation. This framework incorporates three critical capabilities: \acrfull{dtn} ensures reliable data transmission despite frequent signal disruptions; Autonomous Navigation delivers real-time positioning through \acrfull{afs}, enabling spacecraft and rovers to operate independently of Earth-based control; and integrated science and safety services providing solar storm alerts and \acrfull{lunasar} capabilities. 

%\begin{figure}[htb]
%    \centering
%    \includegraphics[width=0.75\linewidth]{figures/LunaNet.png}
%    \caption{Conceptualized LunaNET Framework\cite{LunaNet}}
%    \label{fig:lunanet}
%\end{figure}

The \acrfull{esa} complements this infrastructure through two initiatives: The \textit{Moonlight} constellation (targeting full operations by 2030\cite{Parsonson2025MoonlightTAS, ThalesAleniaSpace2025MoonlightPress}) will provide navigation services across the lunar surface, with prioritized coverage of the South Pole while maintaining interoperability with LunaNet's \acrshort{afs} standards. And second, further enhancing accuracy, the \textit{NovaMOON} surface station -- proposed for integration on ESA's \textit{Argonaut} lander later this decade -- could serve as an anchor node. 

However, before a lunar navigation system can be deployed, it is paramount to understand the underlying principles of \acrfull{pnt}. Uncertainties of clocks and thus signal delays in the couple of nanoseconds translate to meters of inaccuracy -- due to the speed of light of approximately \SI{1}{ft/ns} or \SI{30}{cm/ns}. 

The next section therefore recounts the fundamental framework for lunar operations and beyond: a unified relativistic time- and reference-frame standard that accounts for both gravitational and kinematic corrections across the solar system.
\vspace{1.5cm}
\begin{figure}[!bth]
    \centering
    \includegraphics[width=0.65\linewidth]{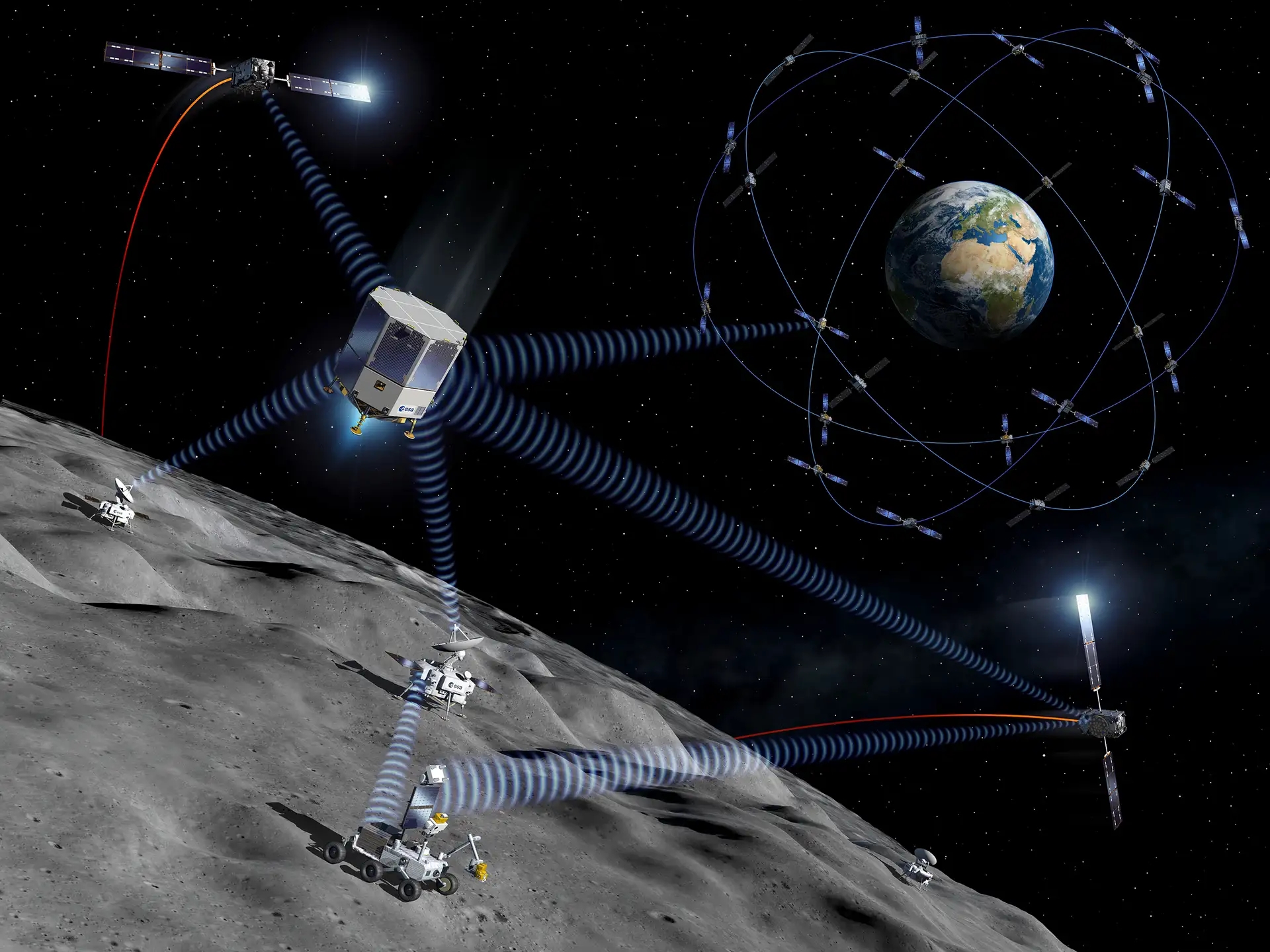}
    \caption{Artist impression of ESA's lunar navigation architecture, which will encompass moon orbiting navigation satellites (Moonlight), geodetic surface stations (like NovaMOON), working in unison with the GNSS constellations around Earth, and serving users like lunar landers and rovers. Image credit: ESA P. Carril\protect\cite{moonlight}.}
    \label{fig:moonlight}
\end{figure}

\chapter{Relativistic Time-Scale Framework}
Newtonian physics treated time as absolute, but \textsc{Einstein}'s relativity revealed time as a dynamic, reference-frame-dependent, relative dimension. This paradigm shift became operationally critical with the advent of space-based astronomy, satellite navigation (e.g. the \acrfull{gps} as the first \acrlong{gnss}), and high-precision astrometry. Modern instruments, like atomic clocks, \acrfull{vlbi} networks and \acrfull{llr}, demand a rigorous relativistic framework to reconcile time measurements across different celestial reference frames. The need for such a framework was formally addressed by the \acrshort{iau}~2000 resolutions\nocite{IAUResolutionsArchive}\cite{iau2000}\cite{Soffel2003iau2000} (especially B1.3-B1.5 and B1.9), which established the \acrfull{bcrs} and \acrfull{gcrs}, which are fully compatible with \textsc{Einstein}'s \acrfull{gr}. The resolutions were further refined in the \acrshort{iau}~2006 Resolutions B2 and B3 (see \cite{IAU2006B2, IAU2006B3}). Further, the IAU~2024 Resolutions II and III give a formal relativistic frame definition and recommendation for the Moon's case (see \cite{IAU2024-ResII, IAU2024-ResIII}). The practical formulas and constants adopted by the geodetic community are collected in the \textit{\acrshort{iers} Conventions 2010 (Technical Notes~36)}~\cite{iers2010}, which this work primarily follows (see Sec.~\ref{ch:theo_iers2010}).

In General Relativity there are two types of time. First, the \textit{proper time}, which is the time an idealized physical clock would measure elapsed between two events on its world-line. Second, the \textit{coordinate time}, which usually refers to the time an idealized clock at rest to a central body and outside said central body's gravity-well would measure. This corresponds to the time a distant observer would experience. Generally, \textit{coordinate time} is a mathematical construct used in relativistic physics to provide a unified time variable across different locations and states of motion within a given coordinate system. It is disconnected from the proper time most clocks would measure. The aforementioned time of a distant observer is just the most natural choice for a coordinate time.

In Sec.~\ref{ch:theory_timeformulas} we outline how the relativistic time transformation as given in the \textit{IERS Conventions} are derived from relativity. Sec.~\ref{ch:eph} discusses ephemerides, orbital elements and simulation tools. We continue detailing how the operational timescales \acrfull{utc} and \acrfull{gpst} are realized (Sec.~\ref{ch:tai_utc} and \ref{ch:gnss_times} respectively). We finish this Chapter on the theoretical background with a brief discussion on clock properties (Sec.~\ref{ch:clock_properties}).
\clearpage
\section{International Astronomical Union Resolutions}\label{ch:theo_iers2010}
The \acrshort{iau}~2000 and IAU~2006 Resolutions\cite{iau2000, Soffel2003iau2000, IAU2006B2, IAU2006B3}, as reflected in the \textit{\acrshort{iers} Conventions 2010}~\cite{iers2010}, distinguish the following four relativistic \textit{coordinate time} scales underlying modern astrometry, geodesy, and satellite navigation.

\textbf{\acrfull{tcb}} is the coordinate time of the relativistic four-dimensional \acrfull{bcrs}, whose origin is the \acrfull{ssb} and whose spatial axes are kinematically non-rotating with respect to distant quasars, which form the \acrfull{icrf}. Clocks synchronized with TCB read the proper time of a clock co-moving with the \acrfull{ssb} but outside the solar system gravity-well.

\textbf{\acrfull{tcg}} is the coordinate time of the relativistic four-dimensional \acrfull{gcrs}, which is a local system of geocentric space-time coordinates. \acrshort{gcrs}s origin is the Earths center of mass, including oceans and atmosphere. It's kinematically non-rotating with respect to the \acrshort{bcrs}. TCG is equivalent to the proper time experienced by a clock at rest in a coordinate frame co-moving with Earth, but without Earth's gravitational influence (so without Earth's gravitational time dilation).

\textbf{\acrfull{tt}} is a uniform time scale an ideal clock on Earth's geoid (a conceptual surface of constant gravitational potential approximating mean sea level) would experience. It is tied to TCG by a defined rate constant $L_G$ and origin $t_0 = \text{JD}\,2443144.5 = \text{1977‑01‑01}\;00{:}00{:}32.184\,\mathrm{TAI}$. \textbf{\acrfull{tai}} is a practical realization of TT, using a global network of very stable atomic clocks and high precision \acrfullpl{pfs} ensuring the SI second. TT is measured in days of \SI{86400}{} SI seconds.

\textbf{\acrfull{tdb}} is a coordinate time, used as a time standard for calculating ephemerides and describing orbits of planets, asteroids and other bodies, as well as spacecraft in the solar system. TDB keeps historical ephemerides (expressed in terms of historical terrestrial time) numerically close by applying a linear scaling of TCB
\begin{equation}\label{eq:TDB}
  \mathrm{TDB} = \mathrm{TCB} - L_B\,(\text{JD}_\mathrm{TCB}-T_0)\times 86400s + \mathrm{TDB}_0
\end{equation}
by the defining constant $L_B$ and defining offsets $T_0 = \text{1977‑01‑01}\,00{:}00{:}00\,\mathrm{TAI}$ and $\mathrm{TDB}_0$ (which was the TDB$-$TCB difference at that epoch). $\text{JD}_\mathrm{TCB}$ is TCB expressed in terms of the \acrfull{jd} -- the count of days since noon on January 1st, 4713 BC on the Julian calendar. By design the difference TDB and TT remains below $2\;\mathrm{ms}$ over several millennia\cite{IAU2006B3}.

\begin{figure}[!htb]
    \centering
    \includegraphics[width=0.6\linewidth]{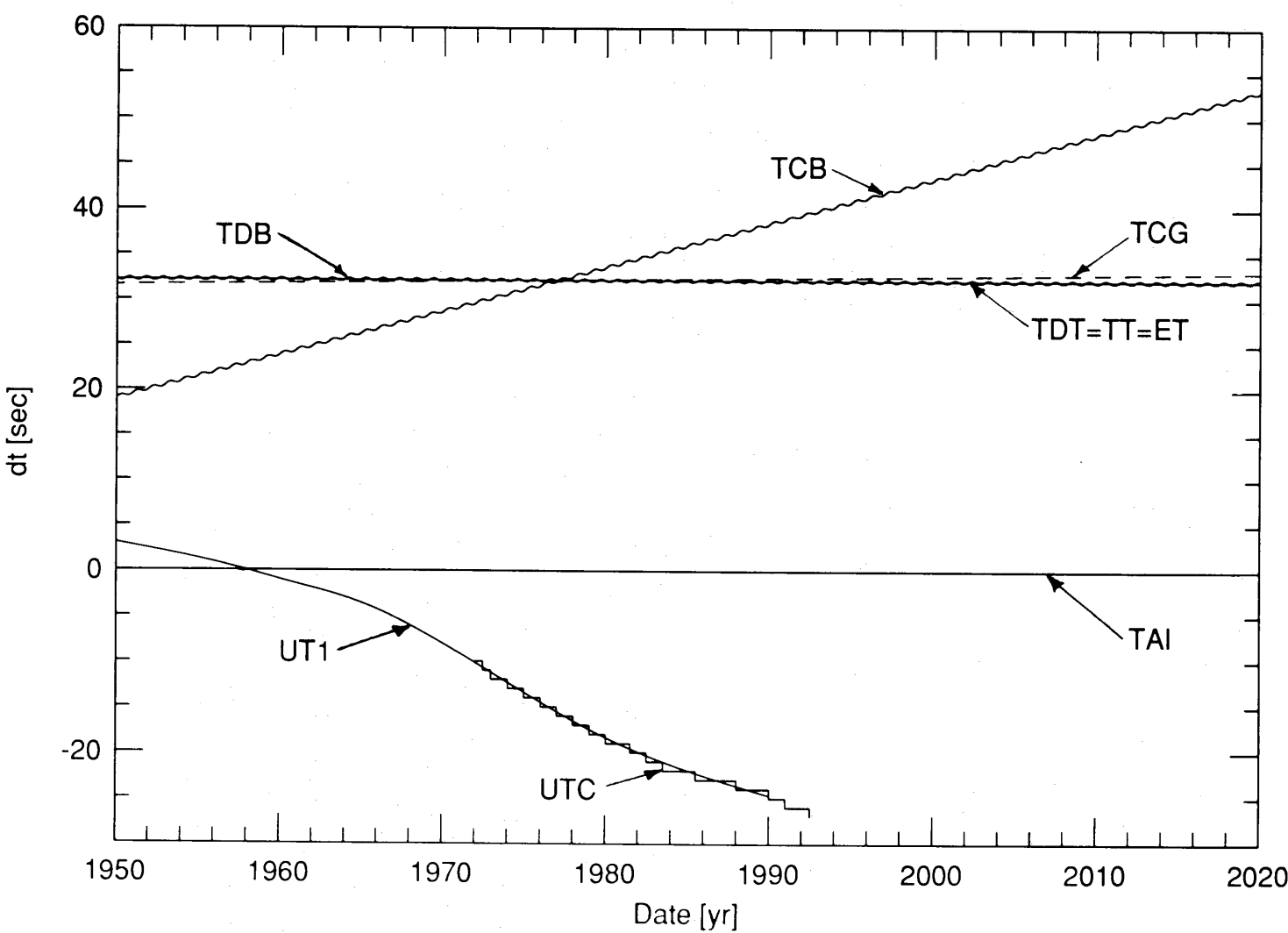}
    \caption{Differences between time scales between 1950 and 2020 with \acrfull{tai} as reference anchor. Periodic terms in TCB and TDB have been exaggerated by a factor of 100 to make them visible. TDB, TCG, and TT are offset from TAI by about 32 seconds. Graphic from \protect\cite{seidelmann}.}
    \label{fig:timescales_to_tai}
\end{figure}

\textbf{Interplay of these timescales:} In order to get a quick understanding how these timescales operate, we can take a look at Fig.~\ref{fig:timescales_to_tai}, where the time difference of these timescales are plotted with respect to the continuous freely running \acrshort{tai}, realizing the SI second. UT1 in this plot is the time tied to Earth's rotation (thus days of 86400 non-SI seconds). As Earth rotation is slowing down, the difference to TAI is compounding over the years. \acrfull{utc} tracks UT1 via leap seconds, such that the civil time established by atomic clocks and their SI-second does not differ from solar time by more than \SI{0.9}{s}. The offset between TT and TAI is by definition 32.184 seconds -- a historical artefact meant to ensure continuity with previous ephemeris time definitions and data. TCB and TDB were chosen to agree with TT at epoch 1977.0. TCB advances faster than TT and TCG, since they are located down a gravitational well, therefore their clocks tick slower than TCB clocks. Similarly, TCG also advances faster than TT and TAI, but to a lesser extent than TCB. TCB's (and TDB's) small oscillations are due to Earth's elliptical orbit around the Sun. 
\clearpage
\vfill
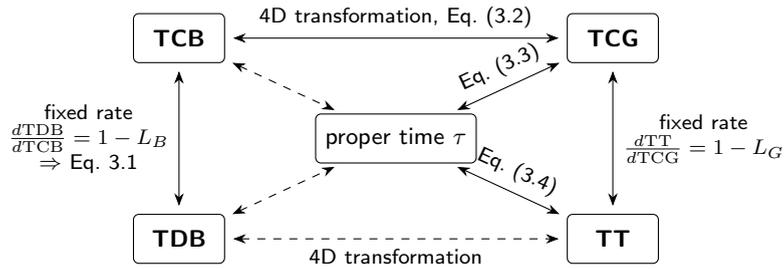
\begin{figure}[!thb]
  \centering
  \begin{tikzpicture}[
        >=Stealth,
        box/.style   = {draw, rounded corners=2pt,
                        minimum width=12mm, minimum height=6.5mm,
                        font=\sffamily\footnotesize},
        note/.style  = {font=\scriptsize\sffamily},
        arr/.style   = {<->, shorten >=3pt, shorten <=3pt},
        darr/.style  = {dashed, arr,},
]

%--- time-scale boxes ---------------------------------------------------------
\node[box] (TCB) {\textbf{TCB}};
\node[box, below=20mm of TCB] (TDB) {\textbf{TDB}};
\node[box, right=45mm of TCB] (TCG) {\textbf{TCG}};
\node[box, below=20mm of TCG] (TT)  {\textbf{TT}};

% proper-time node centred beneath the top row
\coordinate (mid) at ($(TCB)!0.5!(TCG)$);
\node[box, below=10mm of mid] (tau) {$\displaystyle\text{proper time } \tau$};

%--- solid arrows with equation labels ---------------------------------------
\draw[arr] (TCB) -- node[note,sloped,above]{4D transformation, Eq.~(\ref{eq:tcb_tcg})} (TCG);

\draw[arr] (TCB) -- node[note,left,align=center]{fixed rate \\ $\frac{d\mathrm{TDB}}{d\mathrm{TCB}}=1-L_B$\\ $\Rightarrow$ Eq.~\ref{eq:TDB}} (TDB);

\draw[darr] (TDB) -- node[note,sloped,below]{4D transformation} (TT);

\draw[arr] (TT) --  node[note,right,align=center]{fixed rate \\ $\frac{d\mathrm{TT}}{d\mathrm{TCG}}=1-L_G$} (TCG);

%--- dashed arrows to proper time -------------------------------------------

\draw[darr] (TCB) -- node[note,sloped,above]{ } (tau);
\draw[darr] (TDB) -- node[note,sloped,above]{ } (tau);

\draw[arr] (TCG) -- node[note,sloped,above]{Eq.~(\ref{eq:tau_TCG})} (tau);
\draw[arr] (TT) -- node[note,sloped,above]{Eq.~(\ref{eq:tau_TT})} (tau);

\end{tikzpicture} 
  \caption{Relations among the principal coordinate time scales and the proper time a satellite around Earth might have. Inspired by \protect\cite{iers2010}.}
  \label{fig:timescales}
\end{figure}
\vfill
For a detailed understanding of how these relativistic timescales relate to each other, we take a look at Fig.~\ref{fig:timescales}. Timescales/nodes connected by solid line arrows have an explicit defining transformation formula given below. Connections with dashed lines are indirectly defined and can be derived from the given formulas through the other edges. The referenced equation 
\begin{equation}\label{eq:tcb_tcg}
\mathrm{TCB}-\mathrm{TCG}=\frac{1}{c^2}\Biggl\{
        \int_{t_0}^{t}\!\left[\frac{v_e^{2}}{2}+U_{\mathrm{ext}}\!\bigl(\vec{x}_e\bigr)\right]\mathrm{d}t
        \;+\;\vec{v}_e\!\cdot\!\bigl(\vec{x}-\vec{x}_e\bigr)\Biggr\} + O(c^{-4})
\end{equation}
expresses the time difference between \acrlong{tcb} and \acrlong{tcg}. A detailed explanation of all the terms will be given in Sect.~\ref{ch:1pn_from_gr}. Equations \ref{eq:tau_TCG} and \ref{eq:tau_TT} -- relating TT and TCG respectively to the proper time $\tau$ onboard an Earth orbiting satellite, are given here in a simplified form,
 where tidal terms have been ignored. For full expressions, see \cite{iers2010}.
\begin{equation}\label{eq:tau_TCG}
    \frac{d\tau}{d\mathrm{TCG}}=1-1/c^2[\mathbf{v}^2/2+U_E(\vec{x})]
\end{equation}
\begin{equation}\label{eq:tau_TT}
    \frac{d\tau}{d\mathrm{TT}}=1+L_G-1/c^2[\mathbf{v}^2/2+U_E(\vec{x})]
\end{equation}
\vfill
\FloatBarrier
\section{From Relativity to proper time and coordinate time}\label{ch:theory_timeformulas}
Before we adapt to the lunar case in accordance with the \acrlong{iau} (Ch.~\ref{ch:tcl}-\ref{ch:orbit_clocks}), we first outline their derivation for Earth and how to interpret each term physically. Section~\ref{ch:1pn_from_gr} constructs the spacetime metric at the \acrlong{1pn} and recovers the terms adopted in the \textit{\acrshort{iers} Conventions}; Sec.~\ref{ch:2pn_from_gr} briefly notes \acrlong{2pn} refinements.

\subsection{Coordinate time with the 1PN metric}\label{ch:1pn_from_gr}
To understand where Eq.~\ref{eq:tcb_tcg} is coming from, we can start from relativistic principles as follows. Here we will use the conventions: Metric Signature $(+,-,-,-),\; x^\mu=(ct,\vec{x})$ where $c$ is the speed of light. In special relativity the so-called space-time-interval has the form 
\begin{equation}
ds^2 = g_{\mu\nu} dx^\mu dx^\nu = c^2 dt^2 - d\vec x^2
\label{eq:metric}
\end{equation}
and is by definition invariant under Lorentz transformations between inertial frames of reference (those in uniform and constant motion). From the space-time-interval $ds^2$ we can compute the proper time $\tau$ as
\begin{equation}
  d\tau = \sqrt{ds^2/c^2} = dt \sqrt{1 - \frac{v^2}{c^2}} \approx dt \left(1 - \frac{v^2}{2c^2} - \frac{v^4}{8c^4} + \cdots \right)\;,
  \label{eq:proper_time}
\end{equation}
where $v^2 \equiv d\vec{x}^2/dt^2$ is the square of velocity. The leading correction is $-v^2/2c^2$ and can be understood as the low-velocity ($v\ll c$) kinematic time-dilation at \acrfull{1pn}  -- meaning up to expansion terms of $1/c^2$ order. The 2nd term $-v^4/8c^4$ is the kinematic time-dilation correction at \acrfull{2pn} -- meaning expansion terms of up to $1/c^4$ order.

Now, when we want to consider the effect of gravity, we assume the central bodies mass distribution as static and spherically symmetric, and that there exists a locally inertial, non-rotating, freely falling coordinate system with origin at the center of mass. The non-approximate solution would be the well-known Schwarzschild metric:
\begin{equation}
    ds^2 = \left(1 - \frac{2GM}{c^2 r}\right)c^2 dt^2 - \left(1 - \frac{2GM}{c^2 r}\right)^{-1} dr^2 - r^2 d\theta^2 - r^2 \sin^2 \theta \, d\phi^2\;.
    \label{eq:schwarzschild}
\end{equation}
For small $U\ll c^2$ (weak field) and under an appropriate transformation to achieve \textit{isotropic coordinates} (e.g. where the spatial part is as close to Euclidean $d\Sigma^2 = dr^2 + r^2 (d\theta^2 + \sin^2 \theta \, d\phi^2)$ as possible, such that angular and radial parts are treated the same) the Schwarzschild metric in Eq.~\ref{eq:schwarzschild} can be rewritten as an approximate solution to Einstein’s field equations as
\begin{equation}
    ds^2 = \left(1 + \frac{2U}{c^2}\right)(cdt)^2 - \left(1 - \frac{2U}{c^2}\right)(dr^2 + r^2 d\theta^2 + r^2\sin^2 \theta \, d\phi^2)\;,
    \label{eq:isotropic-metric}
\end{equation}
where $U=U(r)= -GM/r$ is the classic Newtonian potential; for real-world accuracy, it may be expanded in terms of spherical harmonics with associated coefficients (multipole moments).

To uncover the effect on proper time, we can rewrite Eq.~\ref{eq:isotropic-metric} by factoring out a $c^2 dt^2$
\begin{equation}
\begin{aligned}
    ds^2 = \left[\left( 1 + \frac{2U}{c^2} \right)- \left(1 - \frac{2U}{c^2}\right) \frac{dr^2 + r^2 d\theta^2 + r^2 \sin^2 \theta \, d\phi^2}{c^2dt^2} \right] c^2dt^2
\end{aligned}
\end{equation}
and simplify by recognizing 
\begin{equation*}
    v^2 = \frac{dr^2 + r^2 d\theta^2 + r^2 \sin^2 \theta d\phi^2}{dt^2}
\end{equation*}
as the velocity in the inertial coordinate system. Then by keeping terms only of order $c^{-2}$, such that the potential term affecting the velocity term is dropped, and after taking the square root of $ds^2$ and expanding, we get for the proper time increment of a moving clock:
\begin{equation}
   d\tau = \sqrt{ds^2/c^2} \approx dt \left ( 1+\frac{U}{c^2}-\frac{v^2}{2c^2}\right)+\mathcal{O}(c^{-4})\;.
   \label{eq:tau_grav_kin}
\end{equation}
The first term here describes the gravitational redshift -- the smaller the potential $U$ (deeper down the gravitational well) where a clock is located, the slower the proper time of that clock elapses compared to the coordinate time. The second term is again the kinematic time dilation, unchanged from above, in this static weak field and at this order. We note that expanding to \acrshort{2pn} would bring squared potential terms and \textit{gravitomagnetic effects} via cross terms $g_{0i}$ into the expression, see Sec.~\ref{ch:2pn_from_gr}.

In case for the solar system, using Eq.~\ref{eq:tau_grav_kin}, we can set $t$ as the coordinate time with respect to the \acrfull{ssb} ($t\rightarrow$\acrshort{tcb}) and $\tau$ as the coordinate time w.r.t. the Earth's geocenter $\vec{x}_e$ ($\tau\rightarrow$ \acrshort{tcg}). Consequently, as we are talking about \textit{coordinate times}, the potential $U$ excludes Earth's own field, such that $U\rightarrow U_{ext}(\vec{x}_e)$ is the external Newtonian potential from all other bodies in the Solar system evaluated at Earths position $\vec{x}_e$. We have
\begin{equation}
\frac{d\tau}{dt}=\frac{d \mathrm{TCG}}{d \mathrm{TCB}} = 1 - \frac{1}{c^2} \left( \frac{v_e^2}{2} + U_{\text{ext}}(\vec{x}_e) \right) + \mathcal{O}(c^{-4})\;.
\label{eq:dtcg_dtcb}
\end{equation}

\begin{figure}[!bht]
    \centering
    \begin{tikzpicture}[scale=2,
        >=Stealth,
        arr/.style   = {thick,->, shorten >=3pt, shorten <=3pt},
        darr/.style  = {dashed, arr,},
]

% Earth
\shade[ball color=blue!20, opacity=0.5] (0,0) circle (0.4);
\node at (0,-0.55) {Earth};
\fill (0,0) circle (0.02);

% Clock position
\coordinate (Clock) at (0.55,0.75);
\draw (Clock) circle (0.04);
\node[above right] at (Clock) {Clock};

% Barycenter
\coordinate (Bary) at (3.5,-0.8);
\draw (Bary) circle (0.04);
\node[below] at (Bary) {Barycenter};

% Vectors
\draw[arr] (Bary) -- (0,0) node[midway,below] {$\vec{x}_e$};
\draw[arr] (Bary) -- (Clock) node[midway,above] {$\vec{x}$};
\draw[arr] (0,0) -- (Clock) node[midway, above left] {$\vec{R}$};

\end{tikzpicture}
    \caption{Geometry of \acrfull{ssb}, Earth and a clock near Earth.}
    \label{fig:clock_geometry}
\end{figure}

When considering coordinate clocks at $\vec{x}$ close to $\vec{x}_e$ (see Fig.~\ref{fig:clock_geometry}), the potential may be expanded as $U_\mathrm{ext}(\vec{x})=U_\mathrm{ext}(\vec{x}_e)-\textbf{a}_\mathrm{ext}\cdot (\vec{x}-\vec{x}_e)+\mathcal{O}(|\vec{x}-\vec{x}_e|^2)$ where $\textbf{a}_\mathrm{ext}\equiv\nabla U_\mathrm{ext}|_{\vec{x}_e}$~\cite{nelson2006}. Evaluated in the \acrshort{1pn} timing relation (Eq.~\ref{eq:dtcg_dtcb}) and defining $\vec{R}=\vec{x}-\vec{x}_e$, one obtains upon integration

\begin{equation}
    \text{TCB} - \text{TCG} = \frac{1}{c^2} \int_{t_0}^{t} \left( \frac{v_e^2}{2} + U_{\text{ext}}(x_e) \right) dt + \frac{\vec{v}_e\cdot \vec{R}}{c^2}  + \mathcal{O}(c^{-4}).
    \label{eq:xdotv_correction}
\end{equation}

Physically the last term is analogous to the $x\cdot v$ term found in the temporal part of the Lorentz transformation (recall from Special Relativity $t'=\gamma (t-\frac{x\cdot v}{c^2})$ where $\gamma=1/\sqrt{1-v^2/c^2}$) meaning comoving-observers with velocity $v_e$ displaced by $R$ from the geocenter are on different surfaces of simultaneity compared to the origin event at the geocenter, as these surfaces are tilted by $v_e/c$, producing a linearly (in distance) dependent time-offset $R/c$. In other words, the last term in Eq.~\ref{eq:xdotv_correction} corrects for the relativity of simultaneity of clocks co-moving with Earth's geocenter at BCRS coordinates $\vec{x}$ as observed from the Barycentric frame.

With Eq.~\ref{eq:xdotv_correction} we now have derived and motivated the \acrshort{iau} definition in Eq.~\ref{eq:tcb_tcg} and explained where each term comes from and what they each represent physically.

\subsection{Beyond 1PN}\label{ch:2pn_from_gr}
To go beyond \acrfull{1pn}, e.g. to \acrshort{2pn}, we start with the general expression for the elapsed proper time for a clock moving along a timelike worldline $x^\mu(t)$: 
\begin{equation}
    d\tau = \sqrt{ds^2/c^2}\;\;\;\;\Rightarrow\;\;\;\; \Delta\tau=\frac{1}{c}\int_{t_0}^{t_1}\sqrt{g_{\mu\nu}\frac{dx^\mu}{dt}\frac{dx^\nu}{dt}}\,dt
\end{equation}
and proceed with a more complete metric $g_{\mu\nu}$ (than Eq.~\ref{eq:isotropic-metric}), which describes an isolated, rotating, stationary and quasi-rigid Earth in \acrshort{2pn}, which has an additional term in $U^2$ in the tt-component, cross-terms in the spatial components and the unchanged spatial part with Kronecker delta $\delta_{ij}$:
\begin{align*}
    g_{00} &=\left(1+2U/c^2+2U^2/c^4\right) \\
    g_{0i} &= 2G\frac{\left(\vec{J_E}\times\vec{r} \right)_i}{c^2\,r^3} \\
    g_{ij} &= -\left(1-2U/c^2 \right)\delta_{ij}\;.
\end{align*}
The cross-term $g_{0i}$ results in the so-called gravitomagnetic clock effect, where $\vec{J_E}$ is the angular momentum vector of Earth and $\vec{r}$ the radial coordinate distance from the origin.

For the purposes of our analysis in the context of lunar time and current space-based clock accuracies, these \acrshort{2pn} terms and effects are so small in magnitude that they can be neglected for timekeeping metrology and navigation in cislunar space~\cite{kopeikin_2024}\cite{geoazur}.

\section{Ephemerides and simulation tools}\label{ch:eph}
This section introduces planetary ephemerides (see Sec.~\ref{ch:eph_what}), explains why different ephemeris families differ and how they are constructed (Sec.~\ref{ch:ephemerides_differences}), reviews Keplerian orbital elements (Sec.~\ref{ch:kepler}), and shows how we use ephemerides in practice via simulation tools (Sec.~\ref{ch:eph_sim}). For a complete review, see \cite{FiengaMinazzoli2024}.

\subsection{What are ephemerides}\label{ch:eph_what}
Ephemerides (from the ancient Greek word for diary or journal) are essentially tables or datasets that provide positions and velocities for solar system bodies (like planets, moons, asteroids, spacecrafts, etc.) as a function of time. In modern days the term also encompasses the entire dynamical framework used to derive these datasets.

In the 19th century, mismatches between prediction and observation pushed astronomy forward: Uranus's anomaly led to Neptune’s prediction; Mercury’s anomalous perihelion advance ultimately required General Relativity. Space-era radar ranging, spacecraft tracking, and \acrshort{llr} turned ephemerides into precision tests of gravity\cite{FiengaMinazzoli2024}. Therefore, as purely analytical models for the motions of the planets reached their limits, ephemerides in their numerical form are now the state-of-art and have been developed and released since the 1970s.

The main families are developed at \acrshort{nasa}'s \acrlong{jpl} (\textbf{DE} Ephemerides), the \acrlong{iaa_ras} (\textbf{EPM} Ephemerides), and at Paris Observatory (\acrshort{imcce}) with \acrlong{oca} (\textbf{INPOP} Ephemerides). Representative references for these ephemeris families are DE\cite{Park2021DE440}, INPOP\cite{Fienga2019INPOP19a}, and EPM\cite{Pitjeva2014EPM}. All three families of ephemeris implement relativistic frameworks and achieve comparable accuracy, with differences mainly in modelling choices and fit strategy.

Modern planetary ephemerides are integrated in days of \acrfull{tdb} and saved as files of \textit{Chebyshev} polynomials fit to the Cartesian positions and velocities of the bodies in the \acrshort{bcrs} and \acrshort{icrf} for efficient high-accuracy interpolation.

\subsection{Differences between ephemeris families}\label{ch:ephemerides_differences}
Dynamical models used for the generation of ephemerides describe the point-mass interactions between all the planets, moons, dwarf-planets, and various asteroids and relativistic effects from the parameterized post-Newtonian~(\acrshort{ppn}) formalism; fitted to an increasingly comprehensive and accurate set of space mission tracking data. 
To be more precise, they share a relativistic \acrfullinv{eihdl} integration framework but differ in specific modelling choices and fitting strategies. 

Variations occur in the small-body perturbation model, such as the number and treatment of \acrfull{mba} (e.g., 343 individually fitted objects in modern DE and INPOP releases) and the handling of \acrfull{tno} or Trojans (e.g., \acrshort{tno} rings in EPM, selected \acrshort{tno}'s plus a ring in DE440, $\sim$500 equal-mass \acrshort{tno} perturbers in INPOP). Differences also arise in the inclusion of additional accelerations (e.g., solar Lense–Thirring frame-dragging, solar oblateness terms), in the size, composition and quality of the observational datasets used for fitting (spacecraft tracking, \acrshort{vlbi}, radar, optical astrometry), and in the parameterization and adjustment procedure itself, such as whether the fit solves for the Sun’s $\mathrm{GM}_\odot$ or historically for the \acrfull{au} (which has been fixed by the \acrshort{iau} in 2012 to a value of \SI{149597870.7}{km}), and how the \acrshort{ssb} is enforced in the integration frame. For comparisons between particular ephemeris releases, you may refer to \cite{FiengaMinazzoli2024, hilton2010comparison}.

\FloatBarrier
\subsection{Keplerian orbital elements}\label{ch:kepler}
\begin{figure}[!htb]
    \centering
    \includegraphics[width=0.50\linewidth]{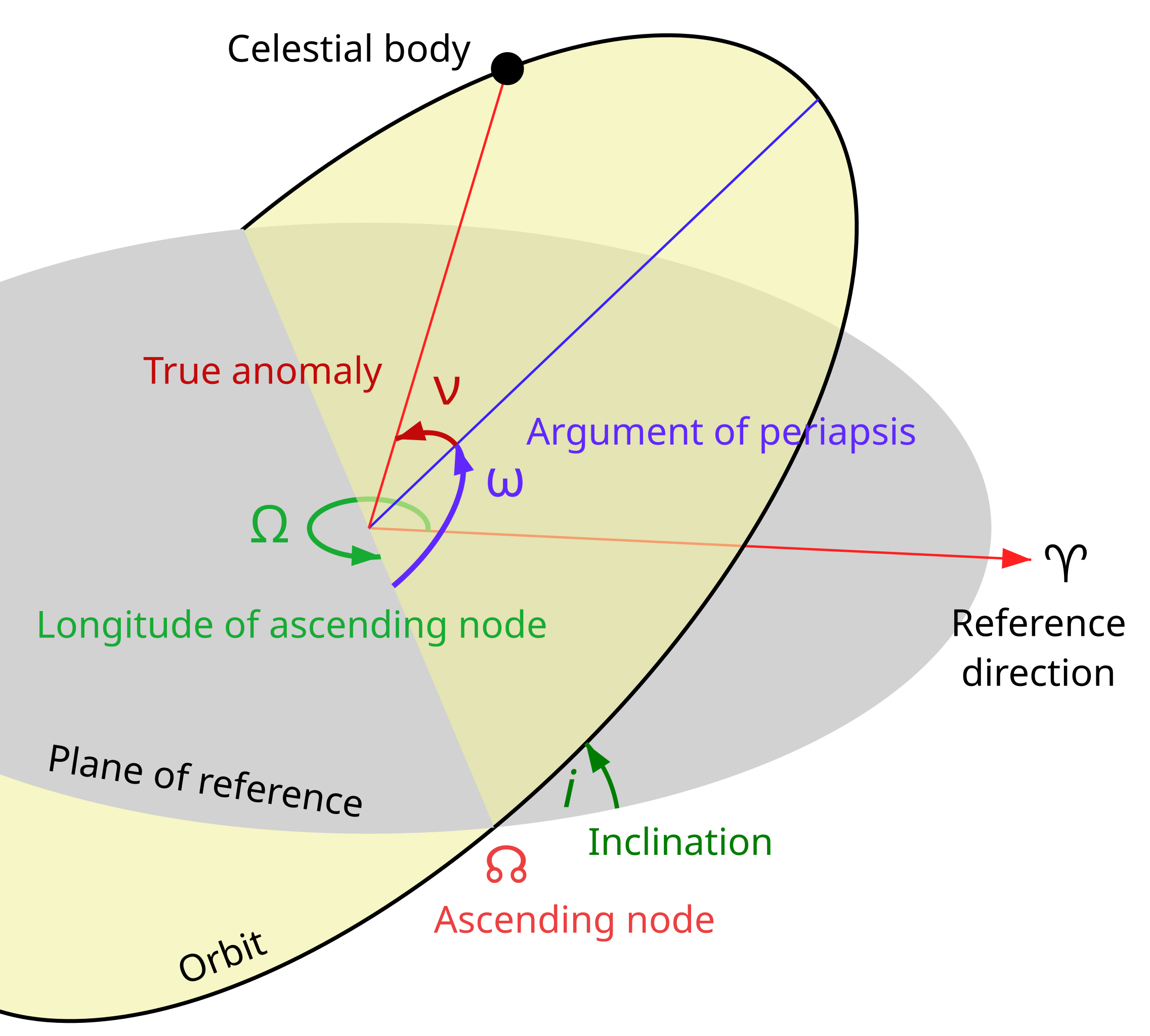}
    \caption{Visualization of the six Keplerian orbital elements. From \protect\cite{wiki_orbit}}
    \label{fig:orbital_elements}
\end{figure}

To describe an orbit in three dimensions, a total of 6 parameters are needed, since there are 6 degrees of freedom. In Cartesian coordinates these correspond to the three $(x,y,z)$-values for position and velocity each. Typically, 6 Keplerian elements are used to describe an object in orbit around a body (with gravitational parameter $\mu = GM$). Two describe the shape of the orbit (eccentricity $e$, semi-major axis $a$), two the orientation of the orbital plane (inclination $i$, right ascension/longitude of the ascending node (\acrshort{raan}) $\Omega$), and two the orientation of the orbit/orbiter within the orbital plane (argument of periapsis $\omega$, true anomaly $\nu$). See Fig.~\ref{fig:orbital_elements} for a useful visualization of these orbital elements and associated symbols.
When solving the Kepler problem (two body problem) and describing its solution, it is also convenient to introduce the so-called eccentric anomaly $E$ and the mean anomaly\footnote{animation showing the anomalies in motion \scriptsize{\url{https://www.youtube.com/watch?v=Mr9t7SLo0I0}}}.

Actual satellite trajectories differ from the idealized Keplerian orbits, because of both gravitational and non-gravitational disturbances. To 0$^\mathrm{th}$ order, Keplerian orbits are accurate, and deviations can be analyzed using perturbation theory. These perturbed orbits are expressed as instantaneous Keplerian orbits, where the orbital elements are time-dependent. These are also referred to as \textit{osculating elements}.

% \FloatBarrier
\subsection{Orbit modeling tools}\label{ch:eph_sim}

Later in this thesis, for the purposes of orbit propagation and the resulting use of planetary ephemerides, we employ the software tool \acrfullinv{godot}~\cite{godot_doc}. \texttt{GODOT} is \acrshort{esa}'s modern flight dynamics framework, developed at the \acrfull{esoc}, for high-precision orbit determination, trajectory design, and mission analysis. It provides a fully relativistic modelling environment for spacecraft and celestial body dynamics. Built on a \texttt{C++} core with a complete \texttt{python} interface, \texttt{GODOT} combines the computational efficiency of compiled code with the flexibility of \texttt{python} scripting, making it suitable both for research and for in-flight operations.

Another tool, that was slated for our work, is \acrfullinv{xhps}~\cite{xhps1}\cite{xhps2} developed by \acrshort{zarm} at University of Bremen. This tool is a \texttt{Matlab} library for high-precision orbit propagation for the motion of artificial satellites around the Earth and the simulation of related dynamical system, high-degree gravity field models, and non-gravitational perturbations such as solar radiation pressure or atmospheric drag. However, for this work -- with regard to proper-time integration of lunar orbiters, the needed clock module was not ready in time yet. Thus \acrshort{xhps} was ultimately not used here.

\section{Earth time laboratories establishing TAI and UTC}\label{ch:tai_utc}

On Earth, international atomic timekeeping is anchored in a network of \acrfullpl{nmi} -- like \acrfullinv{nist} in the USA, or \acrfullinv{ptb} in Germany -- and other designated time laboratories, each operating its own ensemble of high-stability atomic clocks (typically Caesium fountains, hydrogen masers). Each lab's ensemble is combined -- via a local time-scale algorithm such as AT1 at NIST\cite{nist_utc_a1} -- into a continuous, free-running atomic timescale TA(k), which serves as the internal reference for laboratory $k$ . The rate of TA(k) is corrected for relativistic effects arising from the laboratory’s height above the Earth's geoid, ensuring that its tick rate corresponds to proper time on the geoid -- as required for \acrfull{tt}.

A comparison of these local timescales between laboratories is achieved by using time transfer techniques such as \acrfullinv{gnss_cv} or \acrfullinv{twstft}. This interlaboratory data is send to the \acrfull{bipm} in Paris, which processes these measurements. 

At the \acrshort{bipm}, the first step in building a global timescale is the computation of the \acrfull{eal} -- a weighted average of all contributing TA(k) timescales. Because \acrshort{eal}’s unit interval does not perfectly match the SI second, the \acrshort{bipm} applies a small frequency correction (on the relative order of $10^{-13}$~\cite{ptb_tai_eal}) to produce \acrshort{tai}. Leap seconds are then inserted into \acrshort{tai} to generate the civil time \acrshort{utc} -- to stay in sync with Earth's rotation and solar day. The \acrshort{bipm} publishes the [UTC-TA(k)] offsets in monthly \textit{Circular T} and in weekly rapid updates, allowing time laboratories to steer their local UTC(k) outputs.
%\footnote{\url{https://www.bipm.org/en/time-ftp/circular-t}}

In this way, approximately 85 contributing laboratories continuously maintain the globally uniform atomic time reference \acrshort{tai} (and thus \acrshort{utc}) tied to the SI second and corrected to the geoid. Distribution of UTC is provided through long-wave radio stations (e.g. DCF77 controlled by \acrshort{ptb} for continental Europe), \acrshort{gnss} signals, and the \acrfull{ntp} through internet connected servers, which are disciplined by \acrshort{nmi} or \acrshort{gnss} clocks.

\FloatBarrier
\section{GNSS system times}\label{ch:gnss_times}
\begin{figure}[!hbt]
    \centering
    \includegraphics[width=0.75\linewidth]{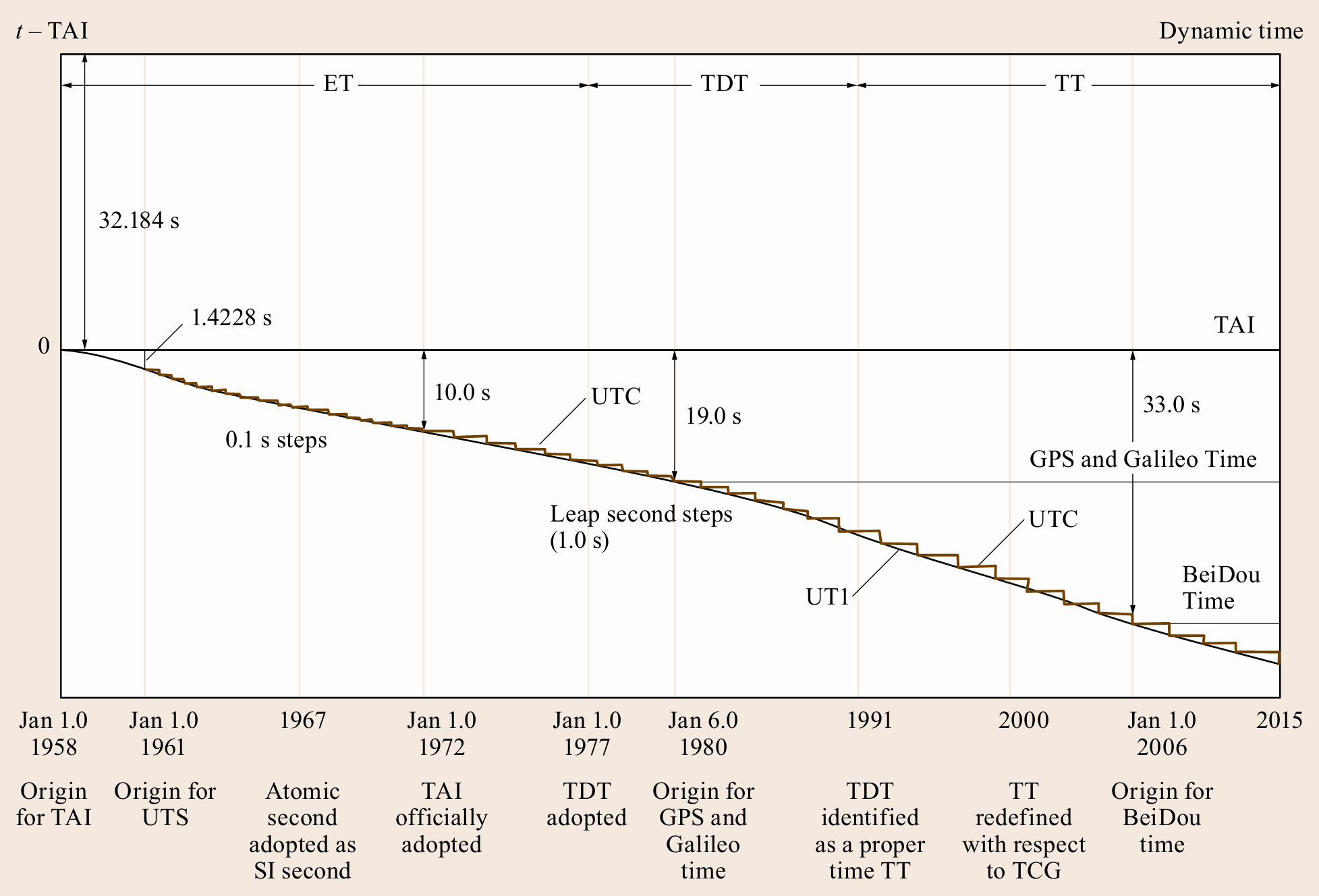}
    \caption{Differences between the GNSS system times and the dynamical timescale discussed prior; leap seconds shown schematically only. From \protect\cite{TeunissenMontenbruck2017}.}
    \label{fig:gnss_timescales}
\end{figure}

Satellite navigation systems determine user positions from distance measurements based on the propagation time of satellite transmitted one-way-signals. This makes them fundamentally dependent on highly accurate clocks and time standards. Each \acrfullinv{gnss} maintains its own system time in order to meet the requirements of internal time synchronization and dissemination. These times-scales are all realized through ensembles of atomic clocks in their ground and space segment and are steered such that they maintain a fixed offset to TAI (see Fig.~\ref{fig:gnss_timescales}).

\acrfull{gpst}, for example, is realized as a composite clock built from atomic clocks within the \acrshort{gps} Control Segment together with the frequency standards aboard \acrshort{gps} \acrfullpl{sv}\cite{TeunissenMontenbruck2017}. Each contributing clock is weighted according to its observed stability, and the resulting ensemble defines the system time, where the offset was set as $\text{GPST}=\text{TAI}-\SI{19}{s}$. Using common-view time transfer, \acrshort{gpst} is steered so that its accuracy from UTC(USNO) -- the \acrshort{utc} realization maintained by the \acrlong{usno} -- remains within \SI{\pm 1}{\mu s}, after accounting for the constant \acrshort{tai} offset and the current total leap seconds. In practice, the offset is much smaller, typically at the level of \SI{20}{ns}\cite{gpst}. This forecast offset is transmitted in the navigation message, enabling users to compute \acrshort{utc} accurately from the broadcasted \acrshort{gpst}.
\vspace{1em}

\FloatBarrier
\section{Clock properties: accuracy vs. stability}\label{ch:clock_properties}
\begin{figure}[!hb]
  \centering
  \begin{subfigure}[t]{0.48\linewidth}
    \centering
    \includegraphics[height=5cm]{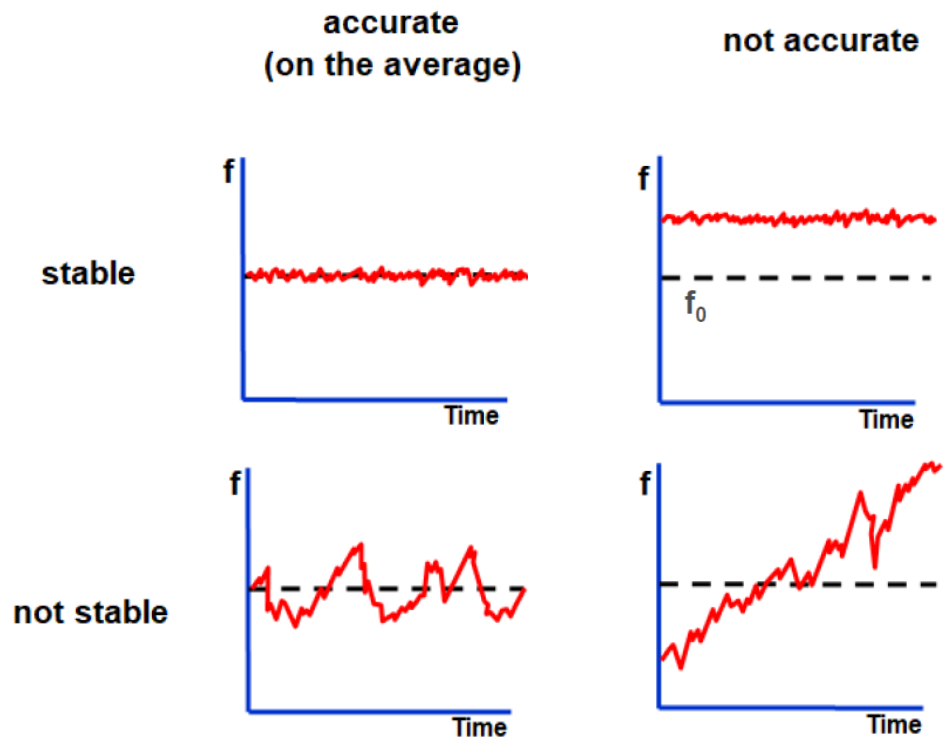}
    \caption{Adapted from \protect\cite{Travagnin2021_CSAC}.}
    \label{fig:accuracy_stability}
  \end{subfigure}\hfill
  \begin{subfigure}[t]{0.48\linewidth}
    \centering
    \includegraphics[height=5cm]{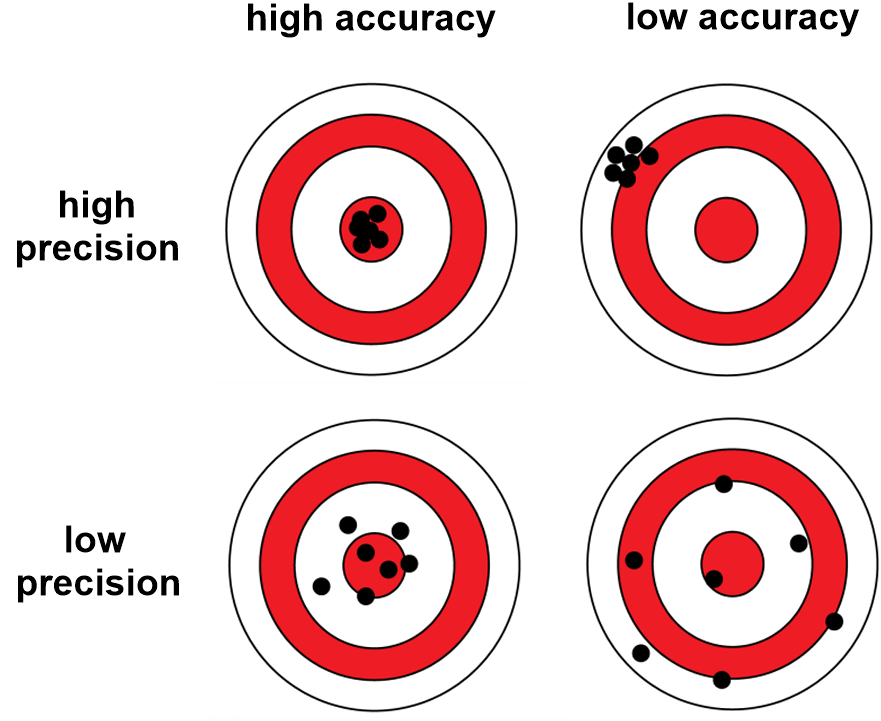}
    \caption{Adapted from \protect\cite{Mkacher2020}.}
    \label{fig:accuracy_precision}
  \end{subfigure}
  \caption{Side-by-side illustrations of (a) Frequency accuracy/stability of clocks and (b) accuracy vs. precision in the dartboard analogy.}
  \label{fig:accuracy_precison_pair}
\end{figure}

When the performance of a clock is evaluated, we primarily look at two characteristics: Accuracy and stability, two
related but distinct concepts; see Fig.~\ref{fig:accuracy_precison_pair}. 

\begin{figure}[!htb]
    \centering
    \includegraphics[width=0.6\linewidth]{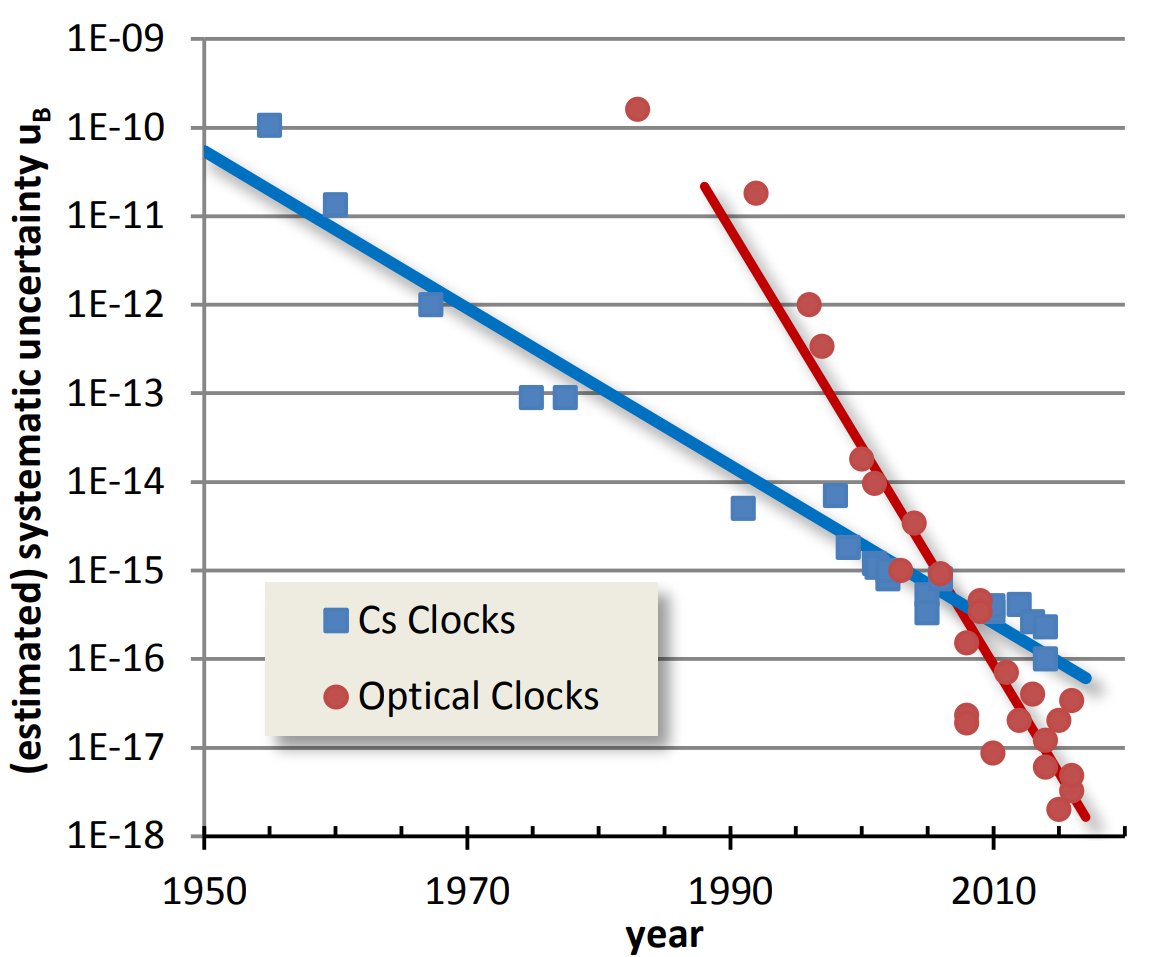}
    \caption{Historical development progress in improving accuracies of Caesium microwave clocks and advanced optical frequency standards. From \protect\cite{Mehlstuebler_2018}.}
    \label{fig:accuracy}
\end{figure}

\textbf{Accuracy} evaluates whether the average tick rate (frequency) is correct. A clock is accurate if, after correcting known systematic effects, its mean frequency matches the SI second, as defined by the ground-state hyperfine transition frequency of the Caesium-133 atom. Systematic effects that shift the frequency of an atomic clock include environmental influences such as magnetic fields (Zeeman effect), electric fields, and blackbody radiation (Stark effect) that shift the energy levels in the atoms, atomic motion causing Doppler shifts, gravitational potential differences (redshift), as well as various instrumental effects. Primary Caesium clocks reach fractional uncertainties near $10^{-16}$, and leading optical clocks below $10^{-18}$, see Fig.~\ref{fig:accuracy}.

\textbf{Stability/Precision} evaluates whether the frequency is repeatable over time. "Precision" in time metrology is quantified as stability, typically by the Allan deviation (sometimes abbreviated \acrshort{adev}), which characterizes root-mean-square fractional-frequency fluctuations dependent on an averaging time/window of length $\tau$.

A clock can be very stable (low noise) yet inaccurate (constant offset), or very accurate on average but noisy from moment to moment, compare with Fig.~\ref{fig:accuracy_stability}. Another helpful picture is a dartboard where accuracy moves the mean of the shots to the bullseye and precision tightens the spread of the shots, regardless of where the mean sits; see Fig.~\ref{fig:accuracy_precision}.

%\vspace{1em}

\begin{figure}[!htb]
    \centering
    \includegraphics[width=0.6\linewidth]{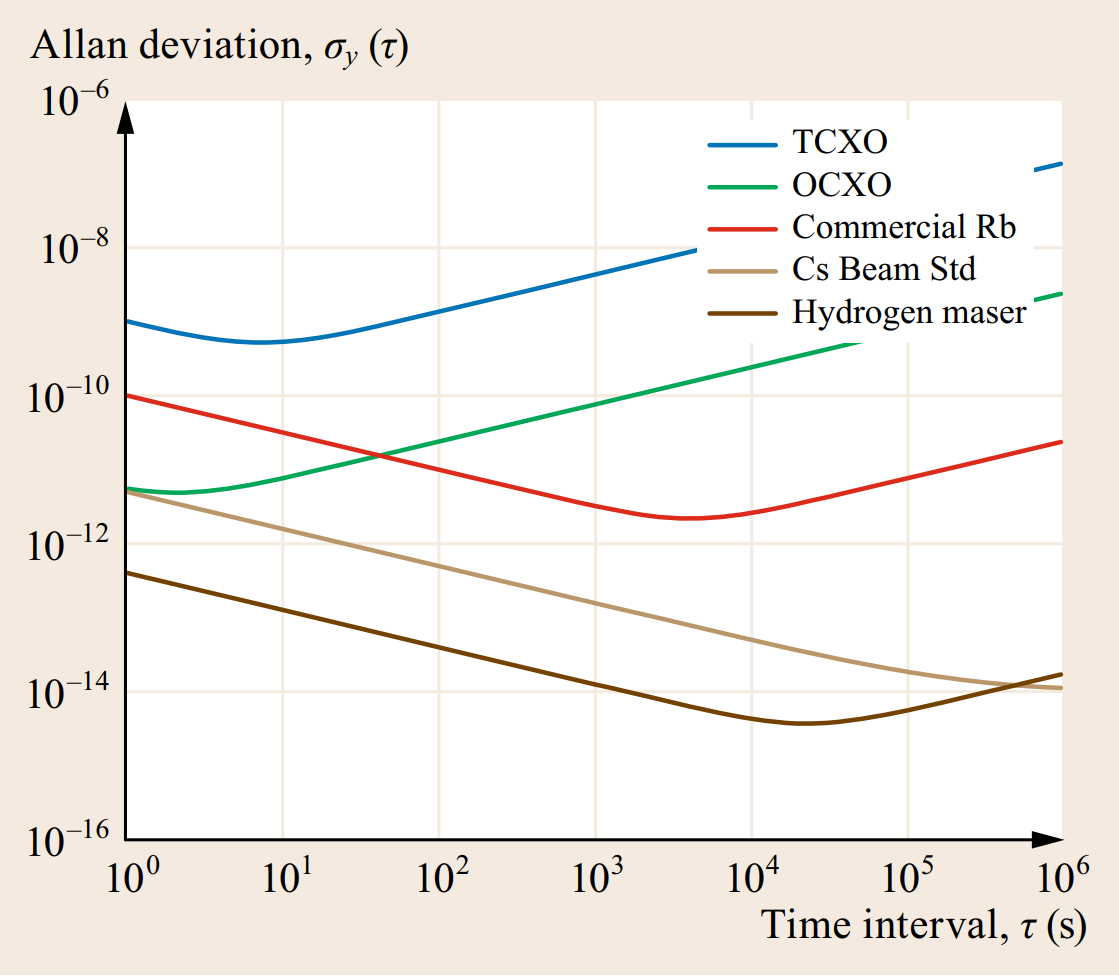}
    \caption{Allan deviation performance plot (describing clock stability/precision) comparing temperature-compensated (\acrshort{tcxo}s) and oven-controlled (\acrshort{ocxo}s) crystal oscillators with atomic frequency standards. From \protect\cite{TeunissenMontenbruck2017}.}
    \label{fig:allan_deviation}
\end{figure}

As has been said, Allan deviation is a metric often used when describing a clock's stability. Real clocks exhibit noise types that change with averaging time $\tau$ -- white frequency noise at short $\tau$, flicker for mid $\tau$, random walk at long $\tau$. Ordinary \textit{variance} can diverge for these non-stationary noises, whereas \textit{Allan deviation} remains well-behaved and, on a log-log plot of $\sigma_y(\tau)$ vs. $\tau$ also reveals the noise type by its slope (power law): $\sigma_y \propto \tau^\alpha$ with $\alpha=-1/2$ characterizes white-frequency-noise, with $\alpha$ around $0$ (constant) we have flicker-frequency-noise, and for $ \alpha=+1/2$ random walk frequency noise. In Fig.~\ref{fig:allan_deviation} the Allan deviation for various clock types is plotted.

In practical timekeeping, high short-term stability is essential for reducing measurement noise in time transfer and for providing a low-jitter reference to steer an ensemble or maintain a local timescale between calibrations. High long-term accuracy is necessary to ensure that the timescale does not drift away from the SI definition. This complementarity explains why international timescales such as \acrshort{tai} are computed from ensembles that combine the best of both worlds: hydrogen masers with excellent short-term stability and primary or secondary frequency standards that provide accuracy.

\vfill
\vfill
\vfill
\chapter{Definition and Computations of Lunar Coordinate Time} \label{ch:tcl}
This chapter first discusses the possible definition of the \acrfull{tcl} in Sec.~\ref{ch:def_TCL}, detailing two formulations, one by Kopeikin et al.\cite{kopeikin_2024} and the other by Fienga et al.\cite{geoazur}. In Section~\ref{ch:tcl_num_comparison}, we compare the numerical results from both approaches. We finish with Sec.~\ref{ch:tcl_finalremarks} with final remarks regarding our comparison and \acrshort{tcl} overall.

\section{Definitions of the TCL transformation}\label{ch:def_TCL}
Knowing the \acrshort{tcg} definition (see Eq.~\ref{eq:dtcg_dtcb} according to the \acrshort{iau} 2000 Resolutions \cite{iers2010}) it is straightforward to define an analogous \acrfull{tcl}, for example, see Kopeikin et al.\cite{kopeikin_2024}. According to this approach, TCL is the coordinate time of a new \acrfull{lcrs} centered at the Moon's center of mass, with its axes kinematically non-rotating with respect to the \acrshort{bcrs}/\acrshort{icrf} axes. The definition of TCL then becomes like in Eq.~\ref{eq:dtcg_dtcb} but with the indices $e$ (for either Earth's position and velocity in the \acrshort{bcrs} coordinate frame) replaced with $L$ for the lunar equivalents, such that TCL is the relativistic time scale of the LCRS. Consequently, for the external potential, rather than summing over all bodies except Earth, the summation is over all bodies except the Moon. To be concrete:
\begin{equation}\label{eq:tcl_tcb}
    TCB-TCL=\frac{1}{c^2}\int_{t_0}^{t} \left ( \frac{v_L^2}{2}+\sum_{A\neq L}\frac{GM_A}{r_{LA}}\right)dt+\frac{\vec{v}_L\cdot \vec{r}}{c^2} +\mathcal{O}(c^{-4})
\end{equation}
with $A$ being an external body to consider (e.g. Earth, Sun, Jupiter, Saturn, and other planets/asteroids ), $r_{LA}=|\vec{r}_{LA}|=|\vec{x}_L-\vec{x}_A|$ being the coordinate distance of Moon and body $A$, $v_L$ being the velocity of the moon in the BCRS, and $\vec{r}=\vec{x}-\vec{x}_L$ the vector connecting the moons center-of-mass position with the clock position. In \cite{kopeikin_2024}, this formula is also given in \acrshort{2pn} rather than here \acrshort{1pn}; however, the extra \acrlong{2pn} terms contribute only by $10^{-8}\,$ns/day -- making them practically unnoticeable, so that they can be ignored for practical purposes.

Now, in order to obtain an expression for purposes of comparing the \acrfull{tcg} with \acrshort{tcl}, rather than with \acrfull{tcb}, Kopeikin massaged TCL-TCB (Eq.~\ref{eq:tcl_tcb}) at \acrshort{1pn} order so that it can be subtracted from TCB-TCG with a handful of terms cancelling. What remains is TCL-TCG. Equation~\ref{eq:tcl_tcb} is first transformed to Earth-centered coordinates and then terms within the sum are Taylor expanded (there known as the expansion in tidal multipoles). The result is an expression that cancels with a lot of terms from Eq.~\ref{eq:tcb_tcg}, resulting in a relatively simple (compared to the here omitted in-between steps) TCL-TCG expression of
\begin{multline}\label{eq:tcl_tcg}
    TCL = TCG- \frac{1}{c^2} \int_{t_0}^{t} \left\{ \frac{v_{\text{LE}}^2}{2} - \frac{\mu_{\text{E}}}{r_{\text{LE}}} - \frac{2\mu_{\text{E}}}{r_{\text{LE}}} + \frac{3}{2} \frac{\mu_{\text{S}}}{r_{\text{ES}}} \left[ (\mathbf{r}_{\text{ES}} \cdot \mathbf{r}_{\text{LE}})^2 - \frac{1}{3} r_{\text{ES}}^2 r_{\text{LE}}^2 \right] \right\} dt \\ - \frac{1}{c^2} (\mathbf{v}_{\text{LE}} \cdot \mathbf{r}_{\text{ES}} - \mathbf{v}_{\text{LE}} \cdot \mathbf{r}_{\text{LE}})\;,
\end{multline}
where $v_{LE}$ is the Moon's velocity with respect to Earth in the \acrshort{gcrs}, the index $S$ refers to the Sun, and $\mu_A=G M_A$ is known as the gravitational parameter of body $A$ -- the product of Newton's gravitational constant $G$ and the body's mass $M_A$. The fact that many terms cancel during the subtraction is consistent with the physical intuition based on \acrlong{gr}'s Equivalence Principle, which states that local experiments cannot detect the orbital motion of a local coordinate system in freefall, and thus the gravitational influence of external bodies can only manifest through tidal terms. Indeed, the leading contributions (the terms inside the integral) of the TCL-TCG transformation are expressed solely in terms of the Moon's geocentric distance $r_{LE}$ and velocity $v_{LE}$ relative to Earth, whereas the terms depending on the Earth's barycentric orbital velocity cancel out. For more details on the derivation, see \cite{kopeikin_2024}.

For the purpose of numerical simulations -- rather than the analytically integrated form in Eq.~\ref{eq:tcl_tcg} -- we can also simply express the relation between \acrshort{tcl} and \acrshort{tcg} through a differential rate form, as done by Fienga et al.\cite{geoazur}. We start with
\begin{equation}
    \frac{dTCL}{dTCB}=1-\frac{1}{c^2} \underbrace{\left ( \frac{v_L^2}{2}+\sum_{A\neq L}\frac{GM_A}{r_{LA}}\right )}_{=:\, \alpha_L} +\mathcal{O}(c^{-4})\;,
\end{equation}
analogous to dTCG/dTCB as seen in Eq.~\ref{eq:dtcg_dtcb}, but for the Moon, rather than the Earth. To arrive at dTCL/dTCG, we can shift our coordinate-system origin from \acrshort{ssb} to the geocenter of Earth, by re-expressing $\alpha_L$ in terms of geocentric velocities/positions and the grav. potential relative to geocenter. This simple translation, also done by \cite{kopeikin_2024} to arrive at Eq.~\ref{eq:tcl_tcg}, is essentially the Galilean coordinate transformation from BCRS to GCRS -- which is sufficient at \acrlong{1pn}, as explained in \cite{kopeikin_2024}. Therefore, we have
\begin{equation}
    \frac{dTCL}{dTCG}=1-\frac{1}{c^2}\alpha_{LE}+\mathcal{O}(c^{-4})
\end{equation}
with
\begin{equation}
\alpha_{LE} = \frac{v_{LE}^2}{2}  + \sum_{A \neq L} \frac{G M_A}{r_{LA}} - \sum_{A \neq E} \frac{G M_A}{r_{EA}} \;.
\label{eq:placeholder_label}
\end{equation}

Fienga et al.\cite{geoazur} then proceeded to also express \acrshort{tcl} in terms of \acrshort{tt}, which is a simple scaling as per $\frac{dTT}{dTCG}=1-L_G$, such that
\begin{align}\label{eq:dtcl_dtt}
    \frac{d(TCL-TT)}{dTT} &= \frac{dTT}{dTCG}^{-1}\left (\frac{dTCL}{dTCG}-\frac{dTT}{dTCG}\right) \notag\\
    &=\frac{1}{1-L_G}\cdot \left ( L_G+\frac{1}{c^2}\alpha_{LE}\right )
\end{align}

Another approach for a Lunar Coordinate Time -- only mentioned here for completeness sake -- is via an intermediate \acrfull{emcrs}\cite{kopeikin_2024}. This route, first formulated for lunar time by Ashby and Patla\cite{Ashby_2024}, introduces local coordinates tied to the Earth-Moon barycenter and expands the external potential in tidal multipoles while keeping the Earth-Moon dynamics explicit. Kopeikin and Kaplan find that this \acrshort{emcrs} construction results in an identical TCL-TCG to the Ashby-Patla model, and also verify consistency with the \acrshort{iau} relativistic framework and note that the \acrshort{iau} based formalism actually allows to accommodate a greater number of terms. See their\cite{kopeikin_2024} Appendix A.

\section{Numerical comparison of different literature results}\label{ch:tcl_num_comparison}
In this section, we cross-check the formulations of Kopeikin et el.\cite{kopeikin_2024} and Fienga et al.\cite{geoazur} by directly comparing their numerical predictions. We first examine the average drift rates in Section~\ref{ch:num_secular_drift}, followed by the periodic components in Section~\ref{ch:num_periodic}.

Kopeikin et al.\cite{kopeikin_2024} evaluate their explicit transformation (Eq.~\ref{eq:tcl_tcg}) over a 10-year span using the DE440 planetary ephemerides with a 0.1-day step; the explicit form also allows for an accompanying analytic treatment. Fienga's implementation numerically integrates the differential-rate form in Eq.~\ref{eq:dtcl_dtt} for TCL-TT using the INPOP21a ephemerides over a 30-year window with a 0.002-day step. The integration is performed simultaneously with the planetary and lunar orbits within the INPOP framework, employing the same order-12 Adams-Cowell integrator (optimized to the Earth-Moon system to sub-mm stability) used for the ephemerides; further details are given in \cite{fienga2006}\cite{fienga2008}. 

Both approaches produce a time-difference signal consisting of an average linear trend (called \textit{secular drift}) plus superimposed oscillations. The secular drift arises mainly from the average gravitational potential and average velocity difference between the two coordinate times, while the remaining periodic components reflect the shape and perturbations on the orbits. 

\FloatBarrier
\subsection{Secular drifts}\label{ch:num_secular_drift}
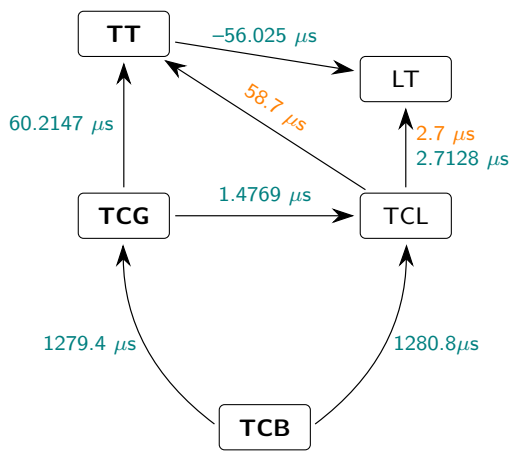
\begin{figure}[!b]
  \centering
  %\documentclass[tikz,border=5pt]{standalone}
%\usetikzlibrary{positioning,arrows.meta,calc}
%\begin{document}
\begin{tikzpicture}[
    box/.style     = {draw, rounded corners=2pt,
                    minimum width=12mm, minimum height=6mm,
                    font=\sffamily\footnotesize, align=center},
    arr/.style     = {-{Stealth[length=3mm,width=2mm]},
                    shorten <=2pt, shorten >=2pt,
                    line width=0.45pt},
    lab/.style     = {font=\scriptsize\sffamily, align=center},
]

% -----------
% Nodes
% ------------
\node[box]                           (TT)  {\textbf{TT}};

\node[box,below=18mm of TT]          (TCG) {\textbf{TCG}};

\node[box,right=25mm of TCG]         (TCL) {TCL};

\node[box,above=12mm of TCL]          (LT)  {LT};

\node[box,below=25mm of $(TCG)!0.5!(TCL)$] (TCB) {\textbf{TCB}};

% ----------------------------------
% Arrows with labels (green text)
% ----------------------------------

% TT  --> LT
\draw[arr] (TT) -- node[lab,above]{\textcolor{teal}{--56.025 $\mu$s}} (LT);

% TCG --> TT
\draw[arr] (TCG) -- node[lab,left]{\textcolor{teal}{60.2147 $\mu$s}} (TT);

% TCL --> LT
\draw[arr] (TCL) -- node[lab,right,align=left]{\textcolor{orange}{2.7 $\mu$s}\\
\textcolor{teal}{2.7128 $\mu$s}} (LT);

% TCB --> TCG
\draw[arr,bend left=26] (TCB.west) to node[lab,left]{\textcolor{teal}{1279.4 $\mu$s}} (TCG);

% TCB --> TCL
\draw[arr,bend right=26] (TCB.east) to node[lab,right]{\textcolor{teal}{1280.8$\mu$s}} (TCL);

%TCG --> TCL
\draw[arr] (TCG) -- node[lab,above]{\textcolor{teal}{1.4769 $\mu$s}} (TCL);

%TCL --> TT
\draw[arr] (TCL) -- node[lab,sloped,above]{\textcolor{orange}{58.7 $\mu$s}} (TT);

\end{tikzpicture}
%\end{document} 
  \caption{Secular drift per day between established and proposed coordinate times. In teal numbers from \protect\cite{kopeikin_2024}, and in orange numbers from \protect\cite{geoazur}. No inconsistencies are found.}
  \label{fig:secular_drifts}
\end{figure}

Kopeikin primarily compared \acrshort{tcl} with \acrshort{tcg}/\acrshort{tcb} (obtaining a secular drift of \SI{1.4769}{\mu s/day} and \SI{1.2808}{ms/day} respectively) and \acrshort{tt} with a proposed lunar surface time \acrfullinv{lt}, whereas Fienga analyzed \acrshort{tcl} to \acrshort{tt}. We summarized the key values in Fig.~\ref{fig:secular_drifts}, which shows the established timescales (TCB, TCG, TT) and the proposed lunar analogues (TCL, LT).

Here, \acrshort{lt} is defined as the lunar equivalent of \acrshort{tt}. TT's rate is realized on the rotating terrestrial geoid -- an equipotential surface of the Earth's gravitational potential best matching mean sea level -- with a reference potential $\Phi_{G}=\SI{6.26368560e7}{m^2/s^2}$. No equivalent equipotential (\textit{"selenoid"}) has yet been internationally adopted for the Moon. Kopeikin~\cite{kopeikin_2024} adopted a literature value $\Phi_L=\SI{2.822336927e6}{m^2/s^2}$ from \cite{selenoid}, derived by least-squares fitting to lunar surface features. Fienga et al. also consider a surface clock, placed at the Moon's mean radius. In both works, the LT-TCL rate difference is directly stated as about \SI{2.7}{\micro s/day}. 
Furthermore, recalling the IERS Conventions\cite{iers2010}, we have TT-TCG defined by a fixed-rate form (as seen in Fig.~\ref{fig:timescales}), such that LT-TCL can also be proposed as a fixed-rate form; plus an analogous TCG-TCB approximation:
\begin{align*}
\text{TT} &= (1 - L_G) \cdot \text{TCG} \\
\text{LT} &= (1 - L_L) \cdot \text{TCL} \\
\text{TCG} &\approx (1-L_C) \cdot \text{TCB}\;,
\end{align*}
where $L_G$ is a defining constant and $L_C$ is the average value for $1-dTCG/dTCB$ as per \acrshort{iers} numerical standards\cite{iers2010}. $L_L$ is derived from $\Phi_L$ as in \cite{kopeikin_2024}. All these can also be expressed as secular drifts (and are also illustrated in Fig.~\ref{fig:secular_drifts}):
\begin{alignat*}{2}
L_G &= \SI{6.969290134e-10}{}  &&= \SI{60.2147}{\mu s/day} \\
L_L &= \Phi_{L}/c^2 &&= \SI{2.7128}{\mu s/day} \\
L_C &\approx \SI{1.48082686741e-8}{} &&= \SI{1.2794}{ms/day}
\end{alignat*}

We find that the secular drift values from both papers are consistent with one another. For example, the TCL-TT drift of \SI{58.7}{\mu s/day} from~\cite{geoazur} is recovered from the values provided by \cite{kopeikin_2024} either via the \acrshort{lt} node ($2.7128+56.025=58.7378$), or through the \acrshort{tcg} node ($-1.4769+60.2147=58.7378$), with signs chosen according to the arrow directions in Fig.~\ref{fig:secular_drifts}. 

\FloatBarrier
\subsection{Analysis of periodic terms}\label{ch:num_periodic}
\begin{figure}[!htb]
    \centering
    \includegraphics[width=14cm]{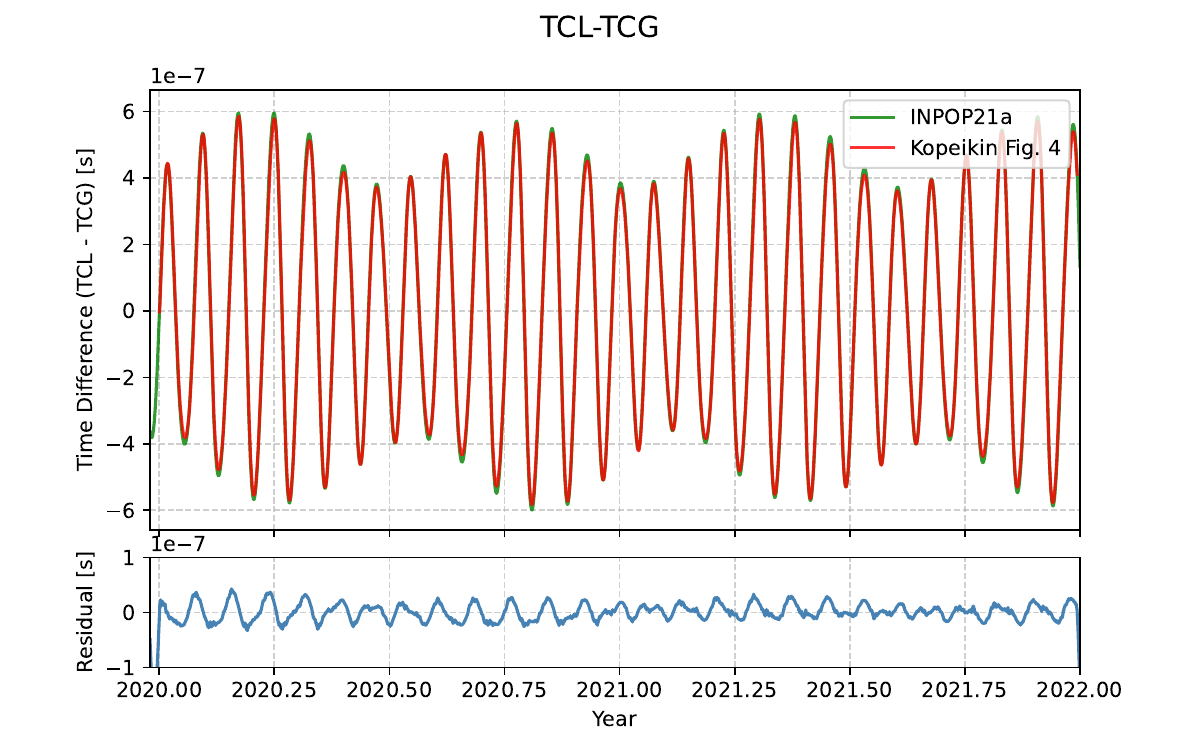}
    \caption{Time difference between TCL and TCG, obtained from the integration of \mbox{INPOP21a} ephemerides between 2019 and 2023 (green), compared with the results from \protect\cite{kopeikin_2024} (red) and the residual between the 2 year TCL-TCG time-series (blue). The secular drift was removed for each dataset. }
    \label{fig:timesignal_tcltcg_comparison}
\end{figure}

To isolate the periodic contributions, the linear \textit{secular drift} of each time-difference series is subtracted. For Kopeikin's TCL-TCG timeseries, we digitised their published curves (Figs. 3-4 of \cite{kopeikin_2024}) to obtain a record suitable for comparison in our plots. Fienga's TCL-TT timeseries data was provided to us by her. To keep the comparison fully consistent with Kopeikin’s plots, we map \acrshort{tt} back to \acrshort{tcg} by scaling the drift-removed series by $(1-L_G)$. We then compare both time series in Fig.~\ref{fig:timesignal_tcltcg_comparison} over the identical 2020-2022 window used in \cite{kopeikin_2024}.

The two waveforms in the time-domain agree closely: both exhibit amplitudes at the \SI{0.6}{\micro s} level, and the difference/residual remains within \SI{0.05}{\micro s} throughout the interval. Minor differences are visible as small, quasi-periodic structures in the residual; these seem consistent with bias of our curve digitization and interpolation rather than a systematic bias in either formulation.

\begin{figure}[!htb]
    \centering
    \includegraphics[width=12cm]{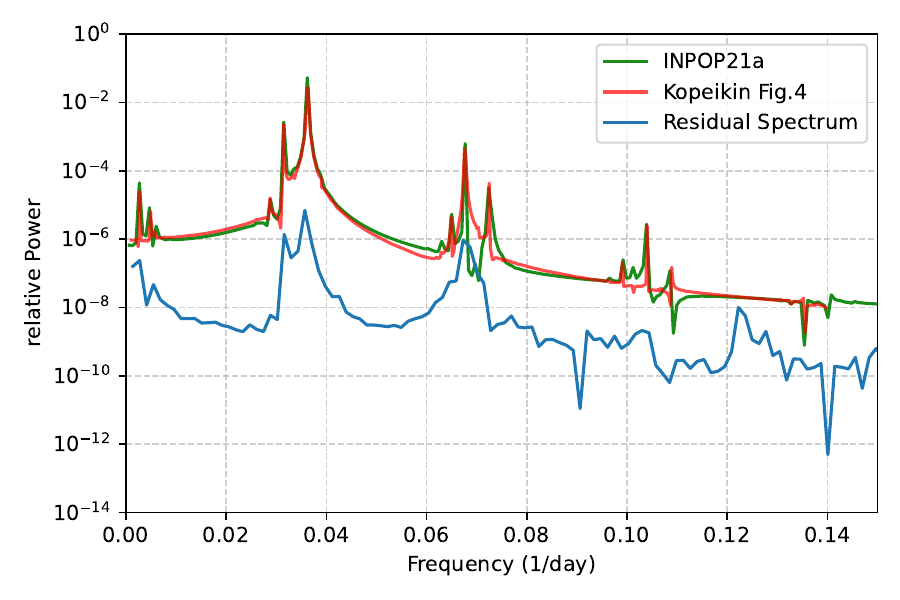}
    \caption{Power spectrum of periodicities for TCL-TCG from the integration of INPOP21a between 2019 and 2023 compared with the results from Fig.4 in \protect\cite{kopeikin_2024} and the spectrum of the residual between the 2-year TCL-TCG time-series.}
    \label{fig:powerspectrum_tcltcg_comparison}
\end{figure}

\begin{figure}[!htb]
    \centering
    \includegraphics[width=12cm]{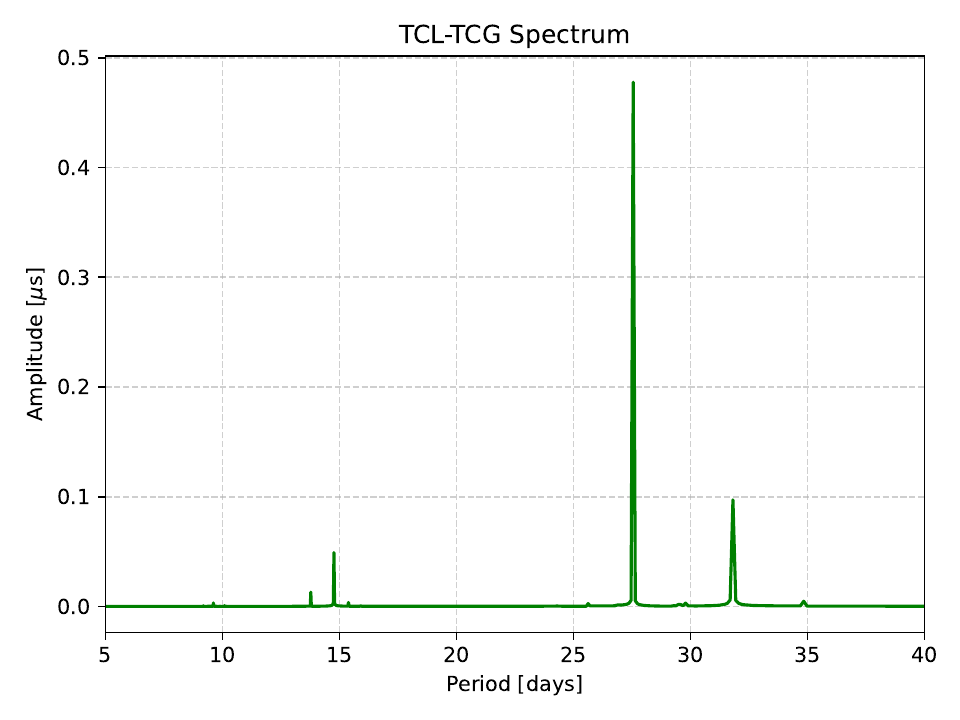}
    \caption{Spectrum of periodicities for TCL-TCG from the integration of INPOP21a between 2000 and 2023 expressed in terms of Period (days).}
    \label{fig:powerspectrum_period}
\end{figure}

Transforming the INPOP21a data to the frequency-domain is done via the \acrfull{fft} -- using the \texttt{python} package \texttt{SciPy} -- see Fig.~\ref{fig:powerspectrum_tcltcg_comparison}. In the recovered power spectrum (for data covering 2019 to 2023) the principal lines closely match with Kopeikin. Discrepancies are likely due to miss-matching signal length and windowing of the datasets. Even small (sub-day) changes in the windowing of the 4 years of selected data result in frequency spectra that look a bit different in shape/curvature. Frequencies appearing in the residual (2 years of data range) are also plotted in Fig.~\ref{fig:powerspectrum_tcltcg_comparison}, and resemble to some extend the main frequencies in the original signals -- while two orders of magnitude weaker.
\FloatBarrier
\newpage\clearpage
Additionally, we present the INPOP21a frequency domain plotted versus period (Fig.~\ref{fig:powerspectrum_period}), computed from the complete 24-year simulation dataset provided to us. To obtain the correctly scaled amplitudes in units of seconds the \acrshort{fft} was normalized with $A_k=(2/N)|X_k|$ where $X_k$ is the \acrshort{fft} of the time-domain dataset $x$ of length $N$.

Table~\ref{tab:freqTCL} lists the periods and amplitudes fit to the period representation. The principal lines match within our uncertainty to periods built from luni-solar arguments $(M,M',D,F)$ -- as proposed and done in \cite{kopeikin_2024}. The lunar-solar arguments are:
{\setlength{\parskip}{0pt}
\begin{itemize}
    \item $M$ -- Moon's mean anomaly (anomalistic month $\approx \SI{27.55}{d}$)
    \item $M'$ -- Earth's mean anomaly (annual phase $\approx \SI{1}{year}$)
    \item $D$ -- Moon's mean elongation (synodic phase $\approx \SI{29.53}{d}$
    \item $F$ -- Moon's argument of latitude (draconic month $\approx \SI{27.21}{d}$)
\end{itemize}
In lunisolar theories and ephemeris expansions, periodic terms appear as integer combinations $n_1M+n_2M'+n_3D+n_4F$ and generate the characteristic frequencies seen in time series and frequency analysis.
}

 The only period predicted by Kopeikin (compare with Table 2 in \cite{kopeikin_2024}) that we do not recover is at $2D-M+M'\approx29.26\,\text{d}$. This analytically predicted amplitude would also be quite low, and likewise was not found by Kopkeikin's numerical simulation (10 years of data). Kopeikin's numerical treatment also could not resolve $M-M'\approx\SI{29.80}{d}$, whereas we could.

\vspace{2cm}

\begin{table}[!htb]
\caption{Major period terms in TCG-TCL and TT-TCL: the first and second columns list the periods and their $\sigma$ values in days, the third column shows the amplitudes, and the last column gives the luni-solar arguments as associated with these periods.}
\label{tab:freqTCL}
\centering
\begin{tabular}{r r r c} \\
Periods & $\sigma$ Period & Amplitude & Luni-solar \\
day & day & $\mu$s &  arguments \\
\hline
 27.5610 & 0.0589 & 0.4778 & M          \\
 13.7775 & 0.0016 & 0.0130 & 2M         \\
  9.1852 & 0.0325 & 0.0005 & 3M         \\
 31.8023 & 0.0488 & 0.0971 & 2D - M       \\
 14.7668 & 0.0083 & 0.0491 & 2D         \\
  9.6136 & 0.0002 & 0.0031 & 2D + M       \\ 
364.8082 & 0.9782 & 0.0136 & M'         \\
173.1869 & 0.1800 & 0.0014 & 2F - 2D    \\
205.7794 & 0.0641 & 0.0057 & 2D - 2M    \\ 
 15.3865 & 0.0014 & 0.0036 & 2D - M'    \\
 14.1953 & 0.0211 & 0.0006 & 2D + M'    \\
 29.8072 & 0.0035 & 0.0030 & M - M'     \\	
 25.6291 & 0.0002 & 0.0026 & M + M'     \\	
 34.8431 & 0.0109 & 0.0047 & 2D - M - M'  \\
\hline
\end{tabular}
\end{table}

\FloatBarrier
\newpage
\section{Further discussion regarding TCL}\label{ch:tcl_finalremarks}
In summary for this chapter, we established that the formulations of Kopeikin et al.\cite{kopeikin_2024} and Fienga et al.\cite{geoazur} are consistent: despite using different ephemerides (DE440 vs. INPOP21a) and implementations (explicit transformation, Eq.\ref{eq:tcl_tcg}, vs. differential-rate form, Eq.\ref{eq:dtcl_dtt}), the resulting time difference signals agree within our uncertainties; see Table~\ref{tab:freqTCL}.

What would merit further study are the periodic components seen in the residual (Fig.~\ref{fig:powerspectrum_tcltcg_comparison}) and why they resemble the original frequencies found in the signals. Because we only had access to Kopeikin’s final plots (rather than the underlying time-difference dataset), an identical processing chain could not be applied. A definitive test would regenerate the DE440 dataset from scratch and repeat the analysis with strictly matched processing (identical time span, sampling, detrending, windowing to the same day-fraction of the start and end day, and normalization) before comparing time series and spectra. Having a residual with a data span longer than 2 years would also be more insightful for frequency analysis. Otherwise, complementary checks (e.g. coherence/cross-spectral analysis with the luni-solar arguments) could help pinpoint the origin of the lines in the residual. Since DE440 and INPOP21a have inherent differences in their construction (as mentioned in Sect.~\ref{ch:ephemerides_differences}), differences in the signals are expected anyway and therefore should manifest in the shape and form of the residual as well.

Finally, the choice of a practical lunar time scale for future missions remains an open question. However, a necessary requirement is that any precision time scale suitable for navigation -- where nanoseconds correspond to meters -- maintains a convenient and consistent link to established SI definitions. For a lunar navigation system, this means that clocks within the \acrshort{lcrs} must either be directly synchronized to the LCRS attached \acrshort{tcl} or advance according to seconds with a known and fixed rate offset from SI seconds as realized by the TCL (similar to how TT and TCG are fixed rate-offset). Since the speed of light $c$ is exact, the choice of time-scale also defines the unit of length (and any other units dependent on the length-scale, like the mass parameter $GM$ fitted for each planet in the solar system ephemerides). As such, any deviation from \acrshort{tcl} implicitly alters the meter as seen in the Moon’s inertial frame, unless a compensating rescaling is applied. Furthermore, any clock system on the Moon that distributes Earth time -- e.g., compensates in some way for the periodic variations shown in Table~\ref{tab:freqTCL} -- would result in seconds that are not constant in cislunar space.

There are three contenders for lunar time-scale realizations: one option is to adopt \acrshort{tcl} alone as the coordinate-time standard (as \cite{Bourgoin2025} suggests); another is to introduce a \acrshort{tt}-like lunar time (LT for the English abbreviation, or TL for the typical French-style abbreviation) tied to a standardized lunar equipotential surface (a \textit{“selenoid”}). A third possibility is to apply a fixed rate scaling to TCL such that the secular drift compared to TCG is removed and only the harmonic terms remain, as suggested by Fienga et al. in \cite{geoazur}. Either way,  we now turn our attention to clocks on the lunar surface.
\chapter{Timing on the Lunar Surface} \label{ch:surface_clocks}
In this chapter, we begin by generating lunar maps that quantify the gravitational redshift experienced by stationary clocks on the Moon's surface (Sec.~\ref{ch:red_shift_maps}). From this we infer the corresponding requirements on clock performance (Sec.~\ref{ch:surface_clock_requirements}). Next we discuss how non-stationary rovers are described and how the Moon Orientation Parameters might have an effect (Sec.~\ref{ch:moon_orentation_impact}). Finally, we conclude with a discussion of further considerations and open ideas (Sec.~\ref{ch:surface_discussion}).

\section{Stationary red-shift maps} \label{ch:red_shift_maps}
We start this Section on the red-shift maps by describing the underlying datasets used (Sec.~\ref{ch:redhshift_data}). We then present the resulting red-shift maps for the Moon (Sec.~\ref{ch:redhshift_mapsmoon}), followed by analogous results for the Earth (Sec.~\ref{ch:redhshift_mapsearth}) as a useful point of reference.

To compute the redshift of clocks placed on the Moon's surface, we employ the \texttt{python} package \texttt{pyshtools}\cite{pyshtools} -- the python interface for \texttt{SHTools}\cite{shtools} -- which provides a wide range of functions and utilities for working with spherical harmonic models. For the basic data workflow, we took inspiration from the \texttt{shtools} tutorial \texttt{jupyter notebook} \textit{Introduction to Gravity and Magnetic Field Classes}\cite{shtools_notebook}. For the further 3D visualisations presented below we use \texttt{plotly}\cite{plotly} -- an open source \texttt{python} library for interactive plotting and graphing.

\FloatBarrier
% === MOON =============
\subsection{Topographical and gravitational data used}\label{ch:redhshift_data}
\begin{figure}[!htb]
    \centering
    \includegraphics[width=0.6\linewidth]{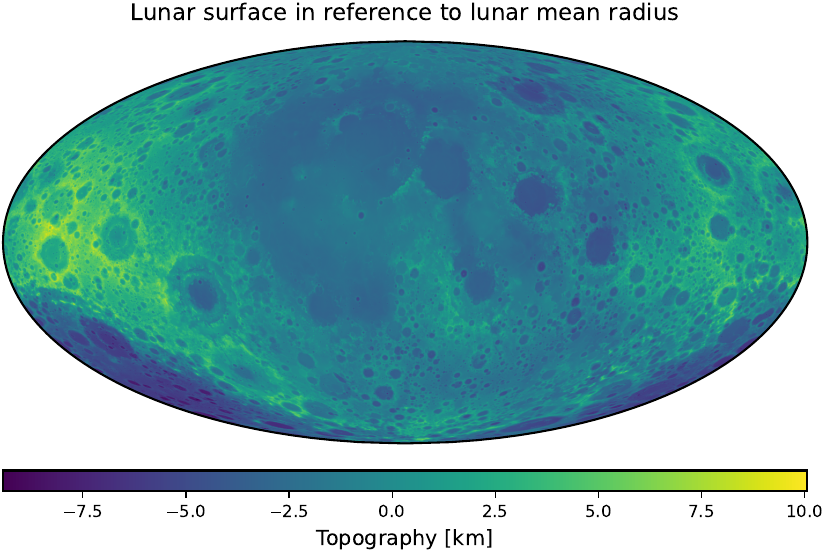}
    \caption{Lunar Topography, obtained from the LDEM128\protect\cite{LDEM128} spherical harmonic model, based on \acrfull{lola} data obtained by the \acrfull{lro} mission and terrain camera data from the Kaguya mission\protect\cite{LDEM128}. Mollweide projection with central longitude chosen as $0^\circ$ -- the Earth-facing side is centered.}
    \label{fig:lunar_topography}
\end{figure}

To construct a map of the gravitational redshift on the lunar surface, both topographic and gravitational field data are required. These datasets are accessed through the \texttt{pyshtools} \textit{datasets} module. As a first step, we load the lunar topography model LDEM128\cite{LDEM128} in the \acrfull{pa} coordinate system. We visualize it using the plotting functions provided by \texttt{pyshtools}, after subtracting the mean lunar radius of $r_\text{moon}=\SI{1738000}{m}$ and projecting the resulting topography in a Mollweide projection, using the optional \texttt{python} package \texttt{cartopy}\cite{cartopy}, see Fig.~\ref{fig:lunar_topography}

Setting up a software environment for this workflow using the Mollweide projection with no package version conflicts proved to be non-trivial. We used an \textit{Ubuntu 24.04} virtual machine running on \acrfullinv{wsl}. We created a dedicated virtual environment with \texttt{python~3.12.3} to host the \textit{Jupyter} server (accessed via our \textit{VS Code} Development Environment). Within this system, the package versions specified were installed using the \texttt{python} package manager \texttt{pip}. It should be noted that this environment was not compatible with the separate \texttt{python} environment used below (Sec.~\ref{ch:orbit_clocks}) with the \texttt{GODOT} simulation tool.

\begin{figure}[!htb]
    \centering
    \includegraphics[width=0.6\linewidth]{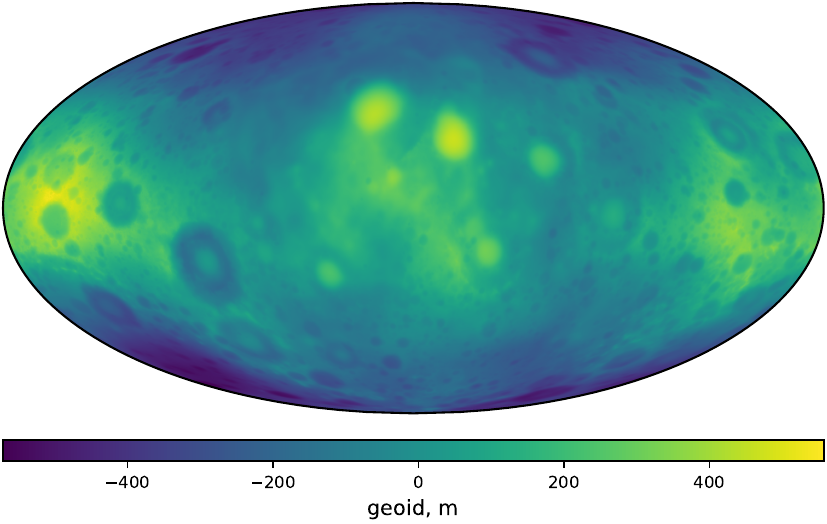}
    \caption{Lunar Geoid (with $u_0=\SI{2.82100e6}{m^2/s^2}$ arbitrarily set), obtained from the GRGM900C\protect\cite{GRGM900C} lunar gravity model in the principal axis coordinate system, from \acrshort{grail} primary and extended mission data.}
    \label{fig:lunar_geoid}
\end{figure}

Next, we load the lunar gravity model GRGM900C\cite{GRGM900C} -- a high-resolution degree-900 model derived from \acrshort{nasa}’s \acrshort{grail} primary and extended mission data, provided in the \acrfull{pa} reference frame consistent with the topography dataset. To visualize and use the gravity field, we compute a lunar geoid (Fig.~\ref{fig:lunar_geoid}) using the \texttt{pyshtools} \texttt{geoid()} method. For this purpose, we adopt an arbitrary reference potential $u_0=\SI{2.82100e6}{m^2/s^2}$, chosen close to the value observed at the mean lunar radius $r_\text{moon}$. The function expands the gravitational potential in a Taylor series on a spherical reference surface of radius $r_\text{moon}$ and solves for the height above this idealized surface to the reference potential $u_0$. The algorithm also accounts for the pseudo-rotational potential through the Moon’s angular rotation rate, and can optionally reference the geoid to a flattened ellipsoid defined by semi-major axis $a$ and flattening $f$ (as done below in Sec.~\ref{ch:redhshift_mapsearth} for the Earth).

\vspace{1em}

\FloatBarrier
\subsection{Generated time-drift maps in 2D and 3D}\label{ch:redhshift_mapsmoon}
As a geoid describes the height of the reference potential $u_0$ above/below an idealized shape (for the Moon, we used a sphere of radius $r_0$), and we have the topography height in reference to the same shape (Fig.~\ref{fig:lunar_topography}), we compute the \textit{orthometric height} as the distance to the selenoid:
\begin{equation}\label{eq:distance_to_geoid}
    h_\text{orth} = h_\text{topo} - h_\text{geoid}
\end{equation}
So, for example, when topographic height and geoid height match, the location is at the reference potential (gravitationally at height zero); and in case the geoid height is \SI{10}{m} smaller (or bigger) than the topographic height, the orthometric height would be \SI{10}{m} (or \SI{-10}{m}).

Since our chosen reference potential $u_0$ does not exactly correspond with the lunar mean radius $r_\text{moon}$, we introduce $r_0= GM_\text{moon}/u_0$, since $U(r_0)=GM_\text{moon}/r_0=u_0$. From Newtonian gravity, we know that the gravitational potential has a $1/r$ dependence, and thus we can write the gravitational potential at the surface as
\begin{equation}\label{eq:surface_potential}
    V_\text{surface} \propto \frac{GM_\text{moon}}{r_ \text{surface}}=\frac{GM_\text{moon}}{r_0}\frac{r_0}{r_\text{surface}}=\frac{u_0\,r_0}{r_0+h_\text{orth}}
\end{equation}
and the fractional time-dilation (red-shift) in reference to the Moon's center of gravity is
\begin{equation}\label{eq:timedilation_from_potential}
    \frac{\Delta t}{t}=\frac{d\text{(LT-TCL)}}{d\text{TCL}}=-\frac{V_\text{surface}}{c^2}
\end{equation}

\begin{figure}[!htb]
    \centering
    \includegraphics[width=0.65\linewidth]{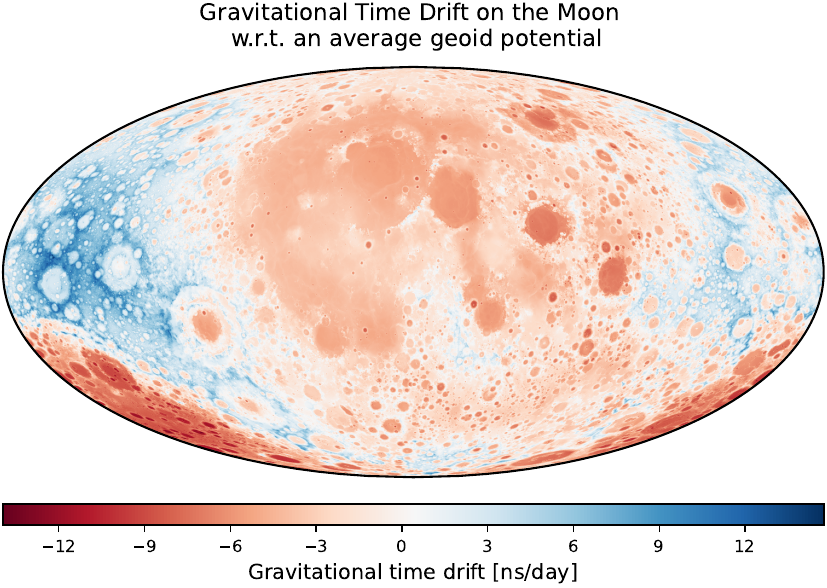}
    \caption{Gravitational time-dilation map of the lunar surface. The geoid potential (where the depicted drift is 0) was arbitrarily chosen as $u_0=\SI{2.82100e+06}{m^2/s^2}$. The low regions (mainly the lunar mare) are color-coded red, as they are red-shifted (clocks run slower), compared to the reference height. The higher regions are blue-shifted, as clocks run faster.}
    \label{fig:redshift_moon_2d}
\end{figure}

Using Equations~\ref{eq:distance_to_geoid}-\ref{eq:timedilation_from_potential}, we generate the following lunar map (Fig.~\ref{fig:redshift_moon_2d}) for the gravitational time-dilation of a static observer on the Moon's surface. The secular drift at the reference potential $u_0$ with $\Delta t/t = \SI{-3.1388e-11}{}=\SI{-2.7119}{\mu s/day}$ has been subtracted. 
We find a time-drift contribution, due to the Moon's gravity and surface topography, of in between about \SI{\pm 15}{ns/day} -- a maximum expected drift of \SI{\sim28.7}{ns/day} from lowest to highest elevation. 
\vspace{1em}

\begin{figure}[htb]
    \centering
    % First row
    \begin{subfigure}[t]{0.48\linewidth}
        \centering
        \includegraphics[width=\linewidth]{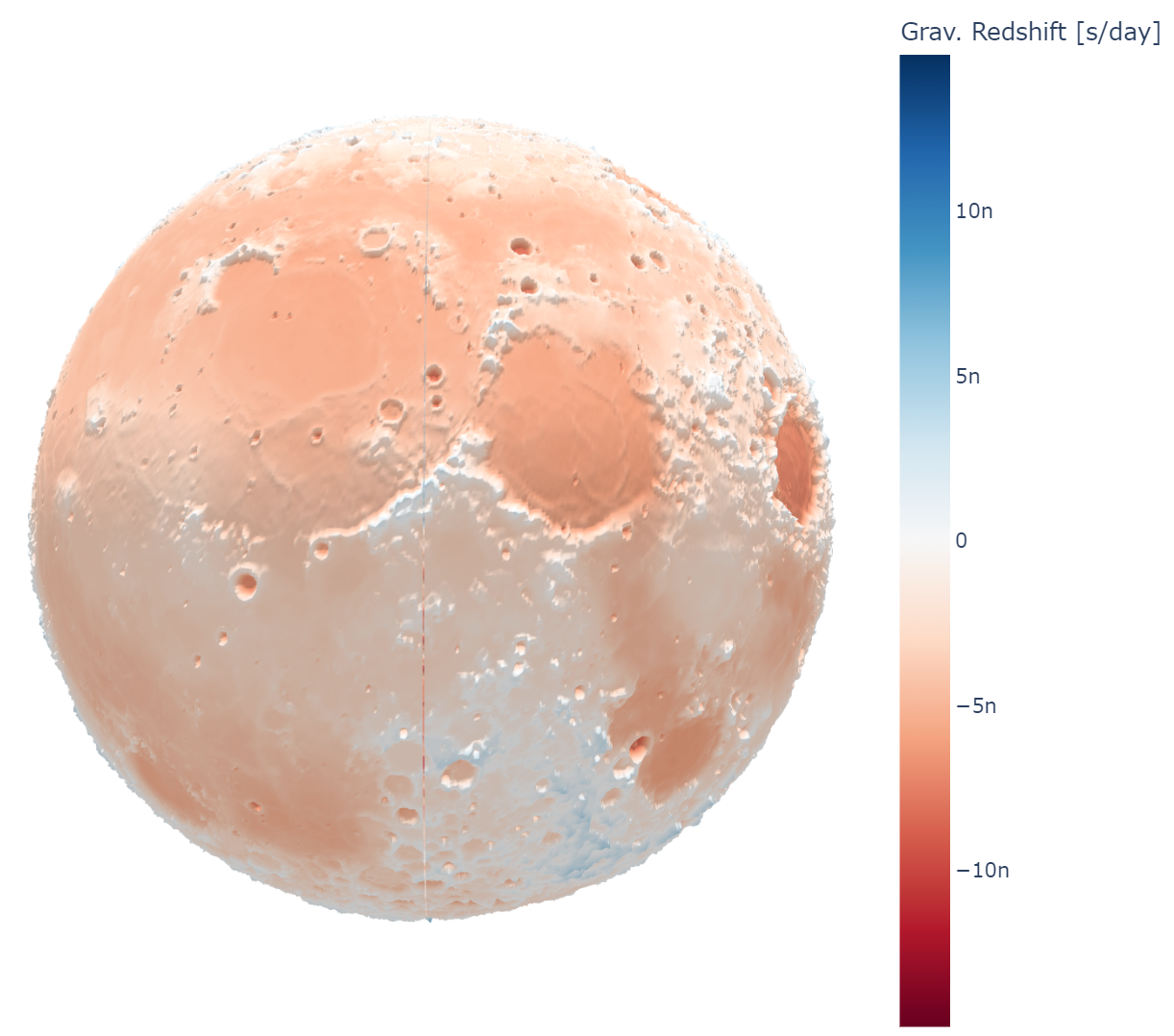}
        \caption{Nearside view}
        \label{fig:moon_nearside}
    \end{subfigure}
    \hfill
    \begin{subfigure}[t]{0.48\linewidth}
        \centering
        \includegraphics[width=\linewidth]{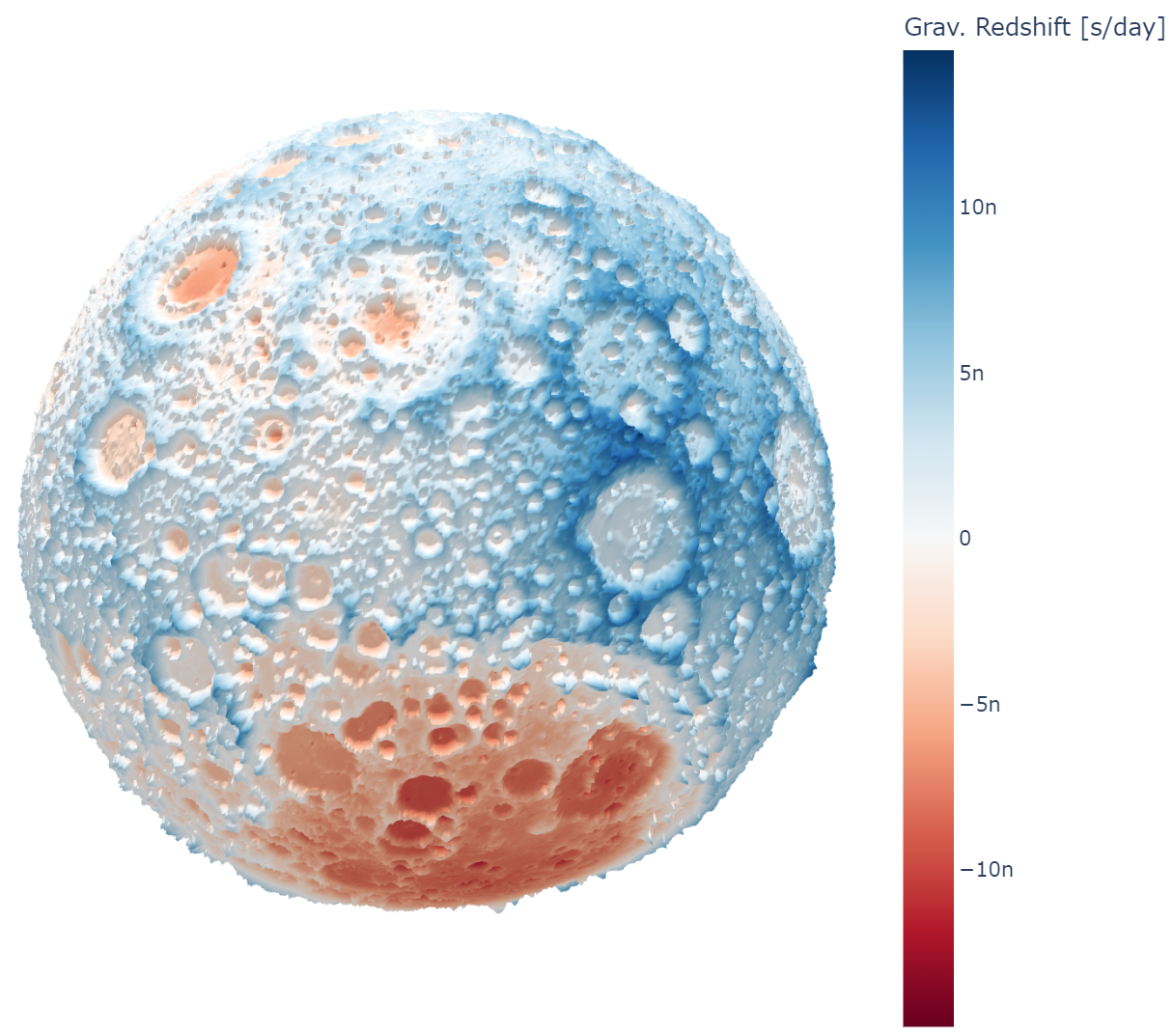}
        \caption{Farside view}
        \label{fig:moon_farside}
    \end{subfigure}
    \vspace{0.5em}
    % Second row
    \begin{subfigure}[t]{0.48\linewidth}
        \centering
        \includegraphics[width=\linewidth]{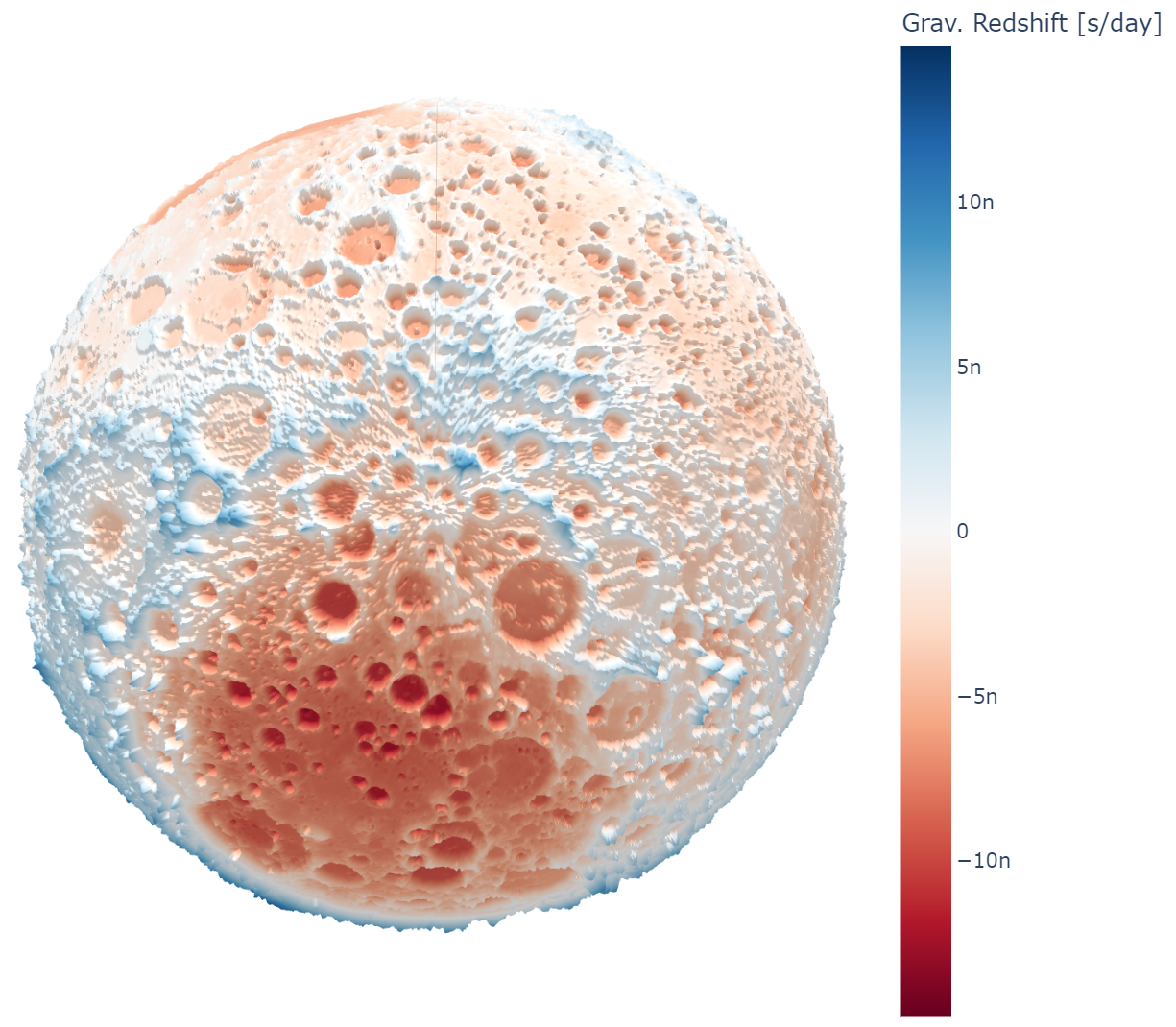}
        \caption{South pole view}
        \label{fig:moon_southpole}
    \end{subfigure}
    \hfill
    \begin{subfigure}[t]{0.48\linewidth}
        \centering
        \includegraphics[width=\linewidth]{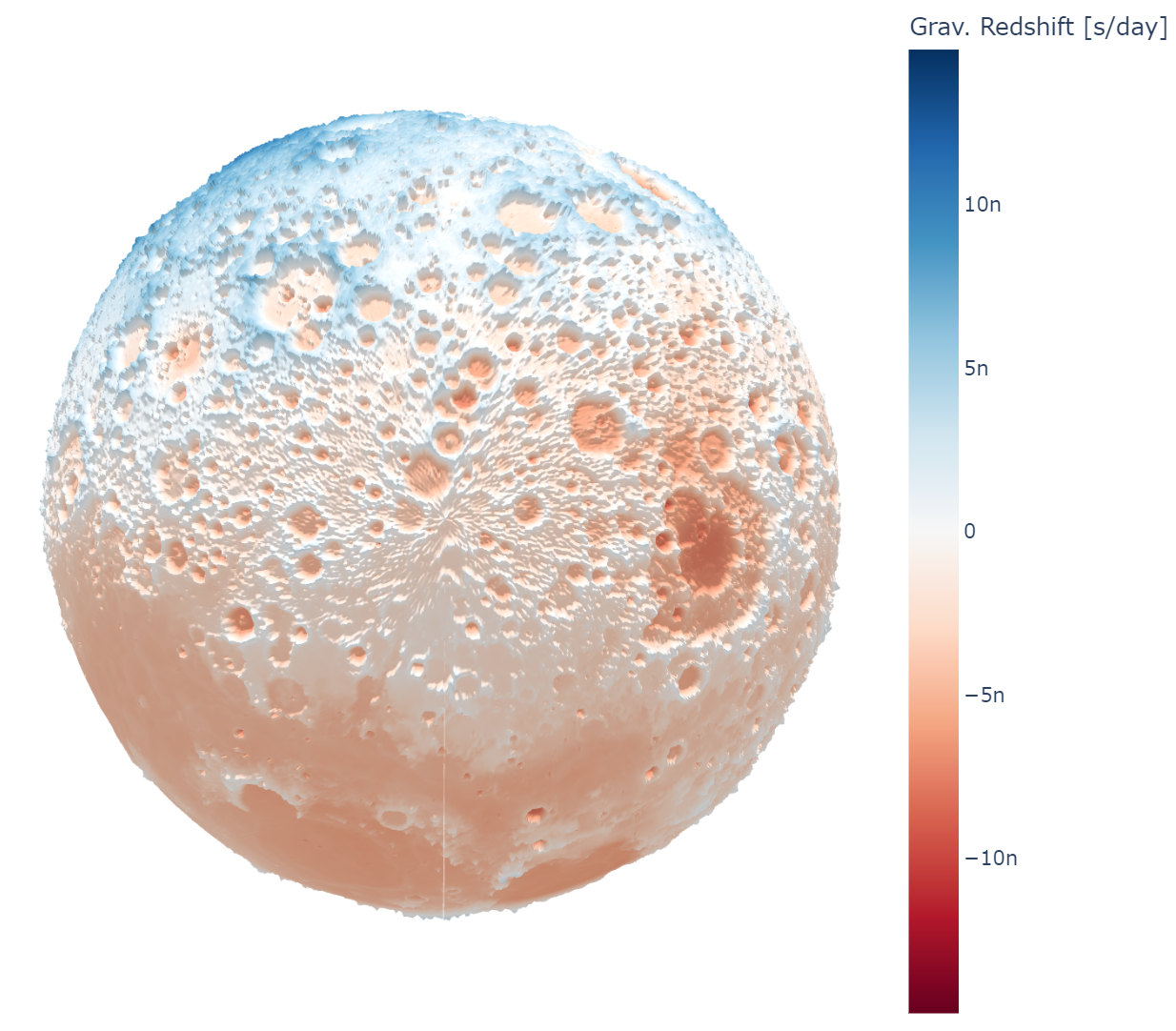}
        \caption{North pole view}
        \label{fig:moon_northpole}
    \end{subfigure}

    \caption{3D visualizations of the Moon's surface gravitational redshift from different perspectives. Variation of \SI{29}{ns} drift per day from lowest to highest regions. }
    \label{fig:moon_4views}
\end{figure}

\newpage
We also depict this gravitational time dilation map as an interactive 3D graphic, implemented with the \texttt{plotly} \texttt{graph\_objects} class, converting the map grid into a 3D Cartesian mesh. Fig.~\ref{fig:moon_4views} shows a couple of views of this interactive representation\footnote{we share the 3D interactive maps on: \scriptsize{\url{https://yanseyffert.github.io/MASS_Thesis_LunarTime/}}}. For an \textit{orthographic} projection centred on the Moon's south pole, see Fig.~\ref{fig:moon_soutpole_othographic} in Appendix~\ref{app:moon_ortho_soud}, which is similar to Fig.~\ref{fig:moon_southpole}.

Compared with a fairly recent study by Bourgoin et al.\cite{Bourgoin2025}, our results are in good agreement, see Appendix \ref{app:moon} for details.

\vfill

\vfill

\newpage
\FloatBarrier
% === EARTH =============
\subsection{Comparison with similar Earth maps}\label{ch:redhshift_mapsearth}
To both validate our method for producing the lunar maps of Sec.\ref{ch:redhshift_mapsmoon} and to establish a useful terrestrial reference, we applied the same pipeline to Earth’s topographic and gravitational field data. Specifically, we used the Earth2014 model\cite{earth2014} together with the EGM2008 gravity field~\cite{egm2008} evaluated up to degree 900; plots of the corresponding topography and geoid (with respect to the WGS84 ellipsoid) are provided in Appendix~\ref{app:earth}. 

\begin{figure}[!htb]
    \centering
    \includegraphics[width=0.65\linewidth]{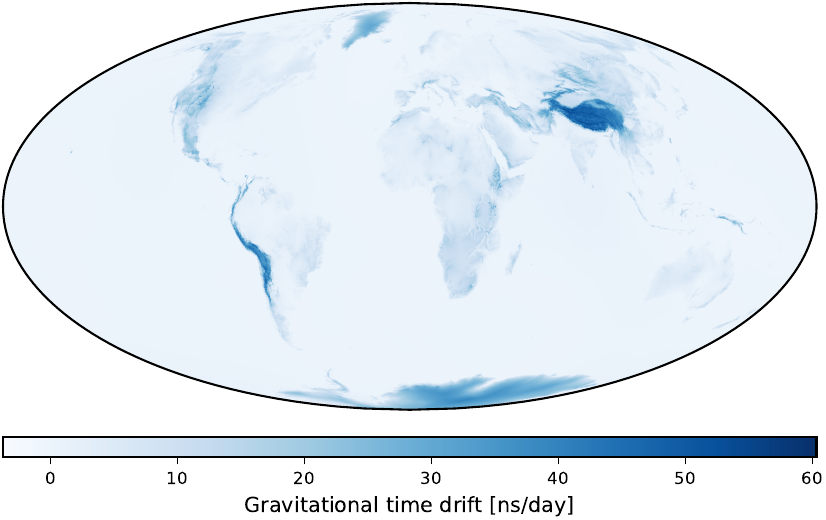}
    \caption{Mollweide projection of Earth's gravitational redshift for stationary clocks on its surface. The surface, where a drift of 0 ns/day would be at the equipotential surface defining the geoid, see Fig.~\ref{fig:earth_geoid} and \ref{fig:earth_potsdam_potato} in the Appendix.}
    \label{fig:redshift_earth_2d}
\end{figure}

We obtain the map, shown in Fig.~\ref{fig:redshift_earth_2d}, that depicts the relative frequency shift of stationary clocks on Earth’s surface (including land, water, and ice). The standard geoid approximating mean sea level was chosen as the zero-drift reference. Relative to this reference, clocks appear blue-shifted (as they run faster). We find a maximum variation of about \SI{64}{ns/day} between the lowest and the highest elevations. 

Notably, the geoid does not always coincide with the actual orthometric sea level height. In the Western Pacific near Indonesia, where subduction zones create a local mass excess, sea level lies about \SI{100}{m} below the geoid; another known mass density anomaly lies in the Indian Ocean, where sea level rises above the geoid, reflecting a low-density mantle beneath the region; see supplemental Fig.~\ref{fig:sealevel_nonzero} in Appendix \ref{app:earth}. 

As viewed from the geoid, we determine a maximum drift of about \SI{60}{ns/day} (corresponding to the Himalaya region with a maximum elevation of \SI{6.38}{km} in our at degree 900 evaluated spherical harmonics dataset) -- only about twice as much as the maximum variation on the Moon of roughly \SI{30}{ns/day} (albeit this is over an elevation change of \SI{19.49}{km}).

\vfill
\begin{figure}[htb]
    \centering
    % First row
    \begin{subfigure}[t]{0.48\linewidth}
        \centering
        \includegraphics[width=\linewidth]{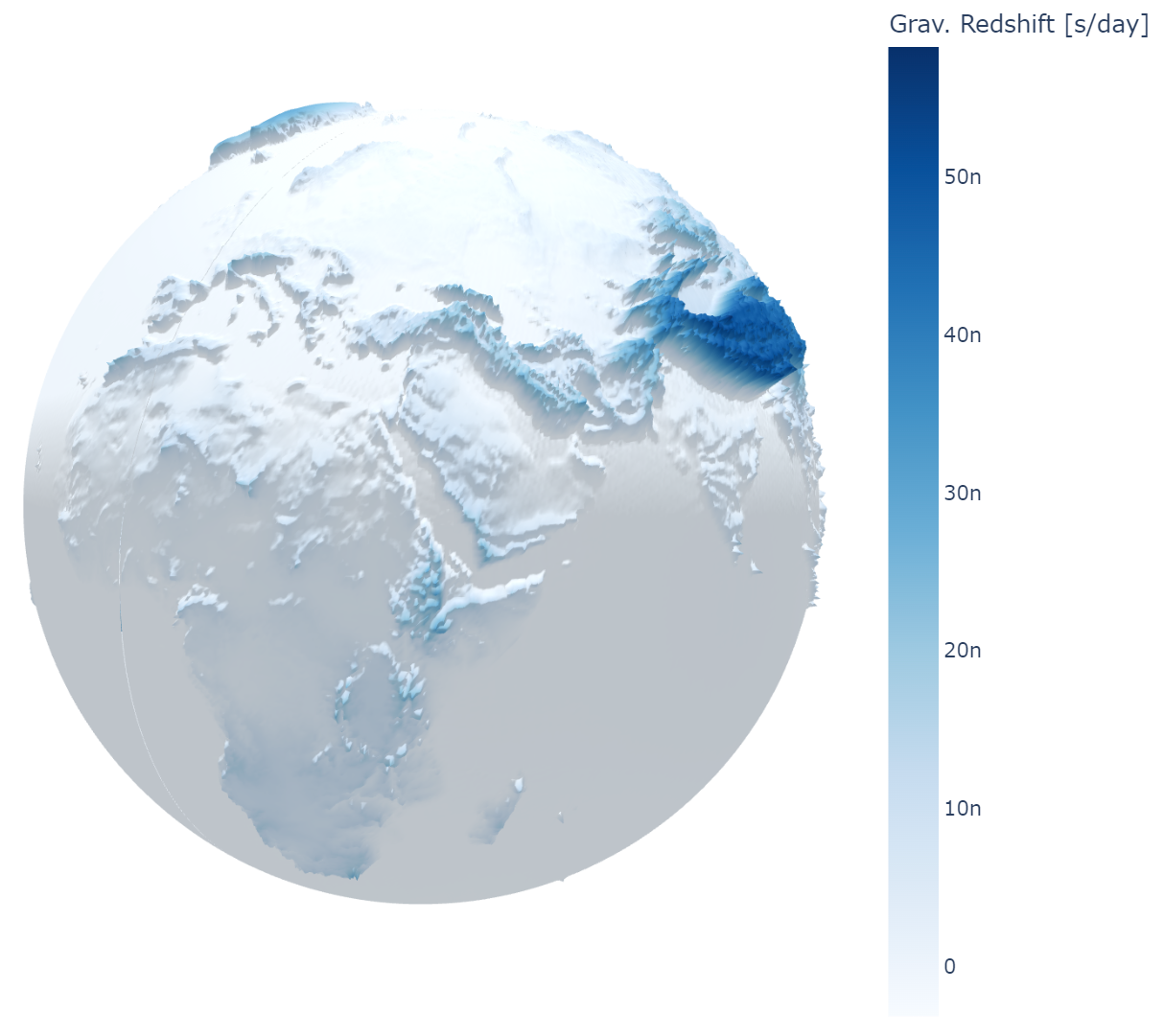}
        \caption{Earth view 1}
        \label{fig:redshift_earth1}
    \end{subfigure}
    \hfill
    \begin{subfigure}[t]{0.48\linewidth}
        \centering
        \includegraphics[width=\linewidth]{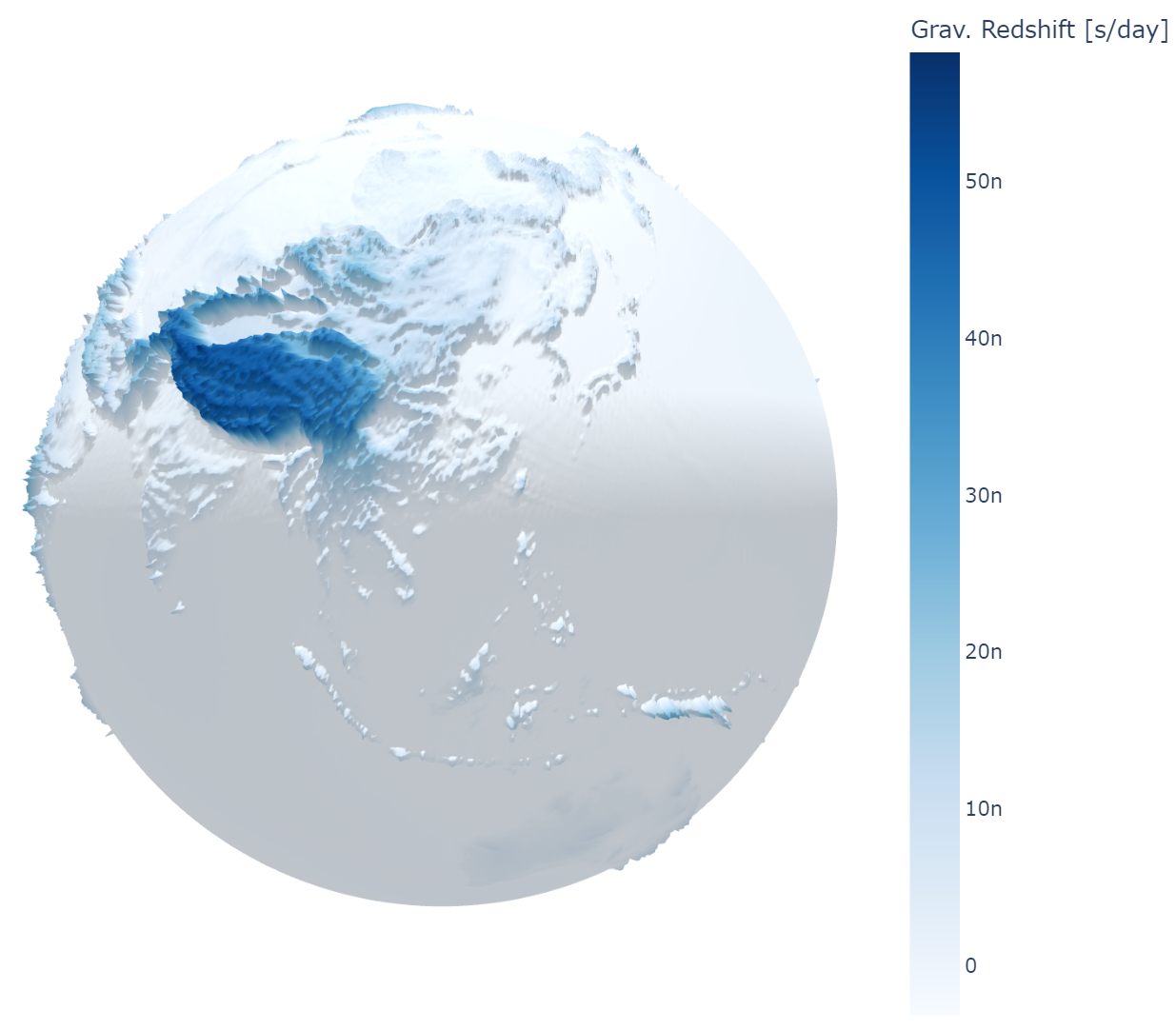}
        \caption{Earth view 2}
        \label{fig:redshift_earth2}
    \end{subfigure}
    \vspace{0.5em}
    % Second row
    \begin{subfigure}[t]{0.48\linewidth}
        \centering
        \includegraphics[width=\linewidth]{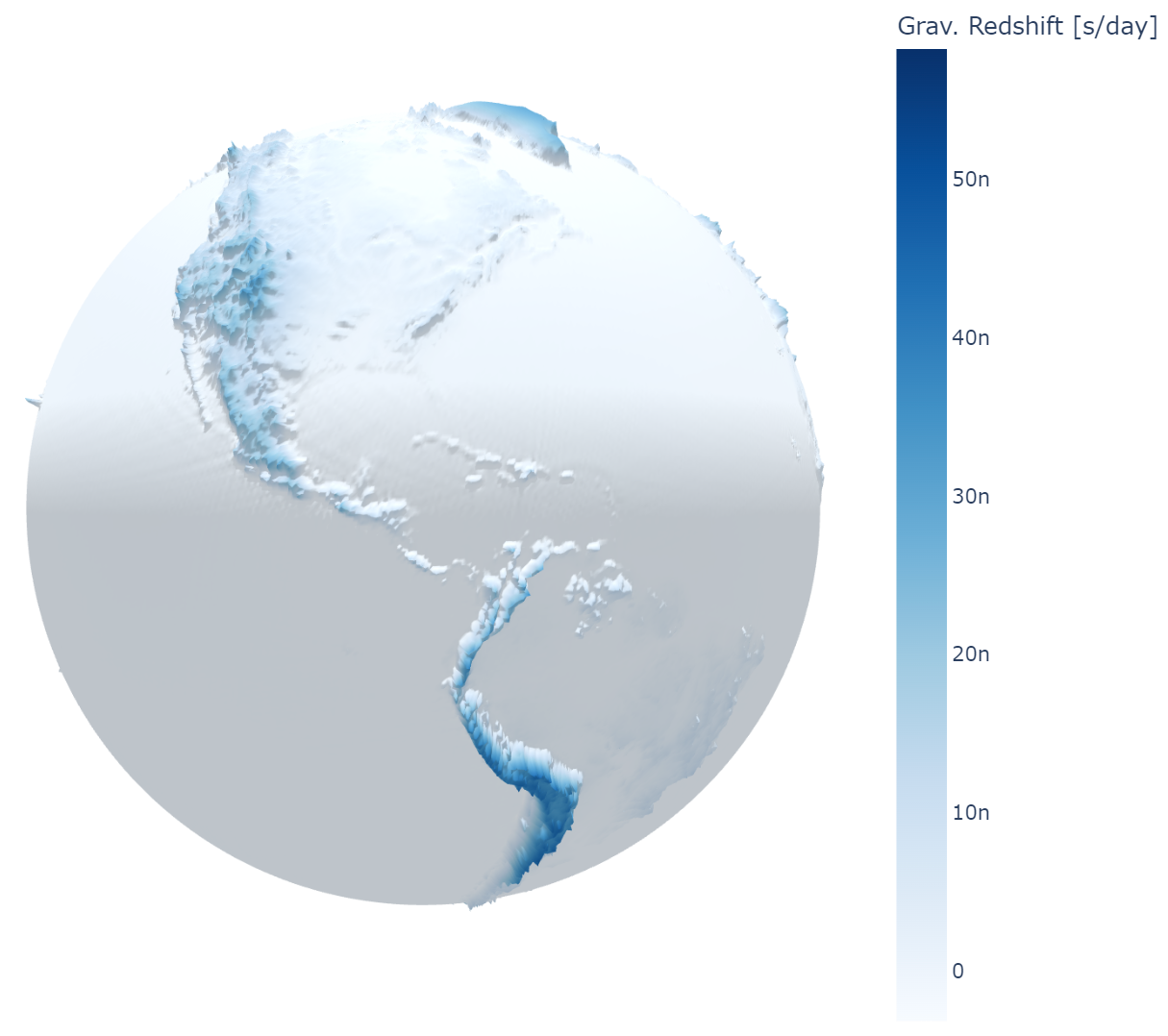}
        \caption{Earth view 3}
        \label{fig:redshift_earth3}
    \end{subfigure}
    \hfill
    \begin{subfigure}[t]{0.48\linewidth}
        \centering
        \includegraphics[width=\linewidth]{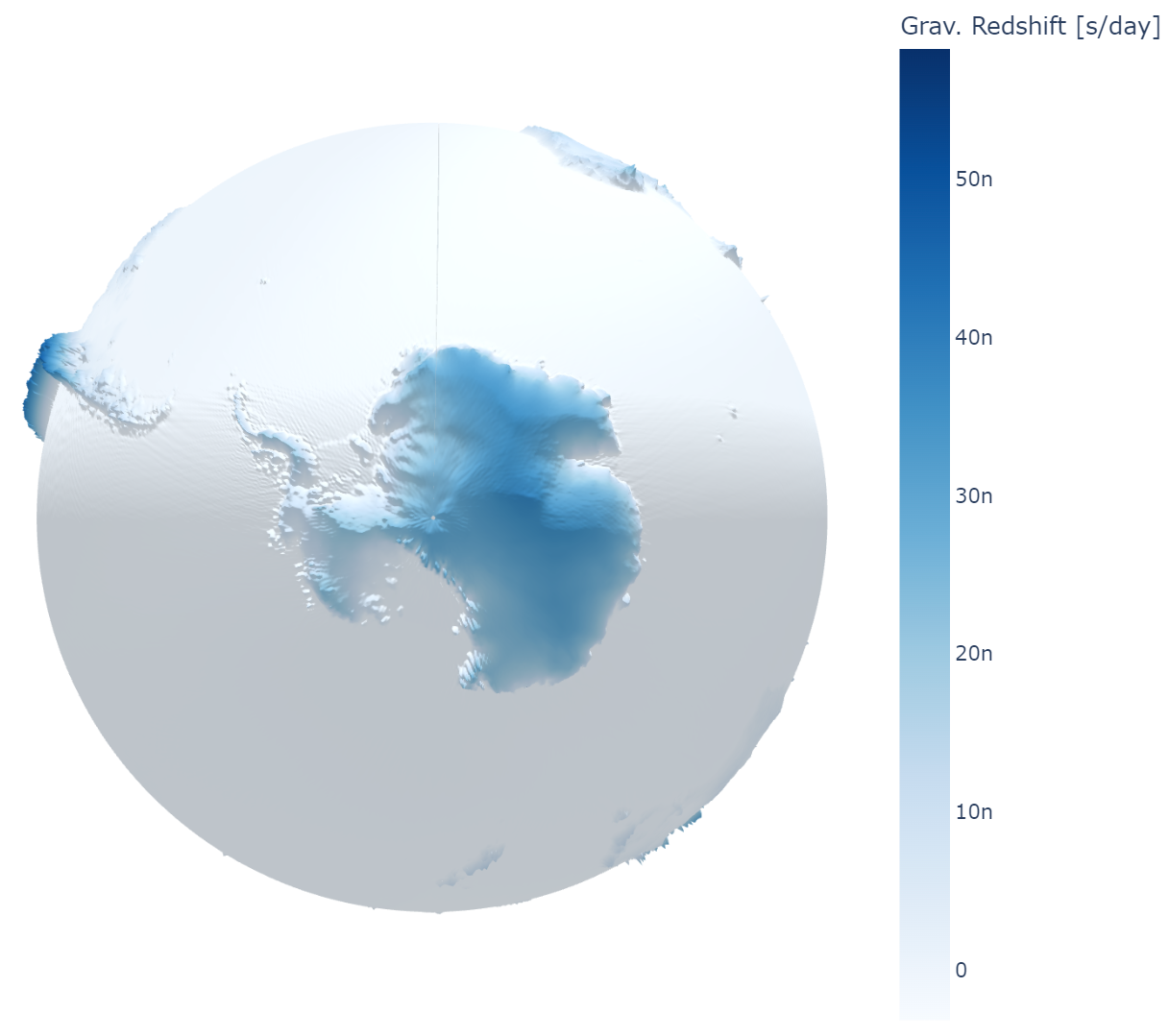}
        \caption{Earth view 4}
        \label{fig:redshift_earth4}
    \end{subfigure}

    \caption{3D visualizations of Earth's gravitational redshift on Earth's surface. A range of about 60 ns/day is observed, from mean sea level to highest elevations.}
    \label{fig:redshift_earth_4views}
\end{figure}

\vfill
\newpage
We also present the Earth gravitational time-dilation map as a 3D graphic in Fig.~\ref{fig:redshift_earth_4views}, which shows several views of this interactive representation\footnote{Earth's interactive map is also available online at: \scriptsize{\url{https://yanseyffert.github.io/MASS_Thesis_LunarTime/}}}.
\vfill

\vfill

% === OTHER =============
\clearpage
\FloatBarrier
\section{Clock requirements}\label{ch:surface_clock_requirements}

Regarding gravitational redshift, state-of-the-art portable optical lattice clocks on Earth can resolve geopotential differences with an uncertainty of \SI{2.6}{m^2 s^{-2}}, equivalent to about \SI{27}{cm} in height \cite{chronometric_leveling_PTB_MUC}. Because the Moon’s surface gravity is only $\sim16.5\%$ of Earth’s ($g_{\mathrm moon}\approx\SI{1.62}{m/s^2}$, $g_\mathrm{earth}\approx\SI{9.81}{m/s^2}$), the same fractional-frequency sensitivity yields height resolutions that are $6\times$ poorer on the Moon. On the other hand, Moon-wide orthometric height variations reach $\sim\SI{18.4}{km}$ (degree/order 900 topography in our evaluation) versus $\sim\SI{6.4}{km}$ for Earth at the same resolution, partly compensationg the gravitational disadvantage.

\begin{figure}[!htb]
    \centering
    \includegraphics[width=0.99\linewidth]{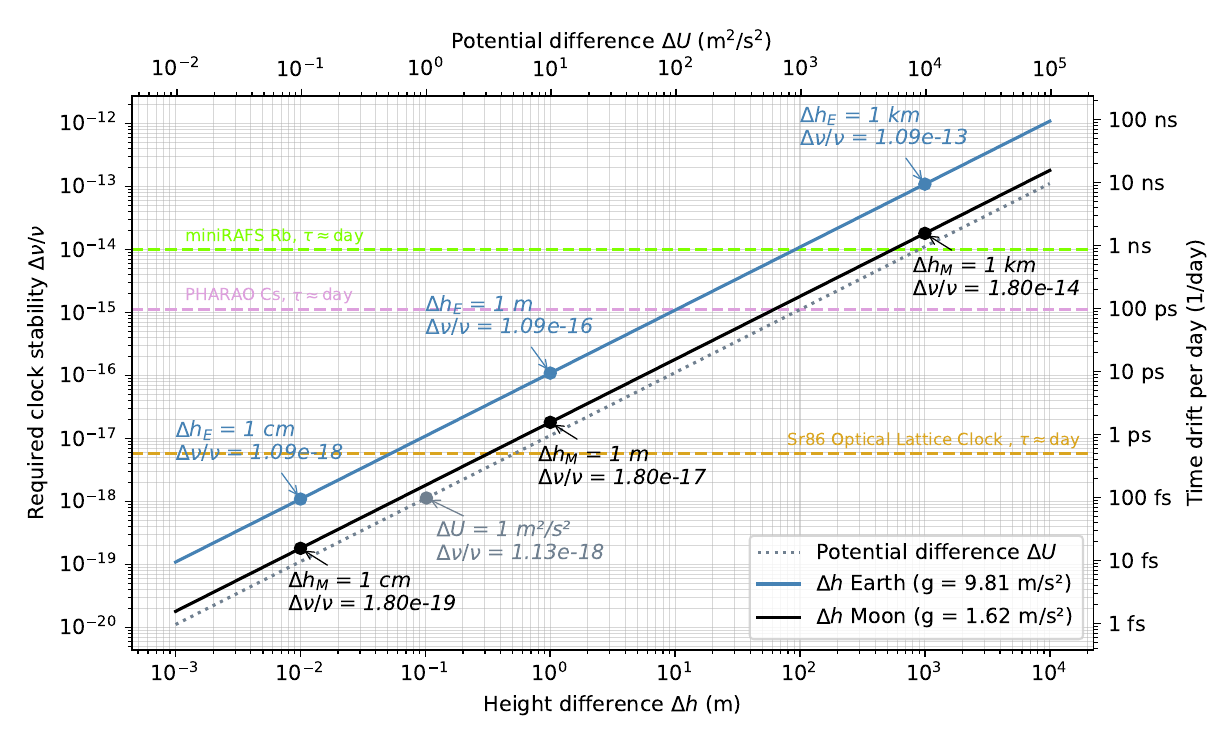} %non-plus with no potential line etc
    \caption{Minimum required clock stability plotted against desired orthometric height resolution (axis below, Earth and Moon lines) or potential difference (axis above, grey dotted line). Frequency stability of selected atomic clocks as horizontal lines for reference. Fractional frequency values are mapped to ns/day values on the right vertical axis.}
    \label{fig:clock_vs_height}
\end{figure}

A rule of thumb linking clock performance to orthometric height resolution is summarized in Table~\ref{tab:orthometric_leveling}. From a given potential-difference resolution $\Delta U$, the required clock stability follows from $\Delta\nu/\nu=\Delta U/c^{2}$; the corresponding height resolution at body $b$ is $\Delta h_b=\Delta U/g_b$. A visual mapping is shown in Fig.~\ref{fig:clock_vs_height}. Experimental campaigns like in \cite{chronometric_leveling_PTB_MUC} typically include a safety margin (typically a coverage factor $2$ for a 95\% two-sided confidence\cite{NIST_CUU_Coverage}) to achieve a statistically significant detection.

\begin{table}[!htb]
    \centering
    \caption{Clock stability vs. potential and resulting orthometric height resolutions.}\label{tab:orthometric_leveling}
        \begin{tabular}{lcll}
        \hline
        \textbf{Clock stability} $\Delta\nu/\nu$ &
        \textbf{Potential} $\Delta U$  &
        \textbf{Height} $\Delta h_\mathrm{earth}$ &
        \textbf{Height} $\Delta h_{\mathrm{moon}}$ \\
        \hline
        $1.1\times10^{-13}$ & $10^4\,\mathrm{m^2/s^2}$ & \SI{1}{km}   & \SI{6}{km}   \\
        $1.1\times10^{-16}$ & \SI{10}{m^2/s^2}  & \SI{1}{m}    & \SI{6}{m}    \\
        $1.1\times10^{-17}$ & \SI{1}{m^2/s^2}   & \SI{10}{cm}  & \SI{60}{cm}  \\
        $1.1\times10^{-18}$ & \SI{0.1}{m^2/s^2} & \SI{1}{cm}   & \SI{6}{cm}   \\
        \hline
        \end{tabular}
\end{table}

\clearpage
Ultimately, the realized clock accuracy depends on stability as a function of averaging time~$\tau$ (see Sec.~\ref{ch:clock_properties} on Allan deviation). Chronometric (orthometric) leveling requires a reference clock and a frequency/time-transfer link; each with its own stability, and the weakest element limits the achievable resolution. In the white-FM regime (random, uncorrelated short-term frequency jitter), a convenient model\cite{nist_handbook_frequency_analysis} is
\[
\sigma_y(\tau)\simeq k\,/\sqrt{\tau/s}\;,
\]
where $k$ is the fractional stability at $\tau=\SI{1}{s}$. On log–log axes this appears as a straight line with slope $-\tfrac{1}{2}$. For the example clocks in Fig.~\ref{fig:clock_vs_height}, we evaluated the provided \acrshort{adev} parameterizations \cite{Excelitas_RAFS_2011,iss_clock,chronometric_leveling_PTB_MUC} at $\tau=\SI{1}{day}\approx\SI{86400}{s}$.

We observe in Fig.~\ref{fig:clock_vs_height} that a miniRAFS atomic clock, as proposed for NovaMoon, should be able to resolve to about \SI{1}{km} height differences, if compared to a similar or better clock with non-limiting link stability. At the current technological limits, a state-of-the-art optical lattice clock could resolve below \SI{1}{m}, if it were possible to get such a clock pair healthy to the Moon.

\FloatBarrier
\section{Moving clocks and rotating frames}\label{ch:moon_orentation_impact}
So far, we have considered coordinate times and clocks stationary with respect to a gravitational field. To extend the discussion to the proper time of dynamically moving clocks, we need a few additional equations that we have so far conveniently avoided -- because the physics of rotating frames is hard and confusing. In Sec.~\ref{ch:propertime_rovers_orbiters}, we discuss the formulas for the proper time of rovers (and orbiters) applied to cislunar space, followed by a discussion of the Sagnac effect and rotating reference frames in Sec.~\ref{ch:sagnac_effec}, before we investigate in Sec.~\ref{ch:moon_orentation_analysis} which role the Moon Orientation Parameters play for surface stations.

\subsection{Proper time for rovers (and orbiters)}\label{ch:propertime_rovers_orbiters}
For the \textit{proper time} of physical clocks we extend the expression given in Eqs.~\ref{eq:tcb_tcg}/ \ref{eq:xdotv_correction} (Earth), and equivalently Eq.~\ref{eq:tcl_tcb} (Moon). These IAU-recommended \acrshort{1pn} expressions given for a \textit{coordinate time} only involve the external potential $U_\text{ext}$. For \emph{proper time}, rather than \textit{coordinate time}, we must additionally include the potential $U$ of the body itself. This yields (referring to \cite{nelson2006}):  
\begin{equation}\label{eq:1pn_propertime}
    \Delta t - \Delta \tau \approx \frac{1}{c^2} \int_{t_0}^{t} \left( \frac{v_L^2}{2}+U_{\text{ext}}(\vec{x}_L)   \right) dt 
    + \frac{1}{c^2} \int_{t_0}^{t} \left( \frac{V^2}{2} +U_\text{L}(\vec{R}) \right) dt 
    + \left. \frac{\vec{v}_L \cdot \vec{R}}{c^2}  \right|_{t_0}^{t}
\end{equation}
Here, $\vec{x}_L$ and $\vec{v}_L$ denote the barycentric position and velocity of the Moon's center-of-mass, while $\vec{R}$ and $\vec{V}=\dot{\vec{R}}$ are the position and velocity of a clock relative to the Moon, see Fig.~\ref{fig:clock_geometry_moon}. 

\begin{figure}[!b]
    \centering
    \begin{tikzpicture}[scale=2,
        >=Stealth,
        arr/.style   = {thick,->, shorten >=3pt, shorten <=3pt},
        darr/.style  = {dashed, arr,},
        varr/.style  = {thick,->, shorten >=1pt, shorten <=4pt, blue},
]

% Moon
\shade[ball color=blue!20, opacity=0.5] (0,0) circle (0.3);
\node at (0,-0.5) {Moon};
\fill (0,0) circle (0.02);

% Clock position
\coordinate (Clock) at (0.55,0.75);
\draw (Clock) circle (0.04);
\node[above right] at (Clock) {Clock};

% Barycenter
\coordinate (Bary) at (3.5,-0.8);
\draw (Bary) circle (0.04);
\node[below] at (Bary) {Barycenter};

% Vectors
\draw[arr] (Bary) -- (0,0) node[midway,below] {$\vec{x}_L$};
\draw[arr] (Bary) -- (Clock) node[midway,above] {$\vec{x}$};
\draw[arr] (0,0) -- (Clock) node[midway, below right] {$\vec{R}$};

% Velocity vectors (blue)
\draw[varr] (0,0) -- ++(-0.9,-0.6) node[midway,above, sloped ] {$\vec{v}_L$};
\draw[varr] (Clock) -- ++(-1.1,-0.3) node[midway,sloped, above] {$\vec{v}=\vec{v}_L+\dot{\vec{R}}$};

\end{tikzpicture}
    \caption{Geometry of \acrfull{ssb}, Moon and a clock near the Moon.}
    \label{fig:clock_geometry_moon}
\end{figure}

During the derivation, terms involving gravitational accelerations (like $\nabla U_\mathrm{ext}\cdot\vec{R}$ and  $\vec{R}\cdot\vec{a}_L$) cancel out, reflecting \textsc{Einstein}'s Equivalence Principle at the heart of \acrshort{gr}. Most terms are familiar from the \textit{coordinate time} case, except for the new second integral, which represents both the time dilation due to the clock’s velocity in the (inertial) \acrshort{lcrs} and the gravitational redshift due to its position in the Moon’s potential. This can be used to integrate the proper time of a surface rover. For orbiting satellites, the same integral can be evaluated as
\begin{equation}\label{eq:tau_sat}
    \frac{1}{c^2} \int_{t_0}^{t} \left( \frac{V^2}{2} +U_L(\vec{R}) \right) dt 
    = \frac{3GM}{2c^2a}\Delta t + \frac{2\,\vec{R}\cdot\vec{V}}{c^2},
\end{equation}
where $\vec{R}\cdot\vec{V}$ is a scalar and can therefore be calculated in either an inertial or rotating frame. This term must not be confused with the term $\vec{v}_L \cdot \vec{R}$ appearing in Eq.~\ref{eq:1pn_propertime}. Assuming an elliptic orbit for the satellite, this last term is typically approximated as  
\begin{equation}\label{eq:sat_eccentric_term}
    \frac{2}{c^2}\vec{R}\cdot\vec{V}
    =\frac{2}{c^2}\sqrt{GMa}\,e\sin{E},
\end{equation}
emphasizing the reliance on the Keplerian orbital elements: semi-major axis $a$, orbital eccentricity $e$, and eccentric anomaly $E$, see Sec.~\ref{ch:kepler} above.

\FloatBarrier
\subsection{Sagnac effect}\label{ch:sagnac_effec}
Synchronizing clocks on a rotating reference frame (such as the Earth's surface, or here the lunar surface) is non-trivial due to path-dependent synchronization effects -- as for any synchronization procedure, electromagnetic signals need to be sent and received. Since a receiver might be moving, while a timing signal is in transit from the transmission location, a greater or lesser distance will be traversed, compared to when the receiver was not moving. In this context, an asynchronization correction -- often called the \emph{Sagnac correction} -- arises. Assuming we have a reciever/rover $r$ fixed on a rotating body with angular momentum $\vec{\Omega}=(0,0,\omega)$, and a sender/satellite $s$ with a known displacement vector $\vec{D}=\vec{x}_r-\vec{x}_s$ at the moment of signal transmission, we can approximate the correction on the signal propagation time as\cite{Ashby_2003}
\begin{align}
    \Delta t_\text{Sagnac} \approx \frac{1}{c^2}\int_s^r(\vec{\Omega}\times \vec{r}_r)\cdot d\vec{D}
    = \frac{2\,\vec{\Omega}}{c^2}\cdot\frac{r_s\times r_r}{2}
    &=\frac{2\,\vec{\Omega}}{c^2}\cdot \vec{A} \;\;=\; \frac{2\omega A_z}{c^2} \label{eq:sagnac_omegaA} \\
    &=\frac{2\,\omega}{c^2}(x_r\,y_s-y_r\,x_s)\;.  \label{eq:sagnac_xyxy}
\end{align}
Here $A_z$ is the equatorial projection of the area swept, in the rotating frame, by the vector from the rotation axis to the light path between the sender at transmission time and receiver. Alternatively, for straight-path signals, as viewed by an inertial frame, expression~\ref{eq:sagnac_xyxy} with cartesian endpoint coordinates applies -- since we assumed $\vec{\Omega}$ is aligned with the $z$-axis. This expression is more computationally efficient\cite{HuFarrell2024} than the vector expressions. To make things more confusing, the Sagnac term can be more generally written as
\begin{equation}\label{eq:sagnac_vd}
    \Delta t_\text{Sagnac} \approx \frac{1}{c}\,\vec{v_r}\cdot\frac{(\vec{x}_r-\vec{x}_s)}{c}=\frac{\vec{v}_r\cdot\vec{D}}{c^2}\;,
\end{equation}
which might remind us of similar looking terms (scalar product of velocity and position vectors), like from the TCG/TCL/$\tau$ definitions w.r.t. TCB (Eq.~\ref{eq:tcb_tcg}, \ref{eq:tcl_tcb} and \ref{eq:1pn_propertime}) or the periodic term in the proper time of a satellite (Eq.~\ref{eq:tau_sat}). While they look similar, they each refer to different velocities and position/distances in various reference frames, and as such must not be confused with each other. Eq.~\ref{eq:sagnac_vd} is motivated in the following way: $D/c$ is the uncorrected for (receiver not moving) signal's time-of-flight. Multiplying this by the receiver's velocity $v_r$ (say, moving away from the transmission location) approximates the additional distance the signal travels in order to "catch up". Divided by the speed of light $c$ results in an associated time delay. Vector notation accounts for more general geometries. From Eq.~\ref{eq:sagnac_vd} and the assumption of a body-fixed receiver ($\vec{v}_r=\Omega\times\vec{r}_r$), the known Sagnac effect formulas from above (Eqs.~\ref{eq:sagnac_omegaA} and \ref{eq:sagnac_xyxy}) are recovered.

To avoid path-dependent inconsistencies for an uninformed observer, clocks are synchronized in the underlying inertial frame~\cite{Ashby_2003}. In practice, GNSS receivers or time laboratories on Earth that compare clocks must correct for the Sagnac effect. For example, a signal circumnavigating the Earth’s equator accumulates a Sagnac delay of $\SI{207.4}{ns}$~\cite{Ashby_2003}, while for a signal travelling from a \acrshort{gps} satellite to a receiver on the equator can result, depending on geometry, in a path-delay of up to \SI{130}{ns} (\SI{26}{m} of inaccuracy).

The physical interpretation of this correction depends on the chosen reference frame. In the rotating Earth-fixed frame, the receiver is stationary and the correction naturally appears as the Sagnac term (Eq.~\ref{eq:sagnac_omegaA}). In the inertial frame, by contrast, the receiver moves during the signal's flight-time due to the frame's rotation, and the correction is instead viewed as an additional propagation delay caused by the increased path length (Eq.~\ref{eq:sagnac_vd}). Thus, while the vocabulary differs -- \textit{Sagnac effect} in the rotating frame vs. for example \textit{velocity correction} in the inertial frame -- the numerical result is identical in both descriptions.

It is worth noting that the Sagnac formulas here are approximations of the actual range, which may, during their derivation, subtract vectors of different reference frames, which in general will not correctly compute the correct range; for a rigorous discussion and the errors imparted, see~\cite{HuFarrell2024}.

\subsection{The role of the Moon Orientation Parameters}\label{ch:moon_orentation_analysis}

As an initial exploration, we want to establish an upper bound on the effect the Moon Orientation Parameters can have. Due to the variation of the orientation parameters, the greatest velocity changes that can impact timing (as observed from the inertial \acrshort{lcrs} frame) are expected at the Moon's equator. First, we discuss the kinematic timing-dilation impact, followed by the maximum possible impact on the Sagnac term.

\paragraph{Kinematic effects}
We load the lunar orientation parameters from the INPOP file \texttt{I4\_AA201119a.tcheb}, which provides \textit{Chebyshev} coefficients for the Euler angles $(\phi,\theta,\psi)$ and their derivatives; in accordance with the \acrshort{iau} 2009 lunar reference frame. This file is part of the \texttt{GODOT} setup used below in Sec.~\ref{ch:comparing_orbit_equations}. These angles describe the rotation from the \acrshort{itrf} into the body-fixed lunar frame. Using the \texttt{calcephy} interface, the orientation state $(\phi, \theta, \psi, \dot\phi, \dot\theta, \dot\psi)$ at a given epoch -- expressed as a \acrfull{jd}, split into integer and fractional part -- is queried via 
\begin{equation*}
(\phi,\theta,\psi,\,\dot\phi,\dot\theta,\dot\psi) = \texttt{peph.orient\_unit}(\texttt{jd\_int}, \texttt{jd\_frac}, \texttt{NaifId.MOON}).
\end{equation*}

\begin{figure}[!htb]
    \centering
    \includegraphics[width=0.8\linewidth]{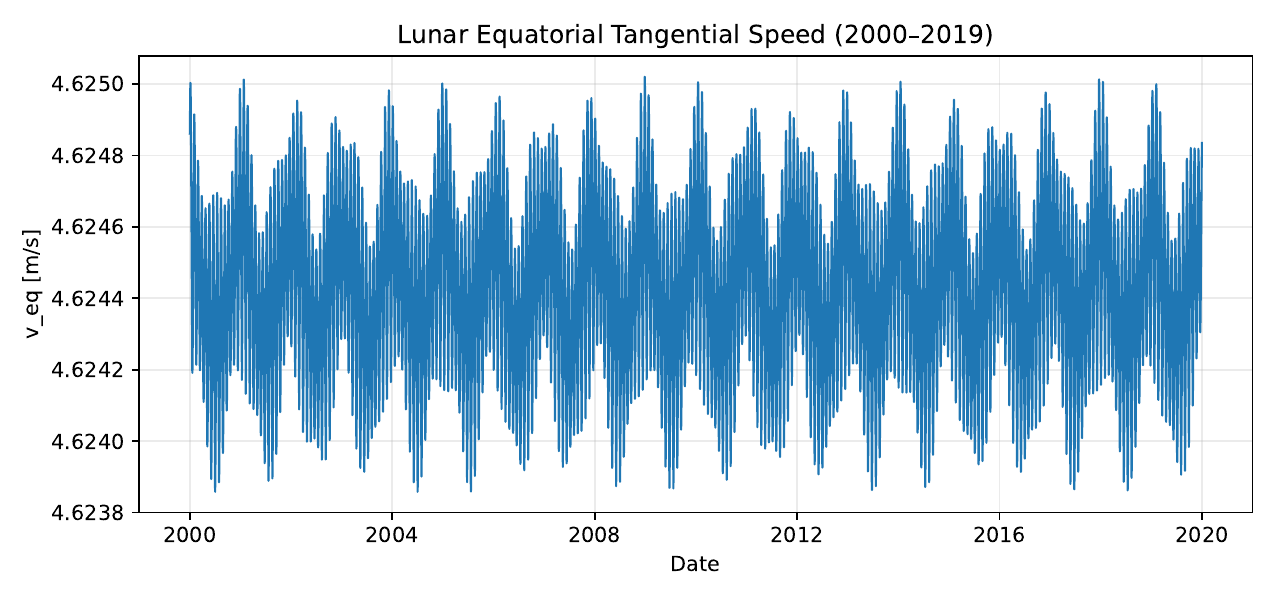}
    \caption{Inertial velocity in the \acrshort{lcrs} frame of clock placed at lunar equator.}
    \label{fig:moon_eq_vel}
\end{figure}

\begin{figure}[!htb]
    \centering
    \includegraphics[width=0.75\linewidth]{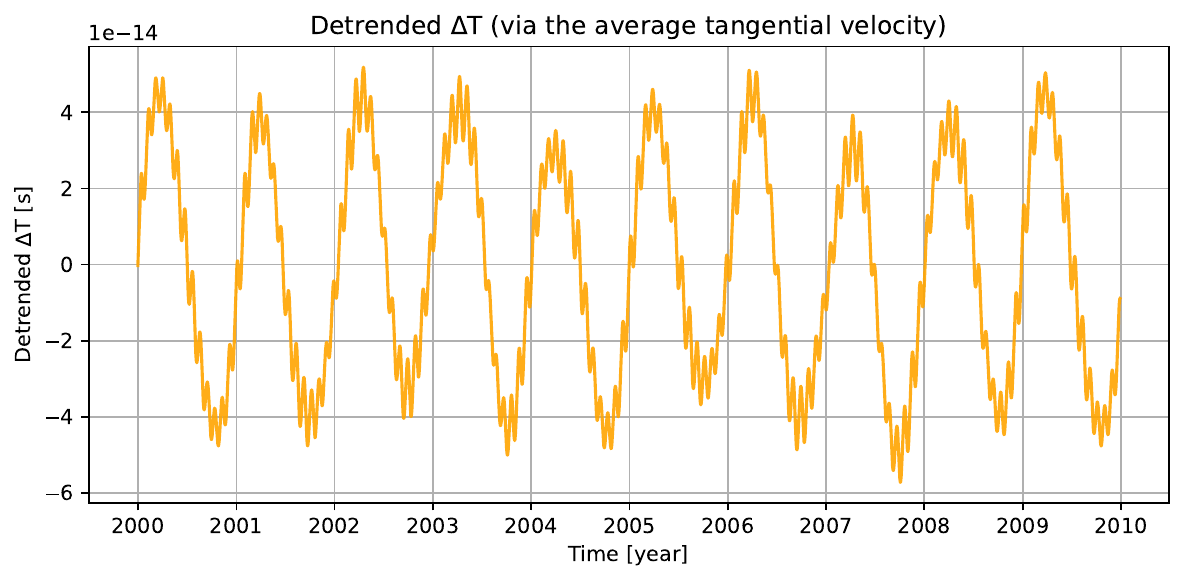}
    \caption{Harmomic components of the kinematic time-dilation at equator, due to the varying Moon orientation parameters. The average drift of \SI{10.28}{ps/day} has been removed.}
    \label{fig:moon_dt_rot_var}
\end{figure}

For rigid body kinematics, the instantaneous angular velocity vector components $\omega_i$ are
\begin{align}
\omega_x &= \dot\theta \cos\psi + \dot\phi \sin\theta \sin\psi, \\
\omega_y &= -\dot\theta \sin\psi + \dot\phi \sin\theta \cos\psi, \\
\omega_z &= \dot\psi + \dot\phi \cos\theta .
\end{align}
We find, as expected, the dominant component as the spin around the $z$-axis,
\begin{equation}
\omega_\text{z} = \dot\psi + \dot\phi \cos\theta \;\approx\; 2.66\times 10^{-6}\,\text{rad/s}\;,
\end{equation}
corresponding to the moons sidereal rotation period of 27.32~days. The equatorial tangential speed then follows as
\begin{equation}
v_\text{eq}(t) = |\omega_\text{z}(t)| \cdot r_\text{moon}\approx\SI{4.624}{m/s}\;,
\end{equation}

with small libration-driven variations, see Fig.~\ref{fig:moon_eq_vel}. From this we integrate the time-drift due to the Moon's rotation as $v_\text{eq}^2/(2c^2)\,\Delta t$, where $\Delta t$ is the interval in seconds between queried epochs. We find an average drift rate of \SI{1.190e-16}{} (\SI{10.28}{ps/day}) and periodic variations, as shown in Fig.~\ref{fig:moon_dt_rot_var}. Over the course of a year, the harmonic terms do not exceed \SI{\pm 50}{fs}. Or \SI{0.1}{ps/6months} as seen in the plot. This is well beyond any modern detection threshold.

\paragraph{Sagnac effect}
The impact of the Moon orientation parameters on the Sagnac effect are also tied to the varying tangential velocity in the \acrshort{lcrs}. The Sagnac delay scales as $v\cdot D / c^{2}$ (Eq.~\ref{eq:sagnac_vd}), so a fractional change in $v$ produces the same fractional change in the delay. Under a worst-case geometry -- an equatorial ground receiver and a transmitter at $D=\SI{70000}{km}$ (\acrshort{nrho}-class distance) -- the baseline Sagnac delay is $v_{\mathrm{eq}}\cdot D/c^{2}\approx\SI{3.6}{ns}$. From Fig.~\ref{fig:moon_eq_vel}, the equatorial speed varies by at most \SI{0.024}{\percent} between extrema, implying a change in the Sagnac term of only $\sim\SI{8.6e-13}{s}$, or equivalently $\sim\SI{0.26}{mm}$ of range error -- a negligible effect. It should be noted that this is only the result if the sender is along the axis of the receiver's velocity (to maximize the dot product). The Sagnac effect is intrinsically suppressed when the sender is above the poles (high-inclination orbits), as would be the case for most orbits we are interested in (see Sec.~\ref{ch:elfos} on orbits).

Refer to \cite{Fateev2015} for a comparative and detailed discussion of Earth's Orientation Parameters and their impact on terrestrial atomic clocks.

\FloatBarrier
\section{Remarks and outlook on further research questions}\label{ch:surface_discussion}

Our approach for computing the gravitational potential for the shown redshift maps is possibly limited to heights near the selenoid. It may be inaccurate, where the mass distributions are not strictly below, but maybe also off to the side, like crater walls. To address this, one could query the acceleration vector at every location on the Moon's topography (e.g., locations on a global grid). This is possible with \texttt{shtools}, but runtimes for multiple points quickly became infeasible. Moreover, because the underlying \texttt{Fortran/C++} kernels are not directly exposed by the \texttt{python} interface, we could not optimize true batch evaluation for large sets of topographic points with varying radii.

Even if such a dense sampling were practical, inferring the potential $U$ (and thus the redshift via $U/c^{2}$) solely from accelerations is ill-posed: since $\vec{a}=\nabla U$, one must perform a path integral from infinity (with $U(\infty)\equiv0$) through the three-dimensional field, not just use values on a single surface. Therefore, a more robust approach -- offering full control and a transparent foundation -- would be to evaluate $U(r,\theta,\lambda)$ directly from the field's multipole moments, i.e. computing the potential at varying radii directly from the coefficient set and then summing effectively encodes an integration from infinity. 

Beyond this, one should, at the same time, account for the Moon's rotation and its impact on the local clock rate. In \acrlong{gr}, the time dilation at a point can be understood operationally as the fractional frequency shift of a photon emitted there in an instantaneously comoving and freely-falling (inertial) frame and then, after propagation through curved space-time, received by a specified observer (e.g. at infinity where spacetime is assumed flat). The photons lost energy (and resulting frequency shift) then automatically includes both gravitational and kinematic time-dilation contributions.

Implementing these aspects and validating that pipeline was beyond the scope of this work.

\chapter{Timing in Orbit} \label{ch:orbit_clocks}
This final chapter -- following the exploration on the Lunar Coordinate Time (Ch.~\ref{ch:tcl}) and clocks on the lunar surface (Ch.~\ref{ch:surface_clocks}) -- focuses on the timing of clocks in lunar orbit. First, the relevant equations are covered in Sec.~\ref{ch:orbit_equations}, then Sec.~\ref{ch:elfos} outlines the characteristics of orbits under consideration. In Sec.~\ref{ch:comparing_orbit_equations} we proceed to simulate these orbits and compare timing results across the different forms that the relevant \textit{proper time} equation takes. The chapter concludes with final remarks and comments in Sec.~\ref{ch:orbits_further_comments}.

\section{Relevant equations for the proper time  of moving clocks}\label{ch:orbit_equations}

Referencing Eq.~\ref{eq:tau_sat} and Eq.~\ref{eq:sat_eccentric_term}, the \acrshort{1pn} corrections to calculate the \textit{proper time} $\tau$ of a cislunar clock from the \textit{coordinate time} $t$ of the \acrshort{lcrs} is:
\begin{align}
    \Delta \tau &= \Delta t-\int_{t_0}^{t} \left( \frac{V^2}{2c^2} +\frac{U_L(\vec{R})}{c^2} \right) dt  \\
    &= \left (1-\frac{3GM}{2c^2a}\right ) \Delta t - \frac{2\,\vec{R}\cdot\vec{V}}{c^2} \\ 
    &=\left (1-\frac{3GM}{2c^2a}\right ) \Delta t -\frac{2}{c^2}\sqrt{GMa}\,e\sin{E},
\end{align}
where $V$ and $R$ are the clock's velocity and position in moon-centered inertial coordinates, and we have the semi-major axis $a$, the eccentricity $e$, and the eccentric anomaly $E$ -- the osculating Keplerian orbital elements associated with the orbiting clock. The Moon's gravitational potential is described as $U_L(\vec{R})=\frac{GM}{|\vec{R}|}$ with $M$ being the lunar mass. To compare with a clock situated close to the proposed \textit{selenoid} at a reference distance $r_0$ from the Moon's center-of-mass (implementing \acrshort{lt}), the resulting equations are as follows:
\begin{align}
    \Delta LT-\Delta \tau &= -\frac{GM}{c^2\,r_0} \Delta t+\int_{t_0}^{t} \left( \frac{V^2}{2c^2} +\frac{U_L(\vec{R})}{c^2} \right) dt  \label{eq:dtau_lander_like}\\
    &= \frac{GM}{c^2}\left (\frac{3}{2a}-\frac{1}{r_0}\right ) \Delta t + \frac{2\,\vec{R}\cdot\vec{V}}{c^2} \label{eq:dtau_orbit}\\ 
    &= \frac{GM}{c^2}\left (\frac{3}{2a}-\frac{1}{r_0}\right ) \Delta t +  \frac{2}{c^2}\sqrt{GMa}\,e\sin{E}\;.\label{eq:dtau_kepler}
\end{align}
These three expressions are the ones we will compare in our implementation in Sec.~\ref{ch:comparing_orbit_equations} below. In this work, we refer to Eq.~\ref{eq:dtau_lander_like} as the \textit{Lander-Like Formula}, Eq.~\ref{eq:dtau_orbit} as \textit{Cartesian-Orbital Formula} and, Eq.~\ref{eq:dtau_kepler} as \textit{Keplerian-Orbital Formula} -- since the first general expression is also applicable to a Lander or Rover, the second one uses Cartesian 3D velocity and positions vectors of the vehicle, and the third equation approximates the vector product using Keplerian orbital elements. The factor $\frac{GM}{c^2\, r_0}$, appearing in each version of the formula, can be thought of as the average time-drift per unit time $t$ of \acrshort{tcl}.

\FloatBarrier
\section{Elliptical Lunar Frozen Orbits}\label{ch:elfos}
Earth-based navigation systems, such as \acrshort{gps}, place \acrfullpl{sv} in near-circular \acrfullpl{meo} to minimize the impact of orbital eccentricity, allowing for a straightforward satellite clock frequency-offset to match the SI second as observed on the ground, as done by the \acrshort{gps} constellation\cite{Ashby_2003}.

However, this approach with circular orbits is generally not feasable for lunar navigation system. Most circular lunar orbits are dynamically unstable because of either the Moon's non-spherical gravity field (\textit{lunar mascons}, large \(C_{20}\), \(C_{22}\), and higher-degree multipole moments) affecting low orbits, or strong third-body perturbations from Earth affecting high orbits~\cite{physorg_LunarOrbits, CERESOLI_2025}. Long-lived orbit designs -- that want to minimize station-keeping maneuvers -- therefore rely on so-called frozen orbits, or cislunar families such as \acrfull{nrho} around the Earth–Moon L1/L2 region.

 The class of so-called \acrfullpl{elfo} are notably stable, which means that their Keplerian elements remain relatively constant over time. These orbits are ideal for near-future lunar navigation satellites, particularly for providing coverage of the lunar polar regions. When an elliptical orbit has its apoapsis positioned above the region most users are in, the satellite will linger longer in that part of the sky (due to the minimum speed at apoapsis and maximum at periapsis), while being farthest from the surface, thus enhancing ground coverage during this orbital phase as well. 
 
\begin{figure}[!htb]
    \centering
    \includegraphics[width=0.75\linewidth]{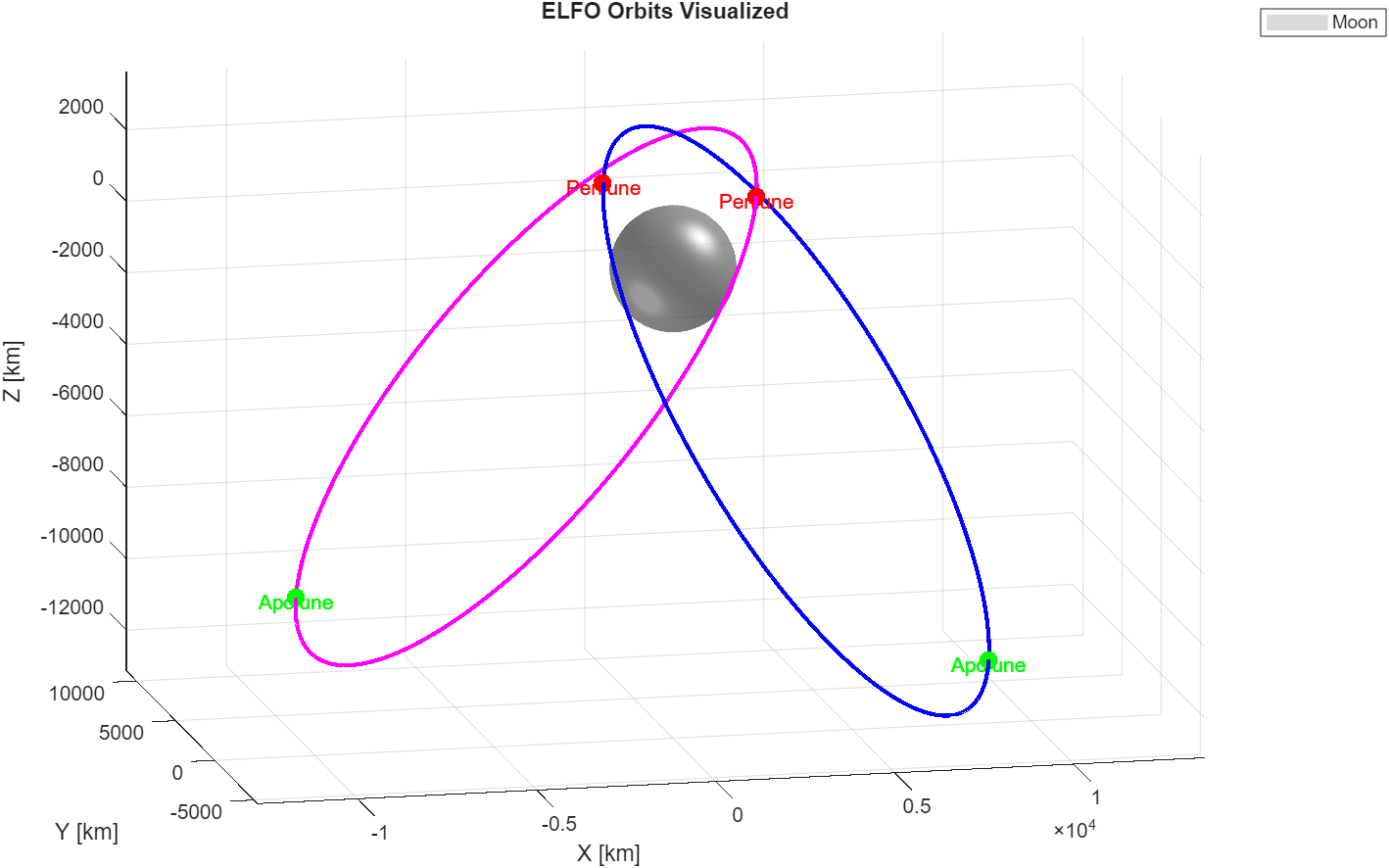}
    \caption{Visualization of the proposed \acrshort{elfo} orbits for a lunar navigation system like Moonlight, using \texttt{MatLab}. Two satellites per orbital plane are envisioned, see Tab.~\ref{tab:orbital_parameters}.}
    \label{fig:orbit_elfo}
\end{figure}

 Figure~\ref{fig:orbit_elfo} illustrates these ELFO orbits\footnote{for an interactive 3D representation see \scriptsize{\url{https://yanseyffert.github.io/MASS_Thesis_LunarTime/}}}. The orbital parameters have been taken from \cite{atlas_2023} and are summarised in Table~\ref{tab:orbital_parameters}. They exhibit an eccentricity of about $0.6$, and thus the varying gravitational potential along these orbits precludes a single constant frequency offset and time-varying relativistic corrections may need to be modeled or steered.

\begin{figure}[!htb]
    \centering
    % First row
    \begin{subfigure}[t]{0.48\linewidth}
        \centering
        \includegraphics[width=\linewidth]{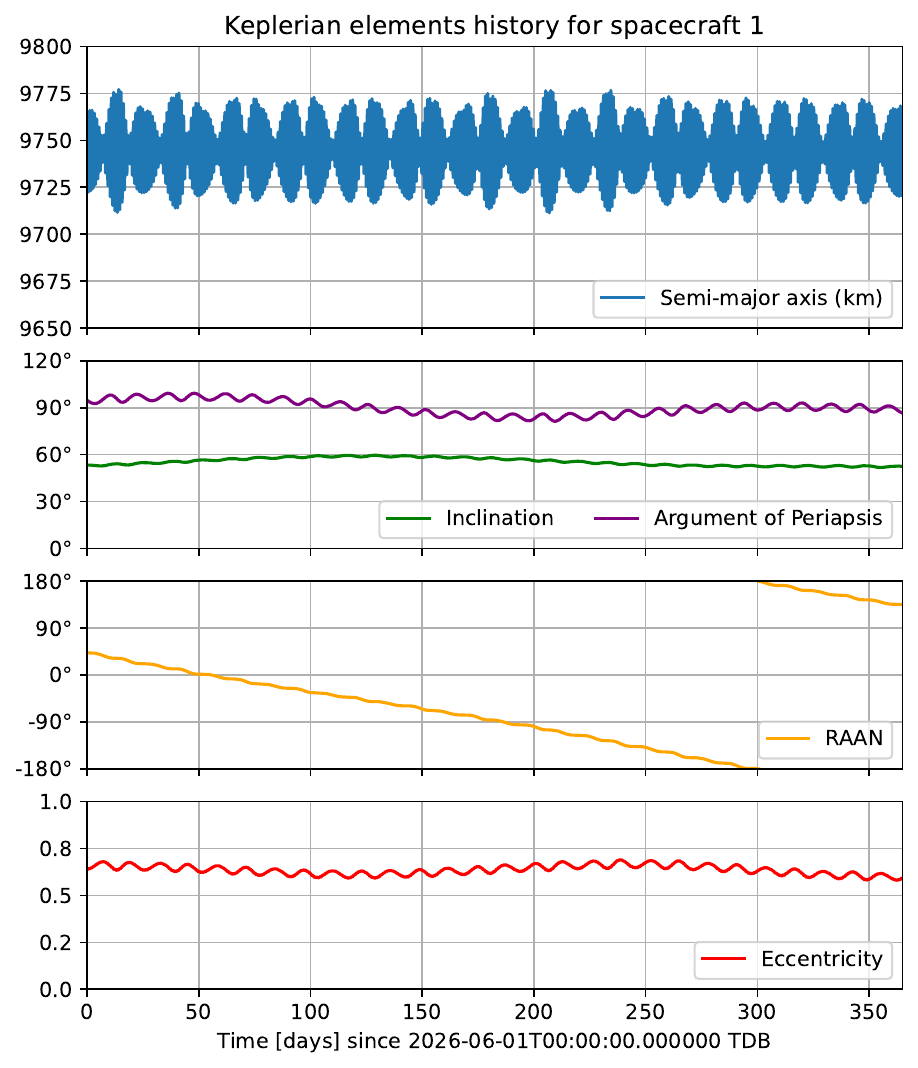}
        \caption{Spacecraft 1}
        \label{fig:kepler1}
    \end{subfigure}
    \hfill
    \begin{subfigure}[t]{0.48\linewidth}
        \centering
        \includegraphics[width=\linewidth]{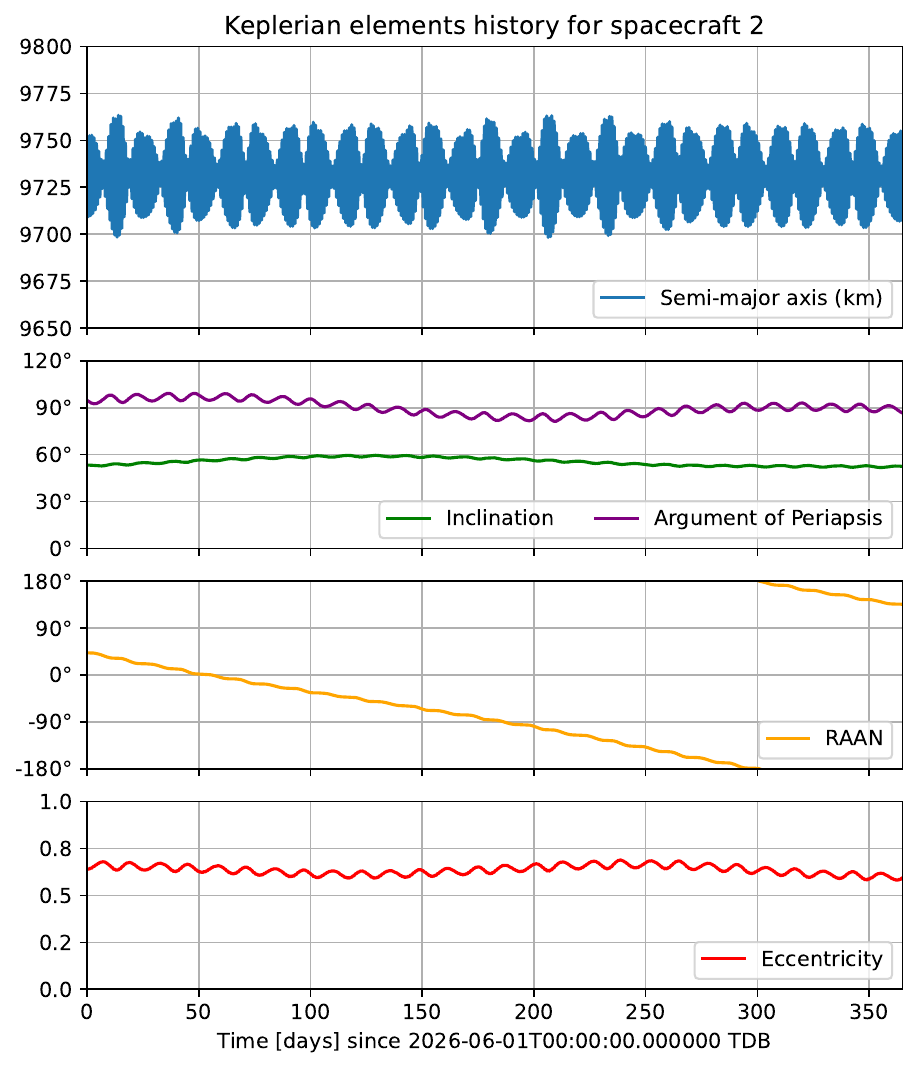}
        \caption{Spacecraft 2}
        \label{fig:kepler2}
    \end{subfigure}

    \vspace{0.5em} % spacing between rows

    % Second row
    \begin{subfigure}[t]{0.48\linewidth}
        \centering
        \includegraphics[width=\linewidth]{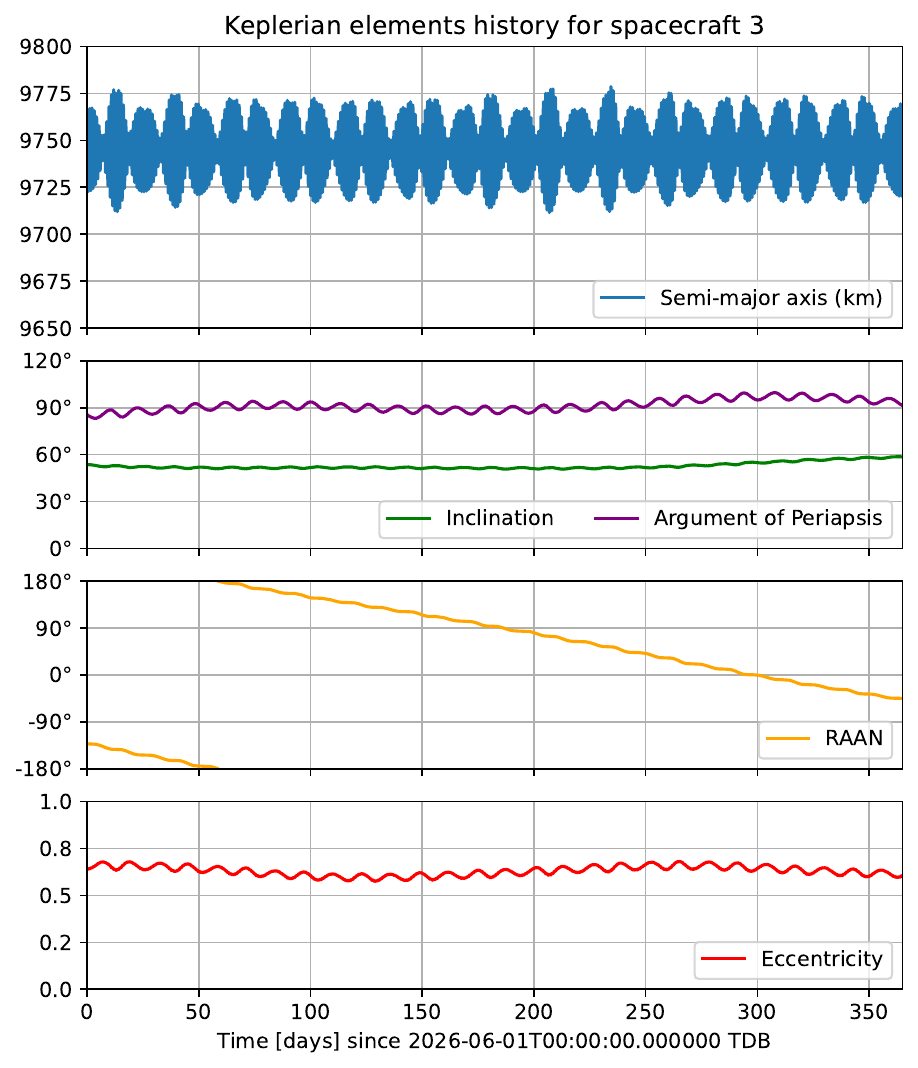}
        \caption{Spacecraft 3}
        \label{fig:kepler3}
    \end{subfigure}
    \hfill
    \begin{subfigure}[t]{0.48\linewidth}
        \centering
        \includegraphics[width=\linewidth]{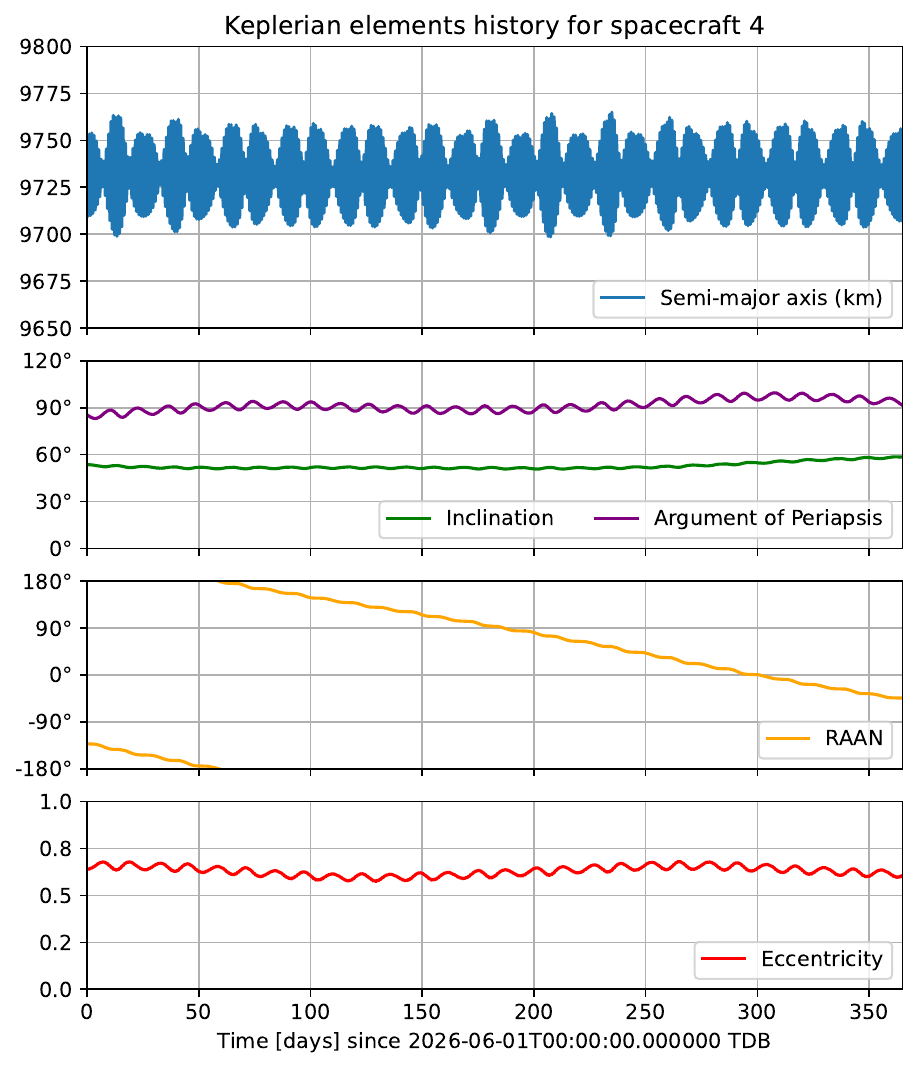}
        \caption{Spacecraft 4}
        \label{fig:kepler4}
    \end{subfigure}

    \caption{Keplerian-element evolution for all four spacecraft. At each propagation step we extract the \emph{osculating} elements from the Cartesian state and track their time histories over one year. As expected for the designed ELFOs, the semi-major axis $a$, eccentricity $e$, inclination $i$, and argument of pericenter $\omega$ remain nearly constant (showing only small periodic variations), while the \acrfull{raan} $\Omega$ exhibits a drift.}
    \label{fig:kepler_grid}
\end{figure}

 To verify their stability in our simulation stack using \acrshortpl{esa} \texttt{GODOT v1.11.0}, we implement these orbits and propagate them for one year (for details on the \texttt{GODOT} setup, see below Sec.~\ref{ch:comparing_orbit_equations}). The orbital elements that we expect to stay stable are the semi-major axis $a$, eccentricity $e$, inclination $i$, and the pericenter argument $\omega$. This implicitly allows the \acrfull{raan} $\Omega$ to drift. This is indeed what we observe, see Fig.~\ref{fig:kepler_grid}.

Previous efforts, which incorporated orbital parameters from an alternative source, exhibited an orbit with abrupt, step-like increases in semimajor axis that ultimately caused the propagator to stall and terminate for runs longer than six months. These pre-failure jumps appeared non-physical, pointing to a modelling inconsistency or numerical issue rather than genuine lunar dynamics. However, after applying the improved initial conditions from \ref{tab:orbital_parameters}, these problems do not occur.

\begin{table}[!htb]
\centering
\caption{Orbital parameters at initial epoch of ELFO constellation from \protect\cite{atlas_2023}.}\label{tab:orbital_parameters}
\begin{tabular}{l|cccc|}
\textbf{Orbital parameter} & \multicolumn{1}{c|}{\textbf{SV1}} & \multicolumn{1}{c|}{\textbf{SV2}} & \multicolumn{1}{c|}{\textbf{SV3}} & \textbf{SV4} \\
\hline
Semimajor axis $a$          & \multicolumn{4}{c|}{\SI{9750.73}{km}}  \\%& \multicolumn{2}{c}{\SI{9750.73}{km}}  \\
Eccentricity $e$            & \multicolumn{4}{c|}{0.6383}            \\%& \multicolumn{2}{c}{0.6383}            \\
Pericenter altitude         & \multicolumn{4}{c|}{\SI{3526.84}{km}}  \\%& \multicolumn{2}{c}{\SI{3526.84}{km}}  \\
Apocenter altitude          & \multicolumn{4}{c|}{\SI{15974.62}{km}} \\%& \multicolumn{2}{c}{\SI{15974.62}{km}} \\
\hline
Inclination $i$             & \multicolumn{2}{c|}{61.96$^\circ$}     & \multicolumn{2}{c|}{54.33$^\circ$}     \\
Argument of pericenter $\omega$ & \multicolumn{2}{c|}{121.7$^\circ$}     & \multicolumn{2}{c|}{55.18$^\circ$}     \\
RAAN $\Omega$               & \multicolumn{2}{c|}{59.27$^\circ$}     & \multicolumn{2}{c|}{277.53$^\circ$}    \\
\hline
True anomaly $\nu$          & \multicolumn{1}{c|}{0$^\circ$} & \multicolumn{1}{c|}{118$^\circ$} & \multicolumn{1}{c|}{0$^\circ$} & 123.42$^\circ$    \\
\hline
Initial Epoch               & \multicolumn{4}{c|}{01-June-2026 00:00 TDB} \\
\end{tabular}
\end{table}

\FloatBarrier
\section{Proper time computations using the different formula forms}\label{ch:comparing_orbit_equations}
As noted above, we use \texttt{GODOT v1.11.0} as the orbit-propagation tool, within an \texttt{python 3.12.3} environment. The simulation is initialized from a \texttt{universe.yaml} configuration file that specifies the solar system's dynamical environment and reference frames. For planetary ephemerides we load \texttt{inpop19a} (Chebyshev file \texttt{athul1.tcheb}), with constants taken from a DE431 kernel (\texttt{gm\_de431.tpc}). The global spacetime is set to \acrshort{bcrs}. 
We define a Moon-fixed body frame \texttt{MoonIAU2009} (\acrshort{iau} 2009 lunar orientation) and a non-rotating frame \texttt{MoonIAU2009\_frozen} referenced to epoch \(\mathrm{TDB}=0.0\), alongside an Earth-fixed \acrshort{itrf} frame with the \acrshort{iers}2000 precession-nutation model (nutation series \texttt{nutation2000A.ipf}). The $N$-body gravity model includes the Sun and all major planets including Earth plus the Moon. Spherical-harmonic gravity models are enabled for Earth (degree $2\times2$, \acrshort{itrf} axes, using \texttt{eigen05c\_80\_sha.tab}) and for the Moon (degree $120\times120$, \texttt{MoonIAU2009} axes, using \texttt{jggrx\_0120a\_sha.tab}). Non-gravitational forces (\acrshort{srp}, drag, tides, thermal) are deactivated. Spacecrafts are mapped from \texttt{trajectory-ELFO-SV\#.yaml} files to internal frame objects (e.g., \texttt{<name>\_Spacecraft\_center}) via a prefix name. 

This setup yields a consistent Moon-centric dynamical environment in which lunar mascon effects are represented by the $120\times120$ lunar field, while third-body perturbations from all major planets are provided by the ephemeris.

We propagate the orbits first and perform all clock-related and diagnostic computations afterwards. For the orbit propagation we ingest the four spacecraft \texttt{trajectory.yaml} files, each of which specifies the initial Keplerian  state and a propagation time interval (from 2026-06-01 to 2027-01-01). For each file we construct a \texttt{trajectory} object bound to the same \texttt{universe.yaml} and invoke \texttt{.compute(partials=False)} to numerically propagate the orbit using the integrator settings contained in the trajectory configuration (we use \textit{adams} with an initial step of \SI{0.01}{s}).

\begin{figure}[!htb]
  \centering

  \begin{subfigure}[t]{0.48\linewidth}
    \centering
    \includegraphics[width=\linewidth]{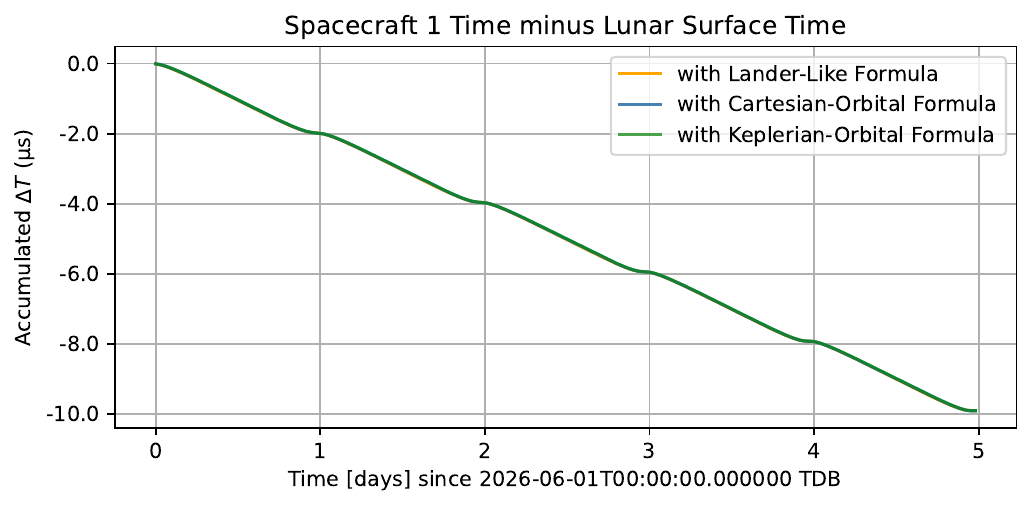}
    \caption{Spacecraft 1}
    \label{fig:orbit_diffT_formulas_sc1}
  \end{subfigure}\hfill
  \begin{subfigure}[t]{0.48\linewidth}
    \centering
    \includegraphics[width=\linewidth]{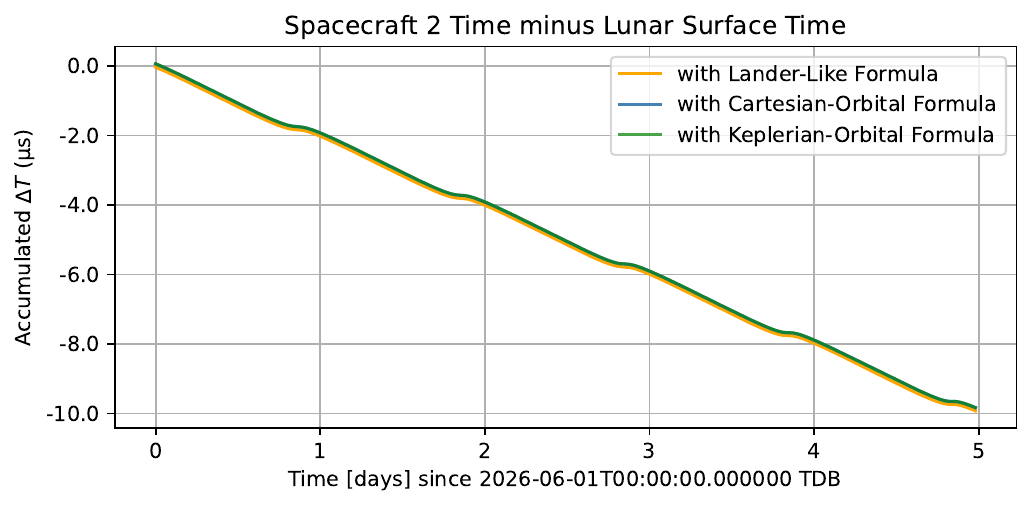}
    \caption{Spacecraft 2}
    \label{fig:orbit_diffT_formulas_sc2}
  \end{subfigure}

  \vspace{0.75em}

  \begin{subfigure}[t]{0.48\linewidth}
    \centering
    \includegraphics[width=\linewidth]{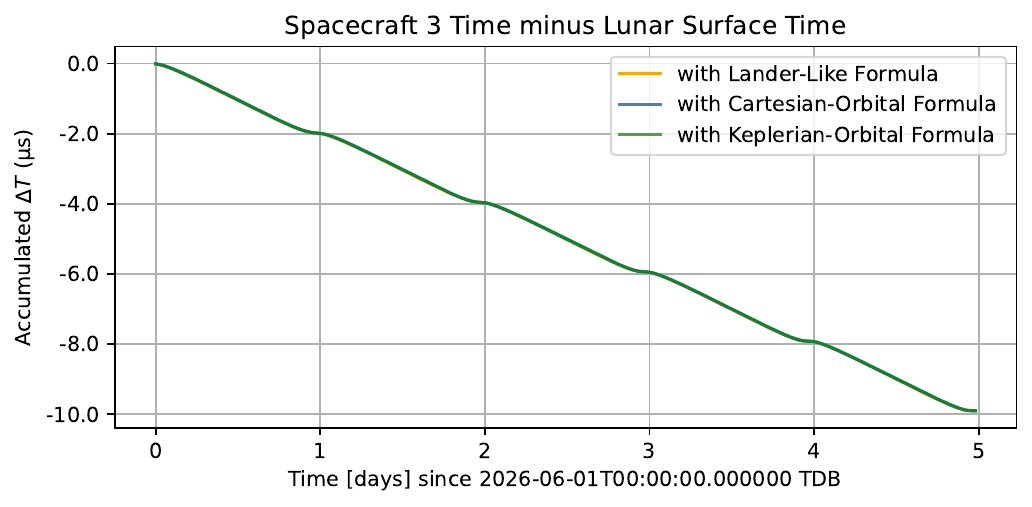}
    \caption{Spacecraft 3}
    \label{fig:orbit_diffT_formulas_sc3}
  \end{subfigure}\hfill
  \begin{subfigure}[t]{0.48\linewidth}
    \centering
    \includegraphics[width=\linewidth]{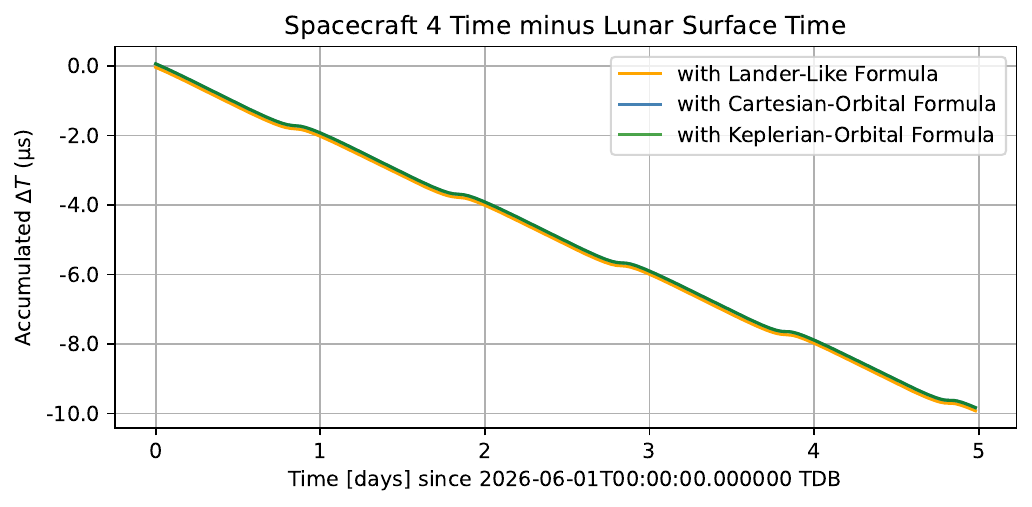}
    \caption{Spacecraft 4}
    \label{fig:orbit_diffT_formulas_sc4}
  \end{subfigure}

  \caption{Accumulated \(\Delta T\) of an orbital \acrshort{elfo} clock, relative to a lunar surface reference clock over five days, evaluated with three formulas.}
  \label{fig:orbit_diffT_formulas}
\end{figure}

\begin{figure}[!htb]
    \centering
    \includegraphics[width=0.75\linewidth]{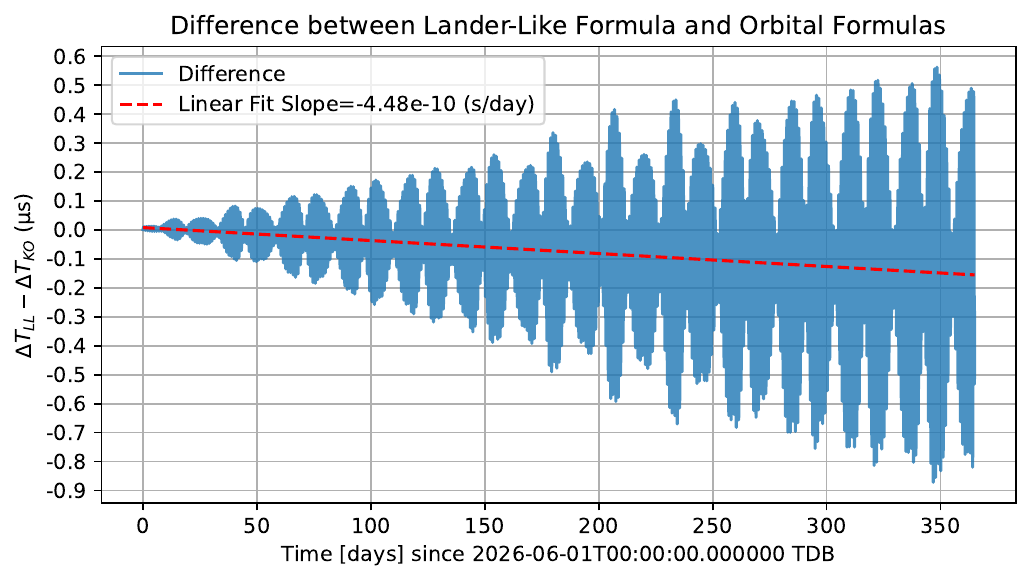}
    \caption{Residual between Lander-Like and the Cartesian/Keplerian-Orbital Formula.}
    \label{fig:dTLLminusKO}
\end{figure}

With the trajectories computed, we proceed to post-processing. We generate a uniform epoch grid with step-size $\Delta t=\SI{0.01}{d}$ (\SI{864}{s}) spanning each trajectory’s start and stop times. At every grid epoch we query the Moon-centered Cartesian state of velocity and position in the rotating frame \texttt{MoonIAU2009} (used in Eq.~\ref{eq:dtau_orbit}), and in the inertial frame \texttt{MoonIAU2009\_frozen}. At each step we compute the momentary Keplerian elements $(a,e,i,\Omega,\omega,\nu)$ from the inertial state vectors and $GM$ using \texttt{GODOT} \texttt{core.astro} module’s \texttt{kepFromCart()}, and then obtain the eccentric anomaly $E$ via \texttt{eccentricFromTrue()} using true anomaly $\nu$ and eccentricity $e$. From these queried properties, we evaluate the three proper-time formulations:
{\setlength{\parskip}{0pt}
\begin{enumerate}[label=\roman*]
    \item The \textit{Lander-Like Formula} accumulates $U(\vec r_i)/c^{2}+\| \vec v_i\|^{2}/(2c^{2})$ times $\Delta t$ for every time step $\Delta t$ on the spacecraft path using states $(\vec r_i,\vec v_i)$ expressed in the inertial lunar frame, and subtracts the result against a surface-clock baseline to obtain the differential time with respect to the lunar surface (see Eq.~\ref{eq:dtau_lander_like}).
    \item The \textit{Cartesian-Orbital Formula} accumulates the already analytically integrated \acrfull{1pn} terms using the instantaneous $(\mathbf r,\mathbf v)$ from the rotating frame (together with the current semi-major axis $a$), and subtracts the result against a surface-clock baseline (see Eq.~\ref{eq:dtau_orbit}). As $\vec r \cdot \vec v$ is a scalar it can be evaluated in either inertial or rotating frames.
    \item The \textit{Keplerian-Orbital Formula} accumulates an analytically approximated version of the \textit{Cartesian-Orbital Formula} in terms of the elements $(a,e,E)$, and then again subtracts the result against a surface-clock baseline (see Eq.~\ref{eq:dtau_kepler}).
\end{enumerate}}
For each spacecraft we store $[\mathrm{\acrshort{mjd}},\,\Delta T_{\text{lander}},\,\Delta T_{\text{cart}},\,\Delta T_{\text{kep}}]$ together with the element histories $[\mathrm{MJD},\,a,e,i,\Omega,\omega,\nu]$. Figure~\ref{fig:orbit_diffT_formulas} presents the outcomes for the initial 5 days of propagation time. As expected, all three formulas display great agreement, in particular are the \textit{Cartesian-Orbital Formula} and the \textit{Keplerian-Orbital Formula} matching; their residual is 0 within our machine precision. The secular drift (visually about \SI{2}{\mu s/day}) relative to the surface reflects the expected gravitational and kinematic frequency redshift along the \acrshort{elfo} trajectories. Only the \textit{Lander-Like Formula} shows a small difference (on the order of $10^{-7}\,\mathrm s$ within the first year, see Fig.~\ref{fig:dTLLminusKO}) to the other two formulas. After around 250 days the residual between the formulas does not exceed \SI{0.7}{\mu s} with an average residual value of \SI{0.1}{\mu s}. The total drift after 250 day is \SI{500}{\mu s}. So this residual is about $10^{-3}$ of the main signal.

\vfill

\vfill

\vfill

\FloatBarrier
As illustrated in Fig.~\ref{fig:orbit_diffT_formulas}, there is an observable mean time-drift between the surface and the orbiter, accompanied by an additional daily periodic oscillation, likely stemming from the eccentric orbit with a mean orbital period of \SI{0.999}{days}. Next, we want to analyze the computed signals further.

\begin{figure}[!htb]
    \centering
    \includegraphics[width=0.75\linewidth]{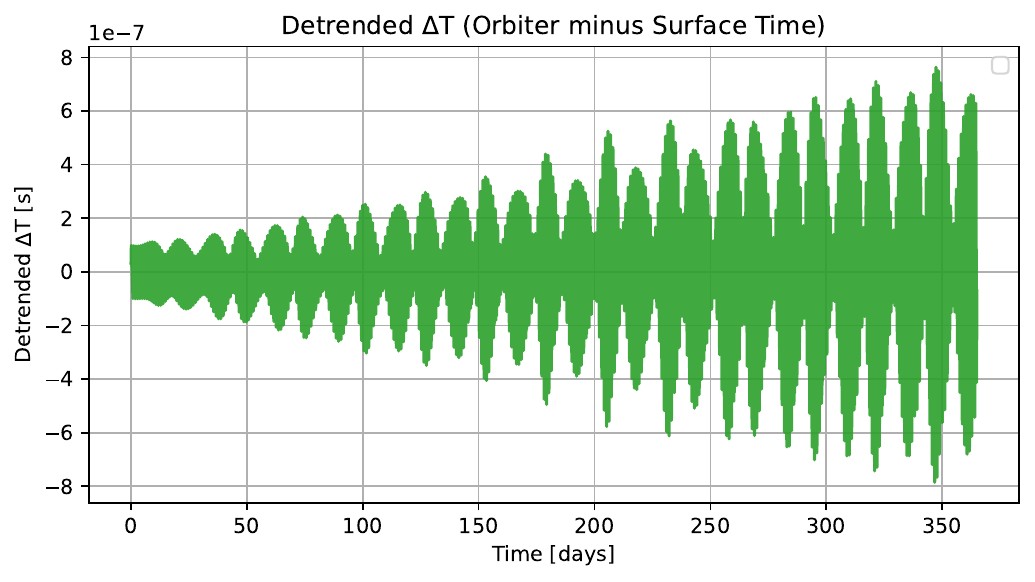}
    \caption{Detrended proper time drift of the Spacecraft 1 \acrshort{elfo} orbiter; obtained from the \textit{Cartesian-Orbit Formula}. Secular drift w.r.t. lunar surface is determined as \SI{-1.987}{\mu s /day}.}
    \label{fig:dt_detrended}
\end{figure}

\begin{figure}[!htb]
    \centering
    \includegraphics[width=0.75\linewidth]{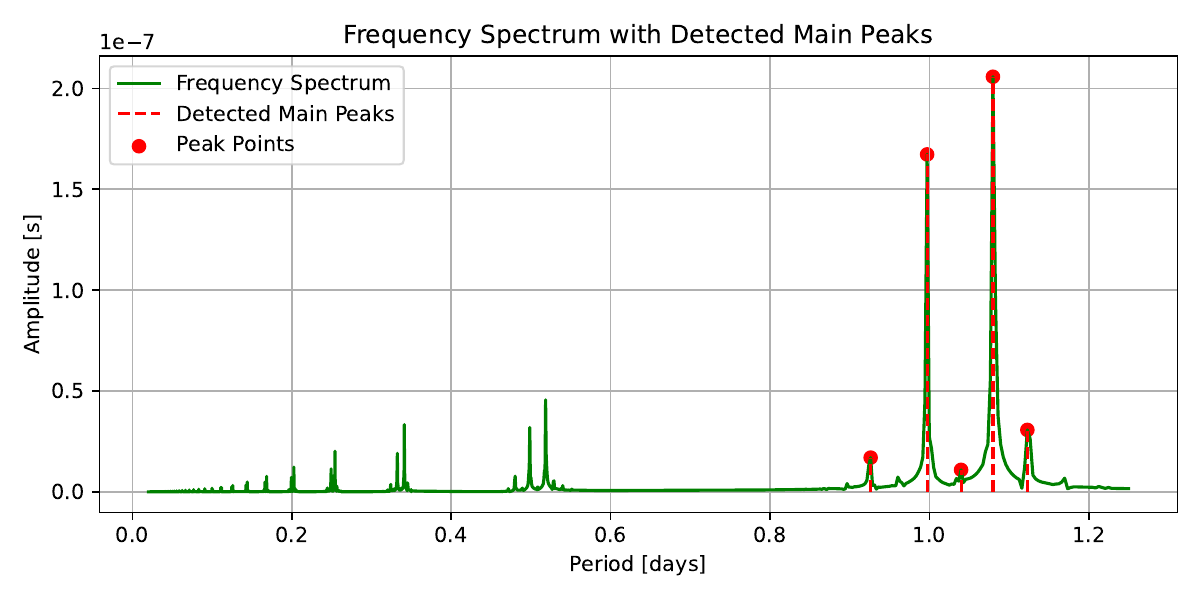}
    \caption{Frequency spectrum of the \acrshort{elfo} orbiter's proper time; obtained from the \textit{Cartesian-Orbit Formula}. Five main peaks near the mean orbital period of \SI{0.999}{days} are identified.}
    \label{fig:dt_periods}
\end{figure}

Conducting a linear fit followed by a frequency analysis on the detrended $\Delta T$ signal of Spacecraft 1 obtained via the \textit{Cartesian-Orbit Formula} results in the plots presented in Fig.~\ref{fig:dt_detrended} and Fig.~\ref{fig:dt_periods}, respectively. The secular drift is determined to be \SI{-1.987}{\mu s/day}. We identify five main peaks centered at $1.123$, $1.080$, $1.040$, $0.997$, and $0.926$ days. The highest amplitudes are \SI{2.06e-07}{} and \SI{1.67e-07}{} seconds.

The first plot reveals harmonics with rising superimposed amplitudes, eerily akin to the residual we examined earlier. This leads us to suspect that problems lie within the evaluation of the \textit{Cartesian-} and \textit{Keplerian-Orbital Formulas}, maybe because both use the semi-major axis $a$ approximated from the inertial state vectors.
\vspace{1em}

\clearpage

\begin{figure}[!htb]
    \centering
    \includegraphics[width=0.75\linewidth]{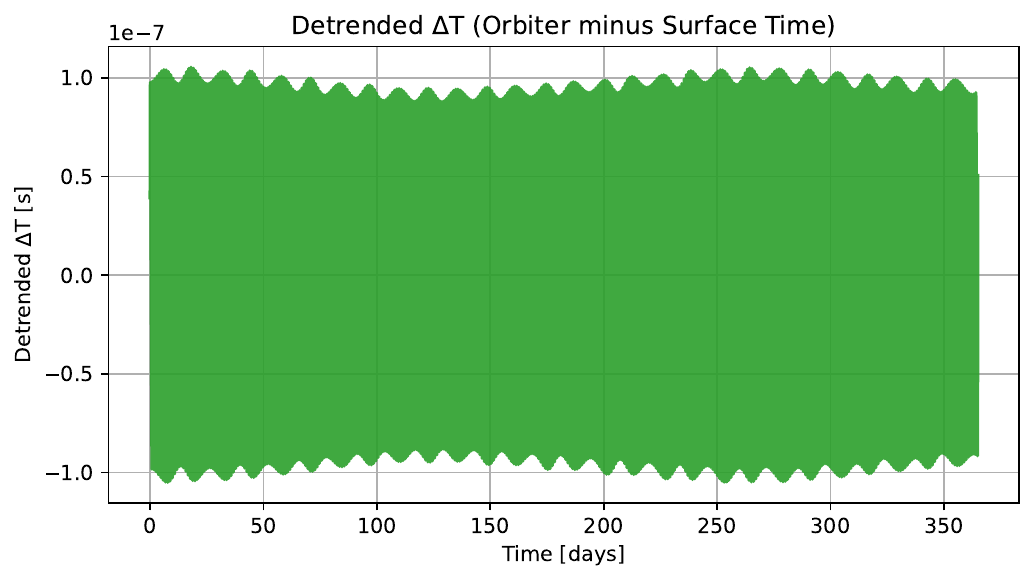}
    \caption{Detrended proper time drift of the Spacecraft 1 ELFO Orbiter; obtained from the \textit{Lander-Like Formula}. Secular drift w.r.t. lunar surface is determined as \SI{-1.987}{\mu s /day}.}
    \label{fig:dt_detrended_LL}
\end{figure}

\begin{figure}[!htb]
    \centering
    \includegraphics[width=0.75\linewidth]{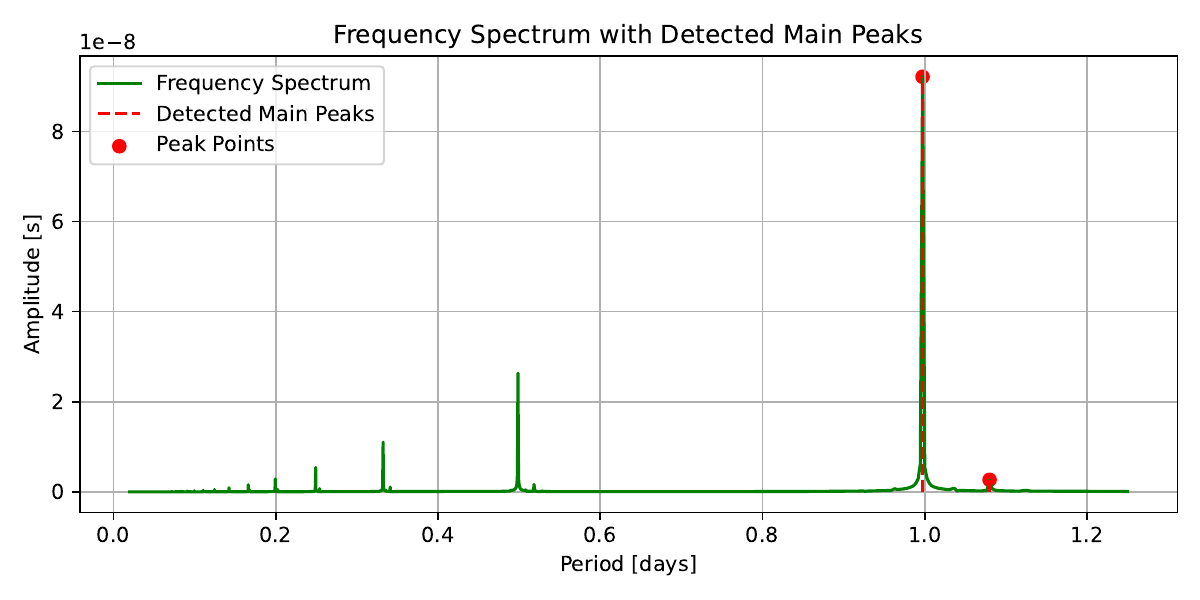}
    \caption{Frequency spectrum of the ELFO orbiters proper time. Two main peaks near the mean orbital period of \SI{0.999}{day} are identified. Copies at half, third, etc. periods.}
    \label{fig:dt_periods_LL}
\end{figure}

We now do the same frequency analysis using the $\Delta T$ signal of Spacecraft 1, but obtained via the \textit{Lander-Like Formula}. We immediately see a smoother detrended signal (see Fig.~\ref{fig:dt_detrended_LL}), with a simpler frequency breakdown (see Fig.~\ref{fig:dt_periods_LL}). We identify two main peaks at periods \SI{1.080}{} and \SI{0.997}{days} and corresponding amplitudes of \SI{2.67e-09}{s} and \SI{9.21e-08}{s}. This matches better with our expectation of \SI{0.999}{days} from the orbital period. For Spacecraft~3, with the \acrshort{elfo} on the other orbital plane, we obtain the same numbers -- just with barely different amplitudes of \SI{2.70e-09}{s} and \SI{9.19e-08}{s}. We are now also able to identify further harmonics with periods of about $28$, $13$ and $9$ days with amplitudes lower than \SI{1}{ns}, see Appendix~\ref{app:elfo_more_harmonics}.


\vspace{1em}
\textbf{We conclude} that our implementation of the \textit{Lander-Like Formula} (Eq.~\ref{eq:dtau_lander_like}) is the best to use for our \acrshort{1pn} proper time simulation implemented using GODOT. The \acrfullpl{elfo} under consideration for lunar navigation satellites exhibits a secular drift of \SI{-1.986}{\mu s/day}, meaning the orbiter's clock runs faster w.r.t. the lunar surface. Periodic variations from the secular drift do not exceed \SI{0.1}{\mu s} and are quite predictably tied to the orbital period of \SI{1}{day}. Harmonic contributions, likely from external gravitational perturbations, of less than \SI{1}{ns} magnitude are identified, but can be ignored because of their minimal effect.

\FloatBarrier
\section{Remarks and outlook on further analysis}\label{ch:orbits_further_comments}
Although the final analysis flow may appear straightforward in hindsight, arriving there required resolving several pitfalls. The technical difficulties and the lessons learnt were:{\setlength{\parskip}{0pt}
\begin{itemize}
  \item \textit{Interpretation of the $\,\vec r\!\cdot\!\vec v\,$ terms and Sagnac:} Early on we confused the kinematic term arising in proper–time/relativistic frequency models with the Sagnac contribution from rotating frames and signal paths. Clarifying their distinct origins and frame and vector dependencies was essential to obtain consistent formulas and to decide when (and where) to include Sagnac corrections.
  \item \textit{Misleading rover equation:} A rover clock equation in \cite{geoazur} proved ambiguous. Re-deriving the expression within our frame conventions avoided mistakes.
  \item \textit{Frame of state vector queries:} Osculating elements and redshift terms must be built from an inertial non-rotating frame. Querying the Moon-fixed frame (\texttt{MoonIAU2009}) for the state-vectors led to huge differences in the formulas; switching to the non-rotating \texttt{MoonIAU2009\_frozen} for orbital element calculation resolved this.
  \item \textit{Orbit propagation robustness:} As already mentioned above, early trajectory propagations timed-out due to suboptimal initial conditions, manifesting as nonphysical step-like jumps in the semi-major axis. Refining the initial states restored convergence.
  \item \textit{\texttt{GODOT} configuration and workflow:} Packing four spacecraft into a single trajectory configuration file did not seem supported, despite efforts. The workaround was to keep one spacecraft per file and use ID/name tagging across the universe mapping. We also observed inconsistent runtimes on a shared server; migrating to a locally installed and updated \texttt{GODOT} (from the official repositories) improved execution.
\end{itemize}
}

As for the directions that future work might go, we can envision the following:
{\setlength{\parskip}{0pt}
\begin{itemize}
  \item \textit{Moon orientation parameters:} Time-varying orientation models can, in principle, feed back into orbital evolution (via the gravity-field orientation) and into proper-time differences. A systematic study should compare otherwise identical runs that differ only in the lunar orientation file and quantify the induced changes in both orbit and $\Delta T$. An inverse problem -- testing whether clock comparisons along lunar orbits constrain the moon orientation parameters -- may be possible, though signals might be too low. But this should be confirmed.
  \item \textit{Comparison with other orbit propagation tools:} For example one could do the same calculations with the \acrshort{xhps} tool from Bremen, as soon as the updated clock-module is available by autumn.
  \item \textit{Longer simulation runs and non-gravitational influences:} Longer simulation runs can be used to better fit and understand the sub \SI{1}{ns/day} harmonics we saw in the \acrfull{elfo}. In our simulation, we also did not consider the kinematic effects of non-gravitational perturbation forces, like \acrfull{srp}, or satellites with control laws for station-keeping. For some time precisions, accounting for these effects might be relevant.
\end{itemize}
}
There could also be an additional relativistic consideration, as for Earth \acrshort{gnss} receivers, targeting cm-level positioning, a higher-order relativistic correction called the \textit{Shapiro delay} is non-negligible in the range model. A radial path in curved spacetime is longer than a flat-space-time would suggest; similar to how the radial coordinate in the Schwarzschild metric does not reflect the distance to the center. Though this is probably more relevant for range calculations, rather than time scales.
\chapter{Summary} \label{ch:conclusion}

In this thesis we have explored and implemented the theoretical foundations of a relativistic lunar time-scale framework at \acrfull{1pn}; compatible with the \acrfullinv{iau} recommendations on solar system reference frames. This is in light of near-term lunar infrastructure projects led by \acrshort{nasa} and \acrshort{esa}, aiming to establish a robust and reliable \acrfull{pnt} architecture for the Moon. Because precise timing underpins navigation, ranging, and scientific experiments, and because the \acrshort{iau} has not yet adopted an international lunar standard, this time-scale framework is essential to explore.

We adopt a layered methodology, building the "layers of time" from the foundations outward. First, we define the \acrfull{lcrs} and its associated coordinate time \acrfullinv{tcl}, compare evaluation methods and their numerical results from the literature, and show their mutual consistency. Next, we move from coordinate time to surface timing by quantifying and visualizing how the Moon's gravitational field and topography affect stationary clocks, via (to the best of our knowledge) not seen before 3D gravitational redshift maps\footnote{made available interactively at \scriptsize{\url{https://yanseyffert.github.io/MASS_Thesis_LunarTime/}}}. Further, we assess detectability with modern clock technology. Finally, we simulate moving and orbital clocks, focusing on stable \acrfull{elfo} envisioned for lunar navigation and communication systems like \acrshort{esa}'s Moonlight constellation.

We find and confirm that \acrshort{tcl} drifts, when compared to the solar system barycentric \acrshort{tcb}, at a secular drift rate of \SI{1280.8}{\mu s/day}; \SI{1.48}{\mu s/day} faster than Earth's coordinate time scale. Drifting faster means ticking slower. At this step, we observe modulations/harmonics of up to \SI{0.5}{\mu s/day} when compared to Earth's coordinate time \acrfullinv{tcg}. We confirm that the Lunar Surface time drifts faster by an additional \SI{2.72}{\mu s/day}. Depending on local elevation this varies by about \SI{\pm 15}{ns/day}. Clocks onboard \acrshort{elfo} satellites are found to run faster than surface clocks; they drift less by \SI{-1.99}{\mu s/day}. Harmonic variations, due to the orbit's eccentricity, do not exceed \SI{0.1}{\mu s/day}. Compared to the time on Earth's surface/geoid, all cislunar time-scales run faster -- with a drift rate ranging from \SI{-56.03}{\mu s/day} to \SI{-58.74}{\mu s/day}, best illustrated in Fig.~\ref{fig:secular_drifts}.

For an orbiter's proper time, we obtained the most reliable simulation results by integrating the \acrfull{1pn} terms with a fixed-step Riemann sum, rather than summing the pre-integrated closed-form expressions. In doing so, we learned a lot about ESA's \acrfullinv{godot}.

Through this theoretical analysis, cross-checks against published results, and dedicated numerical simulations, this thesis aims to support the ongoing development of a reliable, relativistically sound lunar timekeeping framework -- essential for deployment of Lunar \acrfullinv{pnt} services in support of the near-future exploration and settlement of the Moon.

\clearpage
\addcontentsline{toc}{chapter}{Acknowledgments}
\chapter*{Acknowledgments} \label{Ack}

First and foremost, I thank my supervisors for their guidance, patience, and steady encouragement throughout this Master's thesis; and the possibility to work on such an interesting and important topic.

I am deeply grateful to everyone who enriched my time in the MASS program. To the friends, acquaintances, and colleagues I met along the way in Rome, Bremen, and Nice, thank you for the conversations, the adventures, and the camaraderie.

Last but not least, my heartfelt thanks go to my father for his unconditional love and unwavering support throughout my life and every endeavour I have chosen to pursue. Thank you for always being there for me.

\vspace*{\fill}
\begin{large}
Yan Seyffert acknowledges support through an "Erasmus Mundus Joint Master
(EMJM)" scholarship Co-funded by the European Union in the framework of the Erasmus+, Erasmus Mundus Joint Master in Astrophysics and Space Science -- MASS. Views and opinions expressed are however those of the author(s) only and do not necessarily reflect those of the European Union or granting authority European Education and Culture Executive Agency (EACEA). Neither the European Union nor the granting authority can be held responsible for them.
\end{large}
\vspace*{\fill}
%% --------------------
%% |   Bibliography   |
%% --------------------
\cleardoublepage
\addcontentsline{toc}{chapter}{References}
\printbibliography

%% -------------------------------------------------
%% |    Acronym definitons  Appendix & Glossary   |
%% -------------------------------------------------

\clearpage
%GPS
\newacronym{gnss}{GNSS}{Global Navigation Satellite System}
\newacronym{gps}{GPS}{Global Positioning System}
\newacronym{gdop}{GDOP}{Geometric Dilution of Precision}

% Orginisations
\newacronym{nasa}{NASA}{National Aeronautics and Space Administration}
\newacronym{jpl}{JPL}{Jet Propulsion Laboratory}
\newacronym{esa}{ESA}{European Space Agency}
\newacronym{jaxa}{JAXA}{Japan Aerospace Exploration Agency}
\newacronym{iau}{IAU}{International Astronomical Union}
\newacronym{iugg}{IUGG}{International Union of Geodesy and Geophysics}
\newacronym{bipm}{BIPM}{Bureau International des Poids et Mesures}
\newacronym{cgpm}{CGPM}{General Conference on Weights and Measures}
\newacronym{cgm}{CIPM}{International Committee for Weights and Measures}
\newacronym{lne-sytre}{LNE-SYRTE}{Laboratoire National de Métrologie et d'Essais – Systèmes de Référence Temps-Espace}
\newacronym{iers}{IERS}{International Earth Rotation and Reference Systems Service}
\newacronym{itur}{ITU-R}{ternational Telecommunication Union Radiocommunication Sector}
\newacronym{esoc}{ESOC}{European Space Operations Centre}
\newacronym{nmi}{NMI}{National Metrology Institute}
\newacronym{ptb}{PTB}{Physikalisch Technische Bundesanstalt}
\newacronym{nist}{NIST}{National Institute of Standards and Technology}
\newacronym{usno}{USNO}{United States Naval Observatory}
\newacronym{iaa_ras}{IAA RAS}{Institute of Applied Astronomy of the Russian Academy of Sciences}
\newacronym{imcce}{IMCCE}{Institut de Mécanique Céleste et de Calcul des Èphémérides}
\newacronym{oca}{OCA}{C\^ote d'Azur Observatory}
\newacronym{zarm}{ZARM}{Zentrum für angewandte Raumfahrttechnologie und Mikrogravitation}

%Systems
\newacronym{icrs}{ICRS}{International Celestial Reference System}
\newacronym{bcrs}{BCRS}{Barycentric Celestial Reference System}
\newacronym{gcrs}{GCRS}{Geocentric Celestial Reference System}
\newacronym{icrf}{ICRF}{International Celestial Reference Frame}
\newacronym{lcrs}{LCRS}{Lunar Celestial Reference System}
\newacronym{emcrs}{EMCRS}{Earth-Moon coordinate reference system (EMCRS)}
\newacronym{pa}{PA}{principal axis}
\newacronym{itrf}{ITRF}{International Terrestrial Reference System}

%Time
\newacronym{jd}{JD}{Julian Date}
\newacronym{mjd}{MJD}{Modified Julian Date}
\newacronym{gmt}{GMT}{Greenwich Mean Time}
\newacronym{et}{ET}{Ephemeris Time}
\newacronym{ut}{UT}{Universal Time}
\newacronym{lt}{LT}{Lunar Time}
\newacronym{utc}{UTC}{Coordinated Universal Time}
\newacronym{tai}{TAI}{International Atomic Time}
\newacronym{tt}{TT}{Terrestrial Time}
\newacronym{tcb}{TCB}{Barycentric Coordinate Time}
\newacronym{tcg}{TCG}{Geocentric Coordinate Time}
\newacronym{tdb}{TDB}{Barycentric Dynamical Time}
\newacronym{tcl}{TCL}{Lunar Coordinate Time}
\newacronym{eal}{EAL}{Échelle Atomique Libre}
\newacronym{twstft}{TW-STFT}{two-way satellite time and frequency transfer}
\newacronym{gnss_cv}{GNSS CV}{Common-View}
\newacronym{gpst}{GPST}{GPS system time}

%Other
\newacronym{ipta}{IPTA}{International Pulsar Timing Array}
\newacronym{elfo}{ELFO}{Elliptical Lunar Frozen Orbit}
\newacronym{ceo}{CEO}{Celestial Ephemeris Origin}
\newacronym{pnt}{PNT}{Positioning, Navigation and Timing}
\newacronym{vlbi}{VLBI}{Very Long Baseline Interferometry}
\newacronym{llr}{LLR}{Lunar Laser Ranging}
\newacronym{ssb}{SSB}{solar system barycenter}
\newacronym{pfs}{PFS}{primary frequency standard}
\newacronym{1pn}{1PN}{1st post-Newtonian order}
\newacronym{2pn}{2PN}{2nd post-Newtonian order}
\newacronym{ppn}{PPN}{parameterized post-Newtonian formalism}
\newacronym{au}{AU}{astronomical unit}
\newacronym{mba}{MBA}{main-belt asteroids}
\newacronym{tno}{TNO}{Trans-Neptunian Objects}
\newacronym{godot}{GODOT}{General Orbit Determination and Optimisation Toolkit}
\newacronym{xhps}{XHPS}{eXtended High Performance satellite dynamics Simulator}
\newacronym{eihdl}{EIHDL}{Einstein–Infeld-Hoffman-Droste–Lorentz Equation}
\newacronym{adev}{ADEV}{Allan deviation}
\newacronym{tcxo}{TCXO}{temperature-compensated crystal oscillator}
\newacronym{ocxo}{OCXO}{oven-controlled crystal oscillator}
\newacronym{fft}{FFT}{Fast Fourier Transformation}
\newacronym{lro}{LRO}{Lunar Reconnaissance Orbiter}
\newacronym{lola}{LOLA}{Lunar Orbiter Laser Altimeter}
\newacronym{grail}{GRAIL}{Gravity Recovery and Interior Laboratory}
\newacronym{lugre}{LuGRE}{Lunar GNSS Receiver Experiment}
\newacronym{lcross}{LCROSS}{Lunar Crater Observation and Sensing Satellite}
\newacronym{dtn}{DTN}{Delay-Tolerant Networking}
\newacronym{afs}{AFS}{Augmented Forward Signals}
\newacronym{lunasar}{LunaSAR}{lunar search-and-rescue}
\newacronym{clps}{CLPS}{Commercial Lunar Payload Services}
\newacronym{gr}{GR}{General Relativity}
\newacronym{ntp}{NTP}{Network Time Protocol}
\newacronym{wsl}{WSL}{Windows Subsystem for Linux}
\newacronym{sv}{SV}{satellite vehicle}
\newacronym{meo}{MEO}{Medium Earth Orbit}
\newacronym{raan}{RAAN}{Right Ascension of the Ascending Node}
\newacronym{srp}{SRP}{Solar Radiation Pressure}
\newacronym{nrho}{NRHO}{near-rectilinear halo orbits}
\printglossary[type=\acronymtype]

\clearpage
%% ----------------------------------
%% |  Style of appendix numbering   |
%% ----------------------------------
% \newcounter{lettersection}
% \renewcommand{\thelettersection}{\Alph{lettersection}}
% \newcommand{\lettersection}[1]{%
%   \refstepcounter{lettersection}%
%   \subsection*{\thelettersection. #1}%
%   \addcontentsline{toc}{subsection}{\thelettersection. #1}%
% }
% \setcounter{section}{0}%
% \setcounter{subsection}{0}%
% \setcounter{figure}{0}%
% \renewcommand\thesection{\Alph{subsection}}%
% \renewcommand\thefigure{\Alph{lettersection}.\arabic{figure}}
% \renewcommand\thetable{\Alph{lettersection}.\arabic{table}}
%% --- End of appenix numbering style ---

\addcontentsline{toc}{chapter}{Appendix}
\appendix
%% --- End of appenix numbering style ---

%\section*{Appendix}

\FloatBarrier
\chapter*{Appendix}

\section{Relative Frequency Drift around Moon's South Pole}\label{app:moon_ortho_soud}
\vfill
\begin{figure}[!htb]
    \centering
    \includegraphics[width=0.8\linewidth]{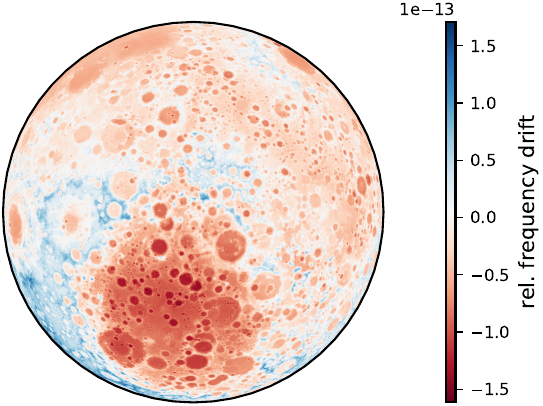}
    \caption{Orthographic projection with the Moon's south pole at the center. The Earth-facing side (longitude 0$^\circ$) points up in this plot. Drift is w.r.t. the selenoid, as defined in Sec.~\ref{ch:red_shift_maps}.}
    \label{fig:moon_soutpole_othographic}
\end{figure}
\vfill

\FloatBarrier
\newpage
\section{Comparing lunar gravitational red-shift result with literature}\label{app:moon}

\begin{figure}[!htb]
    \centering
    \includegraphics[width=0.8\linewidth]{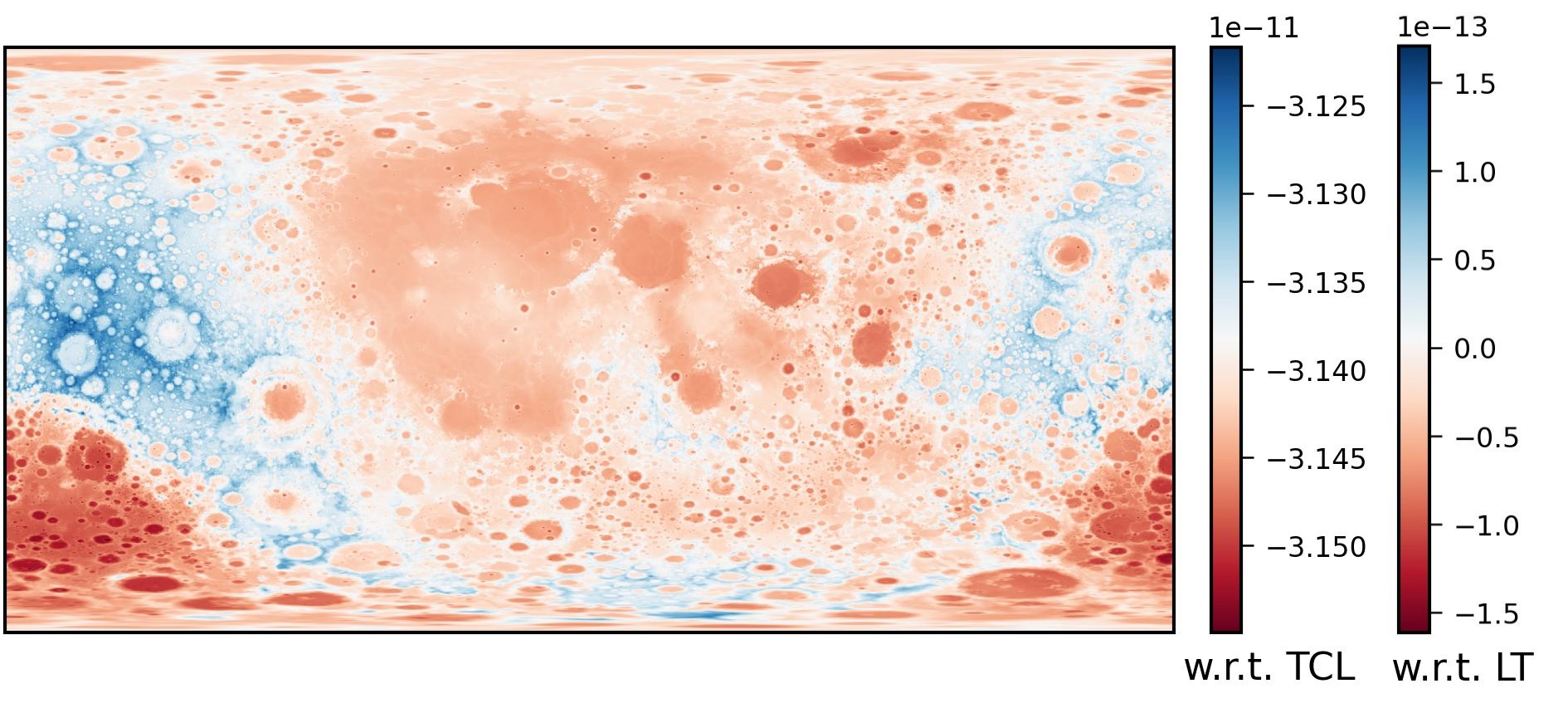}
    \caption{Map of gravitational time-dilation of lunar surface, rendered such that it is directly comparable to the map by Bourgoin et al. (Fig.~\ref{fig:moon_compared_to}). LT is a rescaled TCL for clocks on a selenoid as defined in Sec.~\ref{ch:red_shift_maps}.}
    \label{fig:moon_for_comparison}
\end{figure}

\begin{figure}[!htb]
    \centering
    \includegraphics[width=0.96\linewidth]{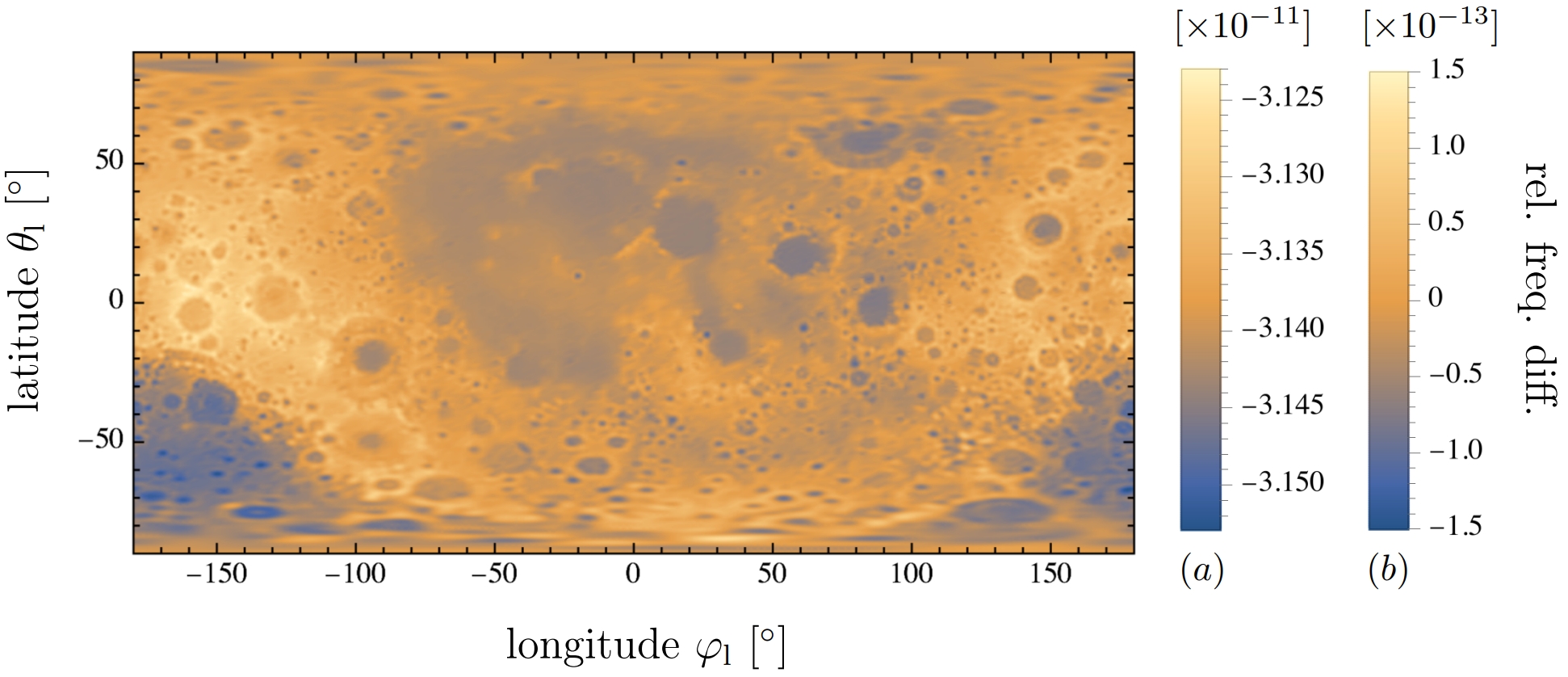}
    \caption{Map of relative frequency difference between a stationary clock on the lunar surface and \textit{(a)} TCL and \textit{(b)} a rescaled TCL. From \cite{Bourgoin2025}. Used for comparison with Fig.~\ref{fig:moon_for_comparison}.}
    \label{fig:moon_compared_to}
\end{figure}

Putting our findings into relative terms with respect to \acrshort{tcl}, we find a mean contribution of \SI{3.1383e-11}{} (\SI{-2.712}{ns/day}) with a symmetric variation of \SI{\pm 1.66e-13}{} (\SI{\pm14.4}{ns/day}), so a maximum difference of \SI{28.7}{ns/day} between lowest and highest locations. Bourgoin et al.\cite{Bourgoin2025} found a main contribution from the monopole of the lunar gravity potential at the level of \SI{3.14e-11}{} and variations due to terms relating to altitude on the order of \SI{1.62e-13}{}. Terms due to the second-degree lunar gravity potential affect by \SI{7.92e-15}{} and terms related to the centrifugal potential contribute with \SI{1.19e-16}{}. In their Appendix they mention amplitude variations reaching \SI{\pm 1.60e-13}{} (\SI{\pm13.8}{ns/day}). There seems to be some variation in the third significant figure they mention (on the $10^{-15}$ scale), but even with the conservative \SI{\pm 1.60e-13}{} figure, we get a maximum difference of \SI{27.6}{ns/day}. Our results thus agree on the \SI{1}{ns/day} (\SI{\pm5e-15}{}) scale. Fig.~\ref{fig:moon_for_comparison} and \ref{fig:moon_compared_to} also match well.
\vspace{2em}

\newpage
\FloatBarrier
\section{Plots of Earth topography and gravity field data}\label{app:earth}

\vspace{2cm}
\begin{figure}[!htb]
    \centering
    \includegraphics[width=0.7\linewidth]{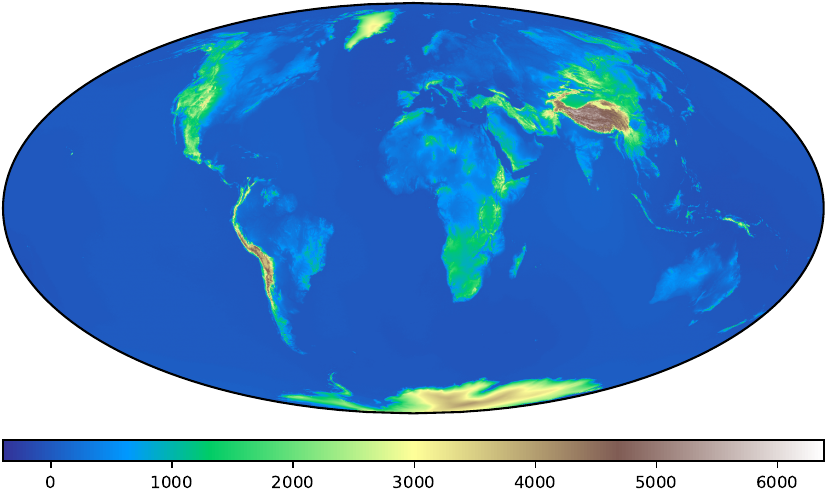}
    \caption{Earth's Topography (including water and ice) with respect to mean sea level, fetched from the Earth2014\cite{earth2014} spherical harmonic model.}
    \label{fig:earth_topography}
\end{figure}
\vfill
\begin{figure}[!htb]
    \centering
    \includegraphics[width=0.7\linewidth]{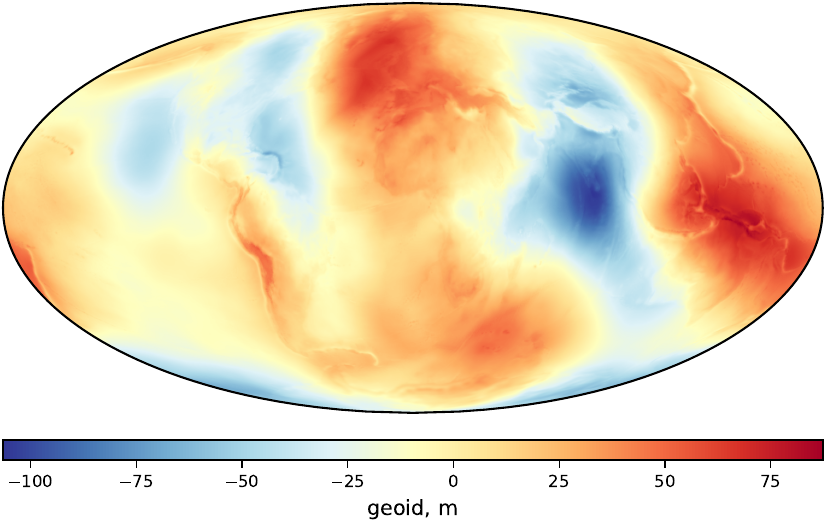}
    \caption{Earth's Geoid (with defining reference potential $u_0=62636852\;\mathrm{m^2/s^2}$) displays how much this equipotential surface deviates from the WGS84 reference ellipsoid. Obtained from the EGM2008\cite{egm2008} Earth gravity model, bases data from altimetry, ground-based measurements, and the satellite GRACE. }
    \label{fig:earth_geoid}
\end{figure}
\vfill

\begin{figure}[!htb]
    \centering
    \includegraphics[width=0.75\linewidth]{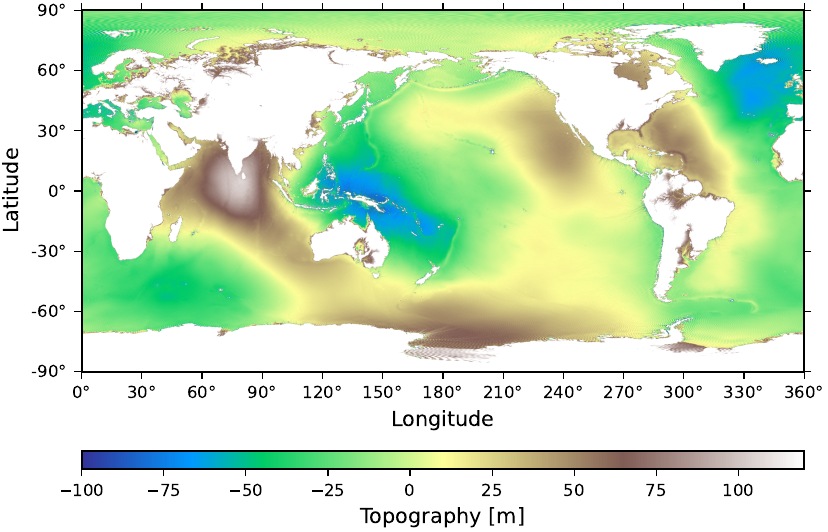}
    \caption{Global map of Earth’s sea level. The surface is referenced to mean sea level (MSL) as provided by the Earth2014 dataset. Deviations reflect variations in the ocean surface relative to MSL. The Indian Ocean and Western Pacific anomalies are visible, indicating regions where sea level lies above or below the geoid, due to local sub-surface density variations.}
    \label{fig:sealevel_nonzero}
\end{figure}

\begin{figure}[!htb]
    \centering
    \includegraphics[width=0.65\linewidth]{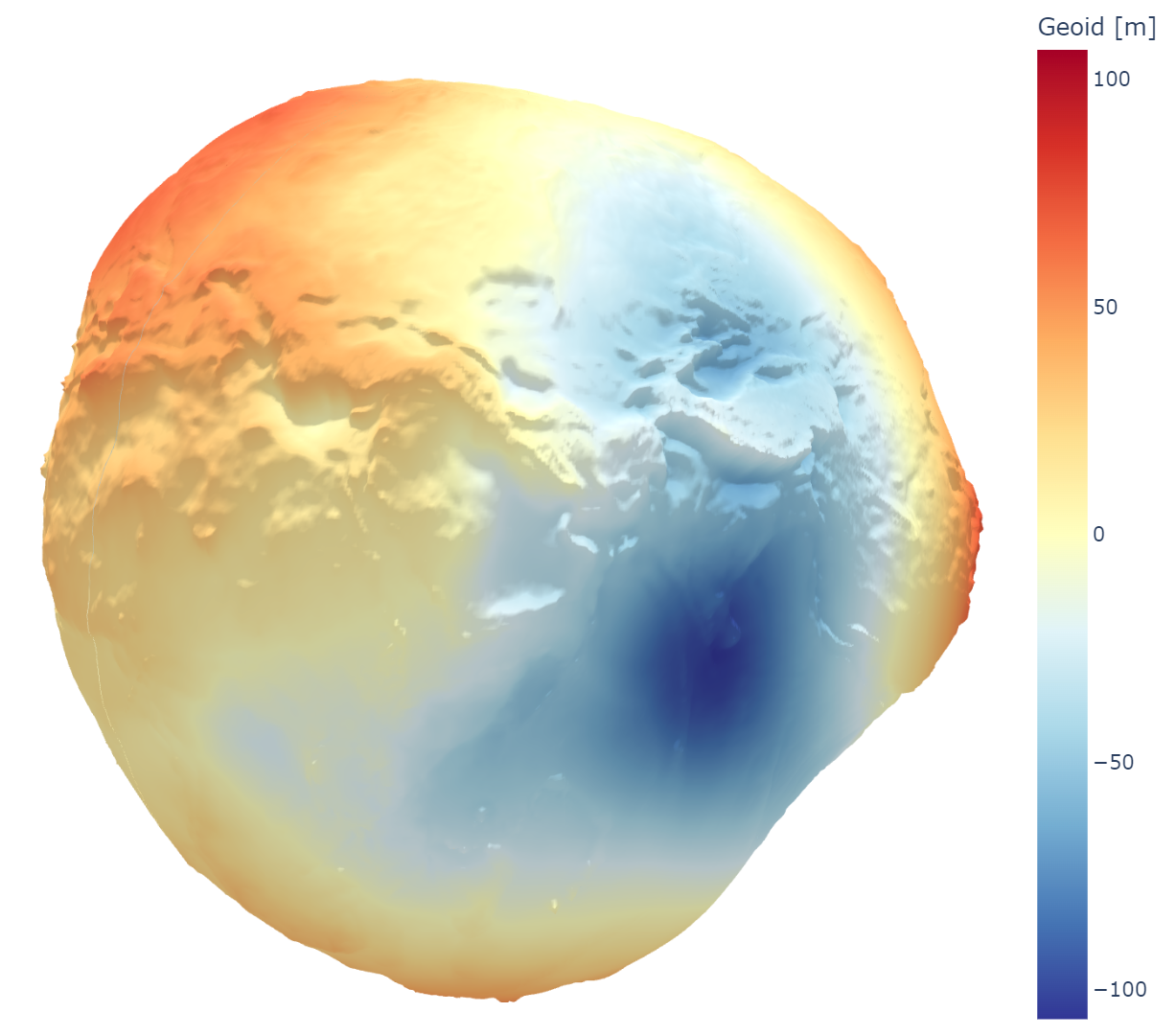}
    \caption{3D Representation of Earth's geoid. Height features are exaggerated by a factor of \SI{10000}{}. This representation (first such obtained by the  Research Centre for Geosciences in Potsdam, Germany) shows an extremely lumpy, uneven surface -- resembling the shape of a potato rather than a smooth sphere or ellipsoid. This earned this plot the nickname "Potsdam Potato". An interactive version is shared at  \scriptsize{\url{https://yanseyffert.github.io/MASS_Thesis_LunarTime/}.}}
    \label{fig:earth_potsdam_potato}
\end{figure}

\clearpage
\FloatBarrier
\section{Additional harmonics found for the proper time of an ELFO orbiter}\label{app:elfo_more_harmonics}
\vfill
\begin{figure}[!htb]
    \centering
    \includegraphics[width=0.85\linewidth]{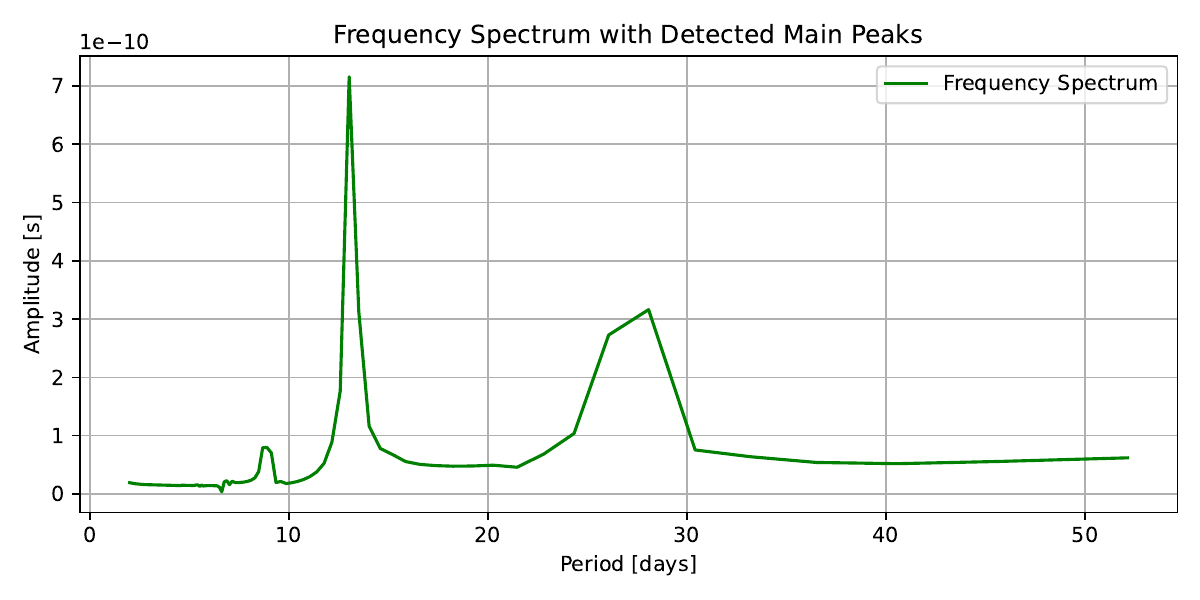}
    \caption{Frequency spectrum of the \acrshort{elfo} orbiter's proper time for periods longer than 2 days. A few more harmonics with periods of roughly 28, 13, 9 days can be identified.}
    \label{fig:dt_periods_LL_extra}
\end{figure}
\vfill

\end{document}